\documentclass[useAMS,usenatbib]{mn2e}
\usepackage{longtable,lscape}
\usepackage{amsmath}
\usepackage[draft]{hyperref}

\usepackage{graphicx}
\usepackage{aas_macros}
\usepackage{amssymb}

\newcommand{\bvri}{\protect\hbox{$BV\!RI$} }

\newcommand{\about}{$\sim\!\!$~}

\newcommand{\be}{\begin{displaymath}}
\newcommand{\ee}{\end{displaymath}}

\def\lsim{\hbox{\rlap{\raise 0.425ex\hbox{$<$}}\lower 0.65ex\hbox{$\sim$}}}
\def\gsim{\hbox{\rlap{\raise 0.425ex\hbox{$>$}}\lower 0.65ex\hbox{$\sim$}}}

\def\arcsec{\hbox{$^{\prime\prime}$}}

\newcommand{\etal}{et~al.\ }

\newcommand{\kms}{km~s$^{-1}$}

\newcommand{\dof}{\rm dof}

\newcommand{\ion}[2]{#1$\;${\small{#2}}\relax}

\newcommand{\totspec}{1298}
\newcommand{\totobj}{582}
\newcommand{\unpubspec}{1159}
\newcommand{\unpubobj}{563}
\newcommand{\phasespec}{914}
\newcommand{\phaseobj}{321}
\newcommand{\lcspec}{770}
\newcommand{\lcobj}{251}
\newcommand{\filtspec}{584}
\newcommand{\filtobj}{199}
\newcommand{\finalspec}{234}
\newcommand{\finalobj}{95}
\newcommand{\mathspec}{\about3} 
\newcommand{\mathobj}{\about18} 
\newcommand{\specperobj}{\about2.2}
\newcommand{\alexnights}{254}

\graphicspath{{/Users/jsilv/BSNIP/lowz_data/}}

\title[BSNIP I: SN~Ia Spectra]{Berkeley Supernova Ia Program I:
  Observations, Data Reduction, and Spectroscopic Sample of \totobj\
  Low-Redshift Type Ia Supernovae} 

\author[Silverman, et~al.]{Jeffrey~M.~Silverman,$^{1,2}$
Ryan~J.~Foley,$^{3,4}$ Alexei~V.~Filippenko,$^{1}$
\newauthor
Mohan~Ganeshalingam,$^{1}$ Aaron~J.~Barth,$^{5}$ Ryan~Chornock,$^{3}$ Christopher~V.~Griffith,$^{1,6}$
\newauthor
Jason~J.~Kong,$^{1}$ Nicholas~Lee,$^{7}$ Douglas~C.~Leonard,$^{8}$ Thomas~Matheson,$^{9}$
\newauthor
Emily~G.~Miller,$^{10}$  Thea~N.~Steele,$^{1,11}$ Brian~J.~Barris,$^{7}$ Joshua~S.~Bloom,$^{1}$
\newauthor
Bethany~E.~Cobb,$^{12}$ Alison~L.~Coil,$^{13}$ Louis-Benoit~Desroches,$^{1,14}$ Elinor~L.~Gates,$^{15}$
\newauthor
Luis~C.~Ho,$^{16}$ Saurabh~W.~Jha,$^{17}$ Michael~T.~Kandrashoff,$^{1}$ Weidong~Li,$^{1}$\dag
\newauthor
Kaisey~S.~Mandel,$^{3}$ Maryam~Modjaz,$^{1,18}$ Matthew~R.~Moore,$^{1}$ Robin~E.~Mostardi,$^{1,19}$ 
\newauthor
Marina~S.~Papenkova,$^{20}$ Sung~Park,$^{1}$ Daniel~A.~Perley,$^{1,21}$ Dovi~Poznanski,$^{1,22}$ 
\newauthor
Cassie~A.~Reuter,$^{1,23}$ James~Scala,$^{1}$ Franklin~J.~D.~Serduke,$^{1}$ Joseph~C.~Shields,$^{24}$
\newauthor
Brandon~J.~Swift,$^{25}$ John~L.~Tonry,$^{7}$ Schuyler~D.~Van~Dyk,$^{26}$ Xiaofeng~Wang,$^{27}$
\newauthor
Diane~S.~Wong$^{1}$ \\
$^{1}$Department of Astronomy, University of California, Berkeley, CA 94720-3411, USA \\
$^{2}$Marc J. Staley Fellow; email JSilverman@astro.berkeley.edu \\
$^{3}$Harvard-Smithsonian Center for Astrophysics, 60 Garden Street, Cambridge, MA 02138, USA \\
$^{4}$Clay Fellow \\
$^{5}$Department of Physics and Astronomy, 4129 Frederick Reines Hall, University of California, Irvine, CA 92697, USA \\
$^{6}$Department of Astronomy and Astrophysics, The Pennsylvania State University, University Park, PA 16802, USA \\
$^{7}$Institute for Astronomy, University of Hawaii, 2680 Woodlawn Drive, Honolulu, HI 96822, USA \\
$^{8}$Department of Astronomy, San Diego State University, San Diego, CA 92182-1221, USA \\
$^{9}$National Optical Astronomy Observatory, 950 North Cherry Avenue, Tucson, AZ 85719-4933, USA \\
$^{10}$University of Pennsylvania, 3451 Walnut Street, Philadelphia, PA 19104, USA \\
$^{11}$Department of Computer Science, Kutztown University of Pennsylvania, Kutztown, Pennsylvania 19530, USA \\
$^{12}$Department of Physics, The George Washington University, Washington, DC 20052, USA \\
$^{13}$Department of Physics, University of California, San Diego, 9500 Gilman Drive, La Jolla, CA 92093, USA \\
$^{14}$Lawrence Berkeley National Laboratory, 1 Cyclotron Road, Berkeley, CA 94720, USA \\
$^{15}$University of California Observatories/Lick Observatory, P.O.~Box 85, Mount Hamilton, CA 95140, USA \\
$^{16}$The Observatories of the Carnegie Institution for Science, 813 Santa Barbara Street, Pasadena, CA 91101, USA \\
$^{17}$Department of Physics and Astronomy, Rutgers the State University of New Jersey, 136 Frelinghuysen Road, Piscataway, NJ 08854, USA \\
$^{18}$Center for Cosmology and Particle Physics, New York University, 4 Washington Place, New York, NY 10003, USA \\
$^{19}$Department of Physics and Astronomy, University of California, Los Angeles, CA 90095, USA \\
$^{20}$Department of Physics and Astronomy, East Los Angeles College, Monterey Park, CA 91754, USA \\
$^{21}$Cahill Center for Astrophysics, California Institute of Technology, Pasadena, CA 91125, USA \\
$^{22}$School of Physics and Astronomy, Tel-Aviv University, Tel Aviv 69978, Israel \\
$^{23}$Department of Physics, Purdue University, West Lafayette, IN 47907-2036, USA \\
$^{24}$Department of Physics and Astronomy, Ohio University, Athens, OH 45701, USA \\
$^{25}$Steward Observatory, University of Arizona, 933 North Cherry Avenue, Tucson, AZ 85721-0065, USA \\
$^{26}$Spitzer Science Center, California Institute of Technology, 1200 East California Boulevard, Pasadena, CA 91125, USA \\
$^{27}$Department of Physics and Tsinghua Center for Astrophysics, Tsinghua University, Beijing 100084, China \\
\dag Deceased 12 December 2011
}

\begin{document}
\date{Accepted  . Received   ; in original form  }
\pagerange{\pageref{firstpage}--\pageref{lastpage}} \pubyear{2012}
\maketitle
\label{firstpage}

\clearpage


\begin{abstract}
In this first paper in a series we present \totspec\ low-redshift ($z
\lesssim 0.2$) optical spectra of \totobj\ Type~Ia supernovae (SNe~Ia)
observed from 1989 through 2008 as part of the Berkeley
SN~Ia Program (BSNIP). 
\filtspec\ spectra of \filtobj\ SNe~Ia have well-calibrated light 
curves with measured distance moduli, and many of the spectra have
been corrected for host-galaxy contamination. Most of the data were
obtained using the Kast double spectrograph mounted on the Shane 3~m
telescope at Lick Observatory and have a typical wavelength range of
3300--10,400~\AA, roughly twice as wide as spectra from
most previously published datasets. 
We present our observing and reduction procedures, and we describe 
the resulting SN Database (SNDB), which will be an
online, public, searchable database containing all of our fully reduced
spectra and companion photometry. In addition, we
discuss our spectral classification scheme (using the SuperNova
IDentification code, SNID; 
Blondin \& Tonry 2007), utilising our newly constructed set of SNID
spectral templates. These templates allow us to accurately
classify our entire dataset, and by doing so we are
able to reclassify a handful of objects as bona fide SNe~Ia and a few
other objects as members of some of the peculiar SN~Ia subtypes.  In
fact, our dataset includes spectra of nearly 90 spectroscopically
peculiar SNe~Ia. We also present spectroscopic host-galaxy redshifts
of some SNe~Ia where these values were previously unknown. The sheer
size of the BSNIP dataset and the consistency of our observation and
reduction methods makes this sample unique among all other published
SN~Ia datasets and is complementary in many ways to the large,
low-redshift SN~Ia spectra presented by Matheson \etal 2008 and
Blondin \etal 2012. In other BSNIP papers in this series, 
we use these data to examine the relationships between spectroscopic
characteristics and various observables such as photometric and
host-galaxy properties.

\end{abstract}




\section{Introduction}\label{s:intro}

Supernovae (SNe) have been integral to our understanding of the cosmos
throughout the history of astronomy --- from demonstrating that the sky
was not unchanging beyond the lunar sphere \citep{Brahe1573} to the
discovery of the acceleration of the expansion of the Universe
\citep{Riess98:lambda, Perlmutter99}.  Type~Ia supernovae (SNe~Ia)
have been particularly useful in recent years as a way to accurately
measure cosmological parameters \citep{Astier06, Riess07, Wood-Vasey07, 
Hicken09:cosmo, Kessler09, Amanullah10,Conley11,Sullivan11,Suzuki12}.
Broadly speaking, SNe~Ia are the result of thermonuclear explosions of 
carbon/oxygen white dwarfs (WDs) (e.g., \citealt{Hoyle60, Colgate69, 
Nomoto84}; see \citealt{Hillebrandt00} for a review). However, we still 
lack a detailed understanding of the progenitor systems and explosion
mechanisms, as well as how differences in initial conditions create
the variance in observed properties of SNe~Ia.  To solve these
problems, and others, detailed and self-consistent observations of
many hundreds of SNe~Ia are required.

The cosmological application of SNe~Ia as precise distance indicators
relies on being able to standardise their luminosity.
\citet{Phillips93} showed that light-curve decline is well correlated
with luminosity at peak brightness for most SNe~Ia, the so-called
``Phillips relation.''  Optical colours have also been used to better
standardise the luminosity of SNe~Ia
\citep[e.g.,][]{Riess96,Tripp98}. Additionally, people have searched
for another, spectroscopic parameter in SN observations which would
make our 
measurements of the distances to SNe~Ia even more precise. 
\citet{Bailey09} and \citet{Blondin11} have decreased the scatter 
in residuals to the Hubble diagram with the help of optical spectra.
\citet{Wang09} obtained an additional improvement by separating their 
sample of SNe~Ia into two groups based on the ejecta velocity near
maximum brightness; they suggested different reddening laws for these 
two samples. Building on this work, \citet{Foley11:vel} found that 
the {\it intrinsic} maximum-light colour of SNe~Ia depends on their 
ejecta velocity.  After accounting for this colour difference, 
the scatter in Hubble-diagram residuals is decreased from 0.19~mag to
0.13~mag for a subset of SNe~Ia. This particular conclusion was
possible only with a large set of spectroscopically observed objects, 
with many of the spectra coming from the sample described in this Paper
\citep[see also][]{Wang09}.


Until now there have been several statistical samples of low-redshift
SN~Ia photometry \citep[e.g.,][]{Hamuy96, Riess99:lc, Jha06, Hicken09,
Ganeshalingam10:phot_paper,Contreras10,Stritzinger11}, but only one
large sample of low-redshift SN~Ia spectra \citep{Matheson08}.  Until the
publication of over 432 spectra of 32 SNe~Ia by \citet{Matheson08},
large samples of SN~Ia spectra were typically constructed by combining
datasets published for individual objects, usually from many different
groups. 

The Berkeley Supernova Ia Program (BSNIP) is a large-scale effort to
measure the properties of low-redshift ($z \lesssim 0.2$) SNe~Ia, focusing
on optical spectroscopy and photometry \citep[see][for the companion
photometry paper to much of the spectroscopic sample presented
here]{Ganeshalingam10:phot_paper}.  One aspect of our strategy for the
last two decades has been to observe as many SNe~Ia as possible in
order to dramatically increase the number of objects with
spectroscopic data.  We have also attempted to obtain good temporal
spectral coverage of peculiar objects as well as objects which were
being observed photometrically by our group. In addition, we
strove to spectroscopically classify all SNe discovered by the 0.76~m
Katzman Automatic Imaging Telescope \citep[KAIT;][]{Filippenko01}.  By
observing and reducing 
our spectra in a consistent manner, we avoid many of the systematic
differences found in previous samples constructed from data obtained
by various groups.

In this paper we present the low-redshift SN~Ia spectral
dataset; more details are given by \citep{Silverman12:thesis}. 
This sample consists of a total of \totspec\ spectra of
\totobj\ SNe~Ia observed from 1989 through the end of 2008. 
A subset of the SNe, along with information about their host galaxies,
is presented in Table~\ref{t:obj} (the full set is available
online --- see the Supporting Information).  Information regarding some of
the SN~Ia spectra in the dataset is listed in Table~\ref{t:spec_info}
(and again the full set is available online --- see the Supporting
Information).
%
%
Many spectra presented in this paper have complementary
light curves from \citet{Hamuy96}, \citet{Riess99:lc}, \citet{Jha06}, \citet{Hicken09},
and \citet{Ganeshalingam10:phot_paper}, which have all been compiled
and fit by Ganeshalingam \etal (in prep.).  Other spectra have
complementary unfiltered light curves given by Wang \etal (in
prep.).

In this paper, we describe our observations and data-reduction
procedure in Sections~\ref{s:obs} and \ref{s:data}, respectively.  We
present our methods of data management and storage in
Section~\ref{s:sndb} and our spectral classification scheme in
Section~\ref{s:classification}. The sample of objects and spectra is
described in Section~\ref{s:sample}, and there we also show our
fully reduced spectra as well as (for the objects with multi-band SN
and galaxy photometry) galaxy-subtracted spectra. Reclassifications 
of a handful of SNe, and previously unknown spectroscopic 
host-galaxy redshifts, are also given. We
discuss our conclusions in Section~\ref{s:conclusions}. Future BSNIP
papers will examine the correlations between spectroscopic properties
and other observables (such as photometry and host-galaxy
properties).

\onecolumn
\scriptsize
\begin{landscape}
\centering
\begin{longtable}{lc@{}c@{}crccccrrrr}
\caption{SN~Ia and Host Information} \label{t:obj} \\[-2ex]
\hline \hline
SN Name & SNID & Host & Host & \multicolumn{1}{c}{$cz_\textrm{helio}$} & $E\left(B-V\right)_\textrm{MW}$ & Discovery & Discovery & Classification & \multicolumn{1}{c}{\# of} & First\hspace{.11in} & Last\hspace{.13in} & JD$_\textrm{max}$ \\
   & (Sub)Type$^\textrm{a}$ & Galaxy & Morp.$^\textrm{b}$ & \multicolumn{1}{c}{(\kms)$^\textrm{c}$} & (mag)$^\textrm{d}$ & Date (UT) & Reference & Reference & \multicolumn{1}{c}{Spec.} & Epoch$^\textrm{e}$ & Epoch$^\textrm{e}$ & Ref.$^\textrm{f}$ \\
\hline
\endfirsthead

\multicolumn{13}{c}{{\tablename} \thetable{} --- Continued} \\
\hline \hline
SN Name & SNID & Host & Host & \multicolumn{1}{c}{$cz_\textrm{helio}$} & $E\left(B-V\right)_\textrm{MW}$ & Discovery & Discovery & Classification & \multicolumn{1}{c}{\# of} & First\hspace{.11in} & Last\hspace{.13in} & JD$_\textrm{max}$ \\
   & (Sub)Type$^\textrm{a}$ & Galaxy & Morp.$^\textrm{b}$ & \multicolumn{1}{c}{(\kms)$^\textrm{c}$} & (mag)$^\textrm{d}$ & Date (UT) & Reference & Reference & \multicolumn{1}{c}{Spec.} & Epoch$^\textrm{e}$ & Epoch$^\textrm{e}$ & Ref.$^\textrm{f}$ \\
\hline
\endhead

\hline \hline
\multicolumn{13}{l}{Continued on Next Page\ldots} \\
\endfoot

\hline \hline
\endlastfoot

SN 1989A & Ia-norm & NGC 3687 & Sbc & 2506 & 0.020 & 1989-01-19 & IAUC 4721 & IAUC 4724 & 1 & $83.80$ & $\cdots$ & 1 \\
SN 1989B & Ia-norm & NGC 3627 & Sb & 728 & 0.030 & 1989-01-30 & IAUC 4726 & IAUC 4727 & 4 & $7.54$ & $152.19$ & 2 \\
SN 1989M & Ia-norm & NGC 4579 & Sb & 1520 & 0.039 & 1989-06-28 & IAUC 4802 & IAUC 4802 & 4 & $2.49$ & $297.42$ & 3 \\
SN 1990G & Ia-norm & IC 2735 & Sab & 10727 & 0.021 & 1990-03-19 & IAUC 4982 & IAUC 4984 & 1 & $\cdots$ & $\cdots$ & $\cdots$ \\
SN 1990M & Ia-norm & NGC 5493 & S0 & 2710 & 0.036 & 1990-06-15 & IAUC 5033 & IAUC 5034 & 5 & $\cdots$ & $\cdots$ & $\cdots$ \\
SN 1990O & Ia-norm & MCG +03-44-3 & Sa & 9192 & 0.095 & 1990-06-22 & IAUC 5039 & IAUC 5039 & 3 & $12.54$ & $54.50$ & 2 \\
SN 1990N & Ia-norm & NGC 4639 & Sbc & 1019 & 0.025 & 1990-06-23 & IAUC 5039 & IAUC 5039 & 5 & $7.11$ & $160.16$ & 2 \\
SN 1990R & Ia-norm & UGC 11699 & Sd/Irr & 4857 & 0.096 & 1990-06-26 & IAUC 5054 & IAUC 5054 & 3 & $\cdots$ & $\cdots$ & $\cdots$ \\
SN 1990Y & Ia-norm & FCCB 1147 & E & 11702 & 0.008 & 1990-08-22 & IAUC 5080 & IAUC 5083 & 1 & $16.78$ & $\cdots$ & 2 \\
SN 1991B & Ia-norm & NGC 5426 & Sc & 2572 & 0.028 & 1991-01-11 & IAUC 5163 & IAUC 5164 & 3 & $\cdots$ & $\cdots$ & $\cdots$ \\
SN 1991K & Ia-norm & NGC 2851 & S0 & 5096 & 0.059 & 1991-02-20 & IAUC 5196 & \citet{Matheson01} & 2 & $\cdots$ & $\cdots$ & $\cdots$ \\
SN 1991M & Ia-norm & IC 1151 & Sc & 2170 & 0.036 & 1991-03-12 & IAUC 5207 & IAUC 5207 & 4 & $18.06$ & $152.09$ & 2 \\
SN 1991O & Ia-91bg & 2MASX J14243792+6545294 & $\cdots$ & $\cdots$ & 0.012 & 1991-03-18 & IAUC 5233 & IAUC 5233 & 1 & $\cdots$ & $\cdots$ & $\cdots$ \\
SN 1991S & Ia-norm & UGC 5691 & Sb & 16489 & 0.026 & 1991-04-10 & IAUC 5238 & IAUC 5245 & 1 & $31.05$ & $\cdots$ & 2 \\
SN 1991T & Ia-91T & NGC 4527 & Sbc & 1736 & 0.023 & 1991-04-13 & IAUC 5239 & IAUC 5251 & 9 & $-10.10$ & $347.19$ & 2 \\
SN 1991am & Ia-norm & MCG +06-37-6 & Sb & 18353 & 0.018 & 1991-07-14 & IAUC 5312 & IAUC 5318 & 1 & $\cdots$ & $\cdots$ & $\cdots$ \\
SN 1991ak & Ia-norm & NGC 5378 & Sa & 3043 & 0.013 & 1991-07-15 & IAUC 5309 & IAUC 5311 & 3 & $\cdots$ & $\cdots$ & $\cdots$ \\
SN 1991at & Ia-norm & UGC 733 & Sb & 12306 & 0.068 & 1991-08-19 & IAUC 5336 & IAUC 5347 & 1 & $\cdots$ & $\cdots$ & $\cdots$ \\
SN 1991as & Ia & [M91k] 224610+0754.6 & $\cdots$ & $\cdots$ & 0.107 & 1991-08-19 & IAUC 5336 & IAUC 5347 & 1 & $\cdots$ & $\cdots$ & $\cdots$ \\
SN 1991ay & Ia-norm & 2MASX J00471896+4032336 & Sb & 15289 & 0.062 & 1991-09-09 & IAUC 5352 & IAUC 5366 & 1 & $\cdots$ & $\cdots$ & $\cdots$ \\
SN 1991bd & Ia-norm & UGC 2936 & Sd/Irr & 3813 & 0.449 & 1991-10-12 & IAUC 5367 & IAUC 5367 & 1 & $\cdots$ & $\cdots$ & $\cdots$ \\
SN 1991bc & Ia-norm & UGC 2691 & Sb & 6401 & 0.071 & 1991-10-12 & IAUC 5366 & IAUC 5366 & 2 & $\cdots$ & $\cdots$ & $\cdots$ \\
SN 1991bb & Ia-norm & UGC 2892 & Sbc & 7962 & 0.331 & 1991-10-13 & IAUC 5365 & IAUC 5365 & 2 & $\cdots$ & $\cdots$ & $\cdots$ \\
SN 1991bf & Ia-norm & MCG -05-56-027 & S0 & 9021 & 0.016 & 1991-11-13 & IAUC 5389 & IAUC 5404 & 1 & $\cdots$ & $\cdots$ & $\cdots$ \\
SN 1991bg & Ia-91bg & NGC 4374 & E & 1061 & 0.037 & 1991-12-03 & IAUC 5400 & IAUC 5403 & 8 & $0.14$ & $161.88$ & 2 \\
SN 1991bh & Ia-norm & [M91o] 024216.2+145713.4 & $\cdots$ & $\cdots$ & 0.102 & 1991-12-07 & IAUC 5401 & IAUC 5404 & 1 & $\cdots$ & $\cdots$ & $\cdots$ \\
SN 1991bj & Ia-02cx & IC 344 & Sb & 5441 & 0.052 & 1991-12-30 & IAUC 5420 & IAUC 5420 & 1 & $\cdots$ & $\cdots$ & $\cdots$ \\
SN 1992G & Ia-norm & NGC 3294 & Sc & 1586 & 0.018 & 1992-02-09 & IAUC 5452 & IAUC 5458 & 8 & $23.41$ & $126.76$ & 2 \\
SN 1992M & Ia-norm & 2MASX J07150996+4525556 & E & 15589 & 0.086 & 1992-02-25 & IAUC 5473 & IAUC 5473 & 2 & $\cdots$ & $\cdots$ & $\cdots$ \\
SN 1992ah & Ia-norm & 2MASX J17374476+1254168 & $\cdots$ & $\cdots$ & 0.149 & 1992-06-27 & IAUC 5559 & IAUC 5559 & 1 & $\cdots$ & $\cdots$ & $\cdots$ \\
SN 1992ap & Ia-norm & UGC 10430 & Sbc & 8958 & 0.009 & 1992-07-29 & IAUC 5573 & IAUC 5601 & 1 & $\cdots$ & $\cdots$ & $\cdots$ \\
SN 1993C & Ia-norm & NGC 2954 & E & 3819 & 0.037 & 1993-01-27 & IAUC 5699 & IAUC 5701 & 3 & $\cdots$ & $\cdots$ & $\cdots$ \\
SN 1993Y & Ia-norm & UGC 2771 & S0 & 5990 & 0.186 & 1993-09-18 & IAUC 5870 & IAUC 5870 & 1 & $28.33$ & $\cdots$ & 4 \\
SN 1993aa & Ia-91bg & APMUKS(BJ) B230046.41-063708.1 & $\cdots$ & $\cdots$ & 0.041 & 1993-09-19 & IAUC 5871 & IAUC 5871 & 1 & $\cdots$ & $\cdots$ & $\cdots$ \\
SN 1993Z & Ia-norm & NGC 2775 & Sab & 1355 & 0.041 & 1993-09-23 & IAUC 5870 & IAUC 5870 & 9 & $28.92$ & $232.69$ & 4 \\
SN 1993ab & Ia-norm & NGC 1164 & Sab & 4176 & 0.155 & 1993-09-24 & IAUC 5871 & IAUC 5877 & 1 & $\cdots$ & $\cdots$ & $\cdots$ \\
SN 1993ac & Ia-norm & CGCG 307-023 & E & 14690 & 0.162 & 1993-10-13 & IAUC 5879 & IAUC 5882 & 2 & $12.68$ & $28.66$ & 2 \\
SN 1993ae & Ia-norm & IC 126 & Sb & 5711 & 0.039 & 1993-11-07 & IAUC 5888 & IAUC 5888 & 2 & $18.99$ & $68.87$ & 2 \\
SN 1993ai & Ia-norm & UGC 3483 & Sbc & 10193 & 0.120 & 1993-12-10 & IAUC 5912 & IAUC 5912 & 1 & $\cdots$ & $\cdots$ & $\cdots$ \\
SN 1993aj & Ia-norm & 2MFGC 09481 & Sb & 23264 & 0.025 & 1993-12-27 & IAUC 5915 & IAUC 5921 & 3 & $\cdots$ & $\cdots$ & $\cdots$ \\
SN 1994B & Ia-norm & [P94a] 081751.35+155320.5 & $\cdots$ & 26682 & 0.044 & 1994-01-16 & IAUC 5923 & IAUC 5923 & 2 & $\cdots$ & $\cdots$ & $\cdots$ \\
SN 1994E & Ia & SDSS J113207.01+552138.0 & $\cdots$ & 19148 & 0.010 & 1994-03-05 & IAUC 5952 & IAUC 5952 & 1 & $\cdots$ & $\cdots$ & $\cdots$ \\
SN 1994J & $\cdots$ & [P94] 095815.51+544427.4 & $\cdots$ & 16788 & 0.013 & 1994-03-05 & IAUC 5971 & IAUC 5974 & 1 & $\cdots$ & $\cdots$ & $\cdots$ \\
SN 1994D & Ia-norm & NGC 4526 & S0 & 447 & 0.023 & 1994-03-07 & IAUC 5946 & IAUC 5946 & 20 & $-12.31$ & $114.73$ & 2 \\
SN 1994Q & Ia-norm & CGCG 224-104 & S0 & 8863 & 0.018 & 1994-06-02 & IAUC 6001 & IAUC 6001 & 3 & $9.68$ & $69.79$ & 2 \\
SN 1994S & Ia-norm & NGC 4495 & Sab & 4551 & 0.017 & 1994-06-04 & IAUC 6005 & IAUC 6005 & 1 & $1.11$ & $\cdots$ & 2 \\
SN 1994T & Ia-norm & CGCG 016-058 & Sa & 10390 & 0.020 & 1994-06-11 & IAUC 6007 & IAUC 6007 & 1 & $33.09$ & $\cdots$ & 5 \\
SN 1994U & Ia-norm & NGC 4948 & Sd/Irr & 1124 & 0.056 & 1994-06-27 & IAUC 6011 & IAUC 6011 & 1 & $\cdots$ & $\cdots$ & $\cdots$ \\
SN 1994X & Ia-norm & 2MASX J00152051-2453337 & Sb & 16668 & 0.013 & 1994-08-15 & IAUC 6056 & IAUC 6068 & 1 & $\cdots$ & $\cdots$ & $\cdots$ \\
SN 1994ab & Ia-norm & MCG -05-50-8 & Sb & 10373 & 0.102 & 1994-09-17 & IAUC 6089 & IAUC 6094 & 1 & $\cdots$ & $\cdots$ & $\cdots$ \\
\hline \hline
\multicolumn{13}{l}{Table abridged; the full table is available online---see Supporting Information.} \\ 
\multicolumn{13}{l}{$^\textrm{a}$See Section~\ref{s:classification} for more information on our SNID classifications.} \\ 
\multicolumn{13}{l}{$^\textrm{b}$Host-galaxy morphology is taken from the NASA/IPAC Extragalactic Database (NED).} \\
\multicolumn{13}{p{9in}}{$^\textrm{c}$Heliocentric redshifts listed are from NED, except the redshift of the host of SN~1991ay (2MASX~J00471896+4032336) comes from IAUC~5366.} \\
\multicolumn{13}{l}{$^\textrm{d}$The Milky Way (MW) reddening toward each SN as derived from the dust maps of \citet{Schlegel98} and includes the corrections of \citet{Peek10}.} \\
\multicolumn{13}{p{9in}}{$^\textrm{e}$Epochs of first and last spectrum are relative to $B$-band maximum brightness in rest-frame days using the heliocentric redshift and date of maximum reference presented in the table. For SNe with photometric information and only one spectrum in our dataset, only a Phase of First Spectrum is listed.} \\
\multicolumn{13}{p{9in}}{$^\textrm{f}$Julian Date of Maximum References: (1) \citet{Tsvetkov90}, (2) \citet{Ganeshalingam10:phot_paper}, (3) \citet{Kimeridze91}, (4) \citet{Ho01}, (5) \citet{Hicken09}.} \\
\end{longtable}
\end{landscape}
\normalsize
\twocolumn


\onecolumn
\small
\begin{landscape}
\begin{center}
\begin{longtable}{lccrcccrrclrccr}
\caption{SN~Ia Spectral Information} \label{t:spec_info} \\[-2.3ex]
\hline \hline
SN Name & UT Date$^\textrm{a}$ & MJD$^\textrm{b}$ & Phase$^\textrm{c}$ & Inst.$^\textrm{d}$ & Wavelength & Res.$^\textrm{e}$ & P.A.$^\textrm{f}$ & Par.$^\textrm{g}$ & Air.$^\textrm{h}$ & See.$^\textrm{i}$ & Exp. & Observer(s)$^\textrm{j}$ & Reducer$^\textrm{k}$ & Flux \\
 & & & & & Range (\AA) & (\AA) & ($^\circ$) & ($^\circ$) & & (\arcsec) & (s) & & & Corr.$^\textrm{l}$ \\
\hline
\endfirsthead

\multicolumn{15}{c}{{\tablename} \thetable{} --- Continued} \\
\hline \hline
SN Name & UT Date$^\textrm{a}$ & MJD$^\textrm{b}$ & Phase$^\textrm{c}$ & Inst.$^\textrm{d}$ & Wavelength & Res.$^\textrm{e}$ & P.A.$^\textrm{f}$ & Par.$^\textrm{g}$ & Air.$^\textrm{h}$ & See.$^\textrm{i}$ & Exp. & Observer(s)$^\textrm{j}$ & Reducer$^\textrm{k}$ & Flux \\
 & & & & & Range (\AA) & (\AA) & ($^\circ$) & ($^\circ$) & & (\arcsec) & (s) & & & Corr.$^\textrm{l}$ \\
\hline
\endhead

\hline \hline
\multicolumn{15}{l}{Continued on Next Page\ldots} \\
\endfoot

\hline \hline
\endlastfoot

SN 1989A & 1989-04-27.000 & 47643.000 & $      83.80$ & 1 & 3450--9000 & 12 & $\cdots$ & $\cdots$ & $\cdots$ & 2 & $\cdots$ & 1,2 & 1 & $0$ \\
SN 1989B$^\textrm{m}$ & 1989-02-15.000 & 47572.000 & $      7.54$ & 2 & 3450--8450 & 7 & 0 & $\cdots$ & $\cdots$ & $\cdots$ & $\cdots$ & 1,3 & 1 & $0$ \\
SN 1989B$^\textrm{m}$ & 1989-02-21.000 & 47578.000 & $      13.53$ & 2 & 3450--7000 & 7 & 0 & $\cdots$ & $\cdots$ & $\cdots$ & $\cdots$ & 1 & 1 & $0$ \\
SN 1989B & 1989-04-27.000 & 47643.000 & $      78.37$ & 1 & 3300--9050 & 12 & $\cdots$ & $\cdots$ & $\cdots$ & 2 & $\cdots$ & 1,2 & 1 & $0$ \\
SN 1989B & 1989-07-10.000 & 47717.000 & $      152.19$ & 1 & 3900--6226 & 12 & $\cdots$ & $\cdots$ & $\cdots$ & 3 & $\cdots$ & 1,2,3 & 1 & $0$ \\
SN 1989M & 1989-07-09.000 & 47716.000 & $      2.49$ & 1 & 3080--10300 & 12 & $\cdots$ & $\cdots$ & $\cdots$ & 2.5 & $\cdots$ & 1,2,3 & 1 & $0$ \\
SN 1989M & 1989-07-10.000 & 47717.000 & $      3.48$ & 1 & 3400--10200 & 12 & $\cdots$ & $\cdots$ & $\cdots$ & 3 & $\cdots$ & 1,2,3 & 1 & $0$ \\
SN 1989M & 1989-12-01.000 & 47861.000 & $      146.76$ & 1 & 4150--9500 & 12 & $\cdots$ & $\cdots$ & $\cdots$ & 1.5 & 900 & 1,2 & 1 & $0$ \\
SN 1989M & 1990-05-01.428 & 48012.428 & $      297.42$ & 1 & 3930--6980 & 12 & 54 & 54 &       2.31 & 2.5 & 2100 & 1,2 & 1 & $0$ \\
SN 1990G & 1990-03-25.000 & 47975.000 & $\cdots$ & 1 & 3932--9800 & 12 & $\cdots$ & $\cdots$ & $\cdots$ & 3 & $\cdots$ & 1,2 & 1 & $0$ \\
SN 1990M & 1990-04-01.000 & 47982.000 & $\cdots$ & 1 & 3932--7060 & 12 & $\cdots$ & $\cdots$ & $\cdots$ & 1.5 & $\cdots$ & 1,4 & 1 & $0$ \\
SN 1990M & 1990-07-17.000 & 48089.000 & $\cdots$ & 1 & 3920--9860 & 12 & $\cdots$ & $\cdots$ & $\cdots$ & 2 & $\cdots$ & 1,2 & 1 & $0$ \\
SN 1990M & 1990-07-31.255 & 48103.255 & $\cdots$ & 1 & 3900--9900 & 12 & 48 & 50 &       4.19 & 1.25 & 1200 & 1,2 & 1 & $0$ \\
SN 1990M & 1990-08-29.171 & 48132.171 & $\cdots$ & 1 & 3940--9850 & 12 & 229 & 49 &       3.85 & 1.25 & 1300 & 1,2 & 1 & $0$ \\
SN 1990M & 1990-08-30.000 & 48133.000 & $\cdots$ & 1 & 6720--9850 & 12 & 229 & 49 &       4.02 & 2.25 & 1300 & 1,2 & 1 & $0$ \\
SN 1990O & 1990-07-17.000 & 48089.000 & $      12.54$ & 1 & 3920--7080 & 12 & $\cdots$ & $\cdots$ & $\cdots$ & 2 & $\cdots$ & 1,2 & 1 & $0$ \\
SN 1990O & 1990-07-31.396 & 48103.396 & $      26.50$ & 1 & 3900--7020 & 12 & 55 & 56 &       2.67 & 1.25 & 1800 & 1,2 & 1 & $0$ \\
SN 1990O & 1990-08-29.252 & 48132.252 & $      54.50$ & 1 & 3900--7020 & 12 & 235 & 54 &       1.52 & 1.25 & 1400 & 1,2 & 1 & $0$ \\
SN 1990N & 1990-07-17.000 & 48089.000 & $      7.11$ & 1 & 3920--9872 & 12 & $\cdots$ & $\cdots$ & $\cdots$ & 2 & $\cdots$ & 1,2 & 1 & $0$ \\
SN 1990N & 1990-07-31.193 & 48103.193 & $      21.25$ & 1 & 3900--9900 & 12 & 55 & 55 &       2.30 & 1.25 & 900 & 1,2 & 1 & $0$ \\
SN 1990N & 1990-08-29.153 & 48132.153 & $      50.11$ & 1 & 3940--9850 & 12 & 234 & 54 &       3.80 & 1.25 & 500 & 1,2 & 1 & $0$ \\
SN 1990N & 1990-08-30.000 & 48133.000 & $      50.96$ & 1 & 6720--9850 & 12 & 234 & 54 &       4.23 & 2.25 & 950 & 1,2 & 1 & $0$ \\
SN 1990N & 1990-12-17.577 & 48242.577 & $      160.16$ & 1 & 3900--7000 & 12 & 151 & 331 &       1.14 & 2.5 & 900 & 1,2 & 1 & $0$ \\
SN 1990R & 1990-07-17.000 & 48089.000 & $\cdots$ & 1 & 3952--7052 & 12 & $\cdots$ & $\cdots$ & $\cdots$ & 2 & $\cdots$ & 1,2 & 1 & $0$ \\
SN 1990R & 1990-07-31.423 & 48103.423 & $\cdots$ & 1 & 3900--7020 & 12 & 32 & 37 &       1.19 & 1.25 & 1200 & 1,2 & 1 & $0$ \\
SN 1990R & 1990-08-29.279 & 48132.279 & $\cdots$ & 1 & 3900--7020 & 12 & 180 & 6 &       1.10 & 1.25 & 2400 & 1,2 & 1 & $0$ \\
SN 1990Y & 1990-08-30.510 & 48133.510 & $      16.78$ & 1 & 3940--7050 & 12 & 171 & 351 &       3.07 & 2.25 & 1200 & 1,2 & 1 & $0$ \\

\hline \hline
\multicolumn{15}{l}{Table abridged; the full table is available online---see Supporting Information.} \\ 
\multicolumn{15}{l}{$^\textrm{a}$If not rounded to the whole day, UT date at the midpoint of the observation.} \\
\multicolumn{15}{l}{$^\textrm{b}$Modified JD (if not rounded to the whole day, modified JD at the midpoint of the observation).} \\
\multicolumn{15}{l}{$^\textrm{c}$Phases of spectra are in rest-frame days using the heliocentric redshift and photometry reference presented in Table~\ref{t:obj}.} \\
\multicolumn{15}{p{9in}}{$^\textrm{d}$Instruments: (1) UV Schmidt (Shane 3~m), (2) Stover Spectrograph (Nickel 1~m).} \\
\multicolumn{15}{p{9in}}{$^\textrm{e}$FWHM spectral resolution as measured from narrow sky emission lines.  If we were unable to accurately measure the sky lines, the average resolution for that instrumental setup is displayed (see Section~\ref{s:obs} for more information regarding our instrumental setups and their average resolutions).} \\
\multicolumn{15}{l}{$^\textrm{f}$Observed position angle during observation.} \\
\multicolumn{15}{l}{$^\textrm{g}$Average parallactic angle (Filippenko 1982) during the observation.} \\
\multicolumn{15}{l}{$^\textrm{h}$Airmass at midpoint of exposure.} \\
\multicolumn{15}{p{9in}}{$^\textrm{i}$Approximate atmospheric seeing as measured from the FWHM of the trace of the SN.  If we were unable to accurately measure the FWHM of the trace, an estimate by the observers of the average seeing from that night is displayed with only one or two significant figures.} \\
\multicolumn{15}{p{9in}}{$^\textrm{j}$Observers: (1) Alex Filippenko, (2) Joe Shields, (3) Michael Richmond, (4) Charles Steidel.} \\
\multicolumn{15}{p{9in}}{$^\textrm{k}$Reducers: (1) Tom Matheson.} \\
\multicolumn{15}{p{9in}}{$^\textrm{l}$Flux Correction: (0) No correction. Negative values indicate that $>5\%$ of the corrected flux is negative.} \\
\multicolumn{15}{p{9in}}{$^\textrm{m}$Observation has unreliable spectrophotometry due to events external to normal telescope operation and data reduction. See Section~\ref{ss:specphot} for more information.} \\
\end{longtable}
\end{center}
\end{landscape}
\normalsize
\twocolumn



\section{Observations}\label{s:obs}

Over the past two decades, our group has had access to several
different telescopes and spectrographs for the purpose of observing
SNe.  The main facility for this study was the Shane 3~m telescope at
Lick Observatory.  During this period, the Shane telescope has had two
low-resolution spectrographs: the UV Schmidt spectrograph until early
1992 \citep{Miller87} and the Kast double spectrograph since then
\citep{Miller93}. Using these instruments we obtained 4.9~per~cent and
72.3~per~cent of our spectra, respectively. We also obtained a handful
of spectra using the Stover spectrograph mounted on the Nickel 1~m
telescope also at Lick.

We have supplemented our Lick Observatory sample with spectra obtained
at the Keck Observatory.  When conditions were not acceptable for our
faint, primary targets (typically in twilight, or during times of
bad seeing or cloudy weather),
we would use one of the 10~m Keck telescopes to obtain spectra of our
relatively bright (typically $R < 18$~mag), nearby SN targets.  We
also obtained many late-time spectra with the Keck telescopes.
16.8~per~cent of our spectra were obtained using the Low Resolution
Imaging Spectrometer \citep[LRIS;][both before and after the addition
of the blue arm]{Oke95}, 3.0~per~cent were obtained using the DEep
Imaging Multi-Object Spectrograph \citep[DEIMOS;][]{Faber03}, and
1.8~per~cent were obtained using the 
Echelle Spectrograph and Imager \citep[ESI;][]{Sheinis02}. 


All of these telescopes were classically scheduled.  We would
typically have 1 night every two weeks on the Shane telescope (near
first and last quarter moon) throughout the year and 4--10 nights per
year with the Keck telescopes (typically 1--2 nights near new moon in
a given lunation).  Recently, we have been allotted a third night per
lunar cycle on the Shane telescope near new moon.  Taking into account
weather and instrument problems, our coverage of any given object is
typically about one spectrum every two weeks.  The telescope
scheduling and observing method are very different from those of
\citet{Matheson08} and \citep{Blondin12}, who observed
fewer SNe~Ia but with a higher cadence for each object (see
Section~\ref{ss:sample_char} for further comparisons of the two
spectral datasets).

All observations of our scheduled time were performed by members of
the BSNIP group and PI Filippenko was present for \alexnights\ nights.
Occasionally, as the result of a swap of time or 
for a particularly interesting object, an observer exterior to the
BSNIP group would observe for our team.  This sometimes resulted in
slight variations in instrument configurations (such as a smaller
wavelength range, for example).  As mentioned above, the bulk of our
data were obtained at the Lick and Keck Observatories where our average
seeing was slightly greater than 2\arcsec\ and slightly greater than
1\arcsec, respectively.



\subsection{Individual Instruments}

\subsubsection{UV Schmidt on the Shane 3~m}

The UV Schmidt spectrograph contained a Texas Instruments $800 \times
300$ pixel charge-coupled device (CCD) and our setup used a slit that was
2--3\arcsec\ wide.  The average resolution of our spectra from this
instrument was \about12~\AA.

\subsubsection{Kast on the Shane 3~m}
 
Until September 2008, the Kast double spectrograph used two Reticon
$1200 \times 400$ pixel CCDs with 27~$\mu$m pixels
and a spatial scale of 0.78\arcsec\ pixel$^{-1}$, with one CCD in each
of the red and blue arms of the spectrograph.  Currently, the blue arm
of Kast uses a Fairchild $2048 \times 2048$ pixel device with 15~$\mu$m pixels,
which corresponds to 0.43\arcsec\ pixel$^{-1}$. For our typical setup, we
would observe with a 300/7500 grating for the red side, a 600/4310
grism for the blue side, and a D55 dichroic. This results in a
wavelength range of 3300--10,400~\AA\ with overlap between the two arms
of 5200--5500~\AA. With our typical slit of 2\arcsec, we achieve a
resolution of \about11~\AA\ and \about6~\AA\ on the red and blue sides,
respectively.
%
%
%

\subsubsection{Stover on the Nickel 1~m}

The Stover spectrograph contains a Reticon $400 \times 1200$ pixel CCD
and 
27~$\mu$m pixels with a spatial scale of 2\arcsec\ pixel$^{-1}$.  Our
setup used a 2.9\arcsec\ wide slit with the 600/4820 grism.  This
yielded an average resolution of \about7~\AA.

\subsubsection{LRIS on the Keck 10~m}

When most of our dataset was obtained, LRIS used a
Tektronix $2048 \times 2048$ pixel CCD with 21~$\mu$m pixels and a
spatial scale of 0.211\arcsec\ pixel$^{-1}$ for the red arm and two $2048
\times 4096$ pixel Marconi E2V CCDs with 15~$\mu$m pixels and a
spatial scale of 0.135\arcsec\ pixel$^{-1}$ for the blue arm.  LRIS
operated with only the red arm until 2000. The original blue-side CCD,
used from 2000 to 2002, was an engineering-grade SITe $2048 \times
2048$ pixel CCD. Our typical setup would use the 400/8500 grating
for the red side, either the 400/3400 or 600/4000 grism for the blue
side, and the D56 dichroic, resulting in a wavelength range of
3050--9200~\AA\ and 3200--9200~\AA\ for the respective grisms. There
was typically an overlap region of 5400--5800~\AA\ and 5400--5700~\AA\
for the 400/3400 and 600/4000 grisms, respectively. With our typical
1\arcsec\ slit, this setup yields resolutions of \about7~\AA\
for the red side, and either \about6.5~\AA\ or \about4.5~\AA\ for the
400/3400 and 600/4000 grisms, respectively, for the blue side.
%
%
%
%
%
%

\subsubsection{DEIMOS on the Keck 10~m}

DEIMOS uses a $2 \times 4$ mosaic of $2048 \times
4096$ pixel CCDs with 15~$\mu$m pixels and a spatial scale of
0.1185\arcsec\ pixel$^{-1}$ for a total detector array of $8192 \times
8192$ pixels.  Our typical setup would use the 600/7500 grating with a
GG455 order-blocking filter, resulting in a wavelength range of
4500--9000~\AA.  We would generally use a 1.1\arcsec\ slit which,
along with our typical setup, would result in a resolution of
\about3~\AA.  Occasionally we would use the 1200/7500 grating instead,
yielding a wavelength range of 4800--7400~\AA\ and a resolution of
\about1.5~\AA.  The slit was tilted slightly to provide better sky
subtraction (see Section~\ref{sss:deimos} for details).
%

\subsubsection{ESI on the Keck 10~m}

ESI has an MIT-LL $2048 \times 4096$ pixel CCD with
15~$\mu$m pixels and an average spatial scale of 0.154\arcsec\
pixel$^{-1}$, with the redder orders having a larger spatial scale than the
bluer orders. Our observations were typically performed in the
echellette mode with a 1\arcsec\ wide slit in $2 \times 1$ binning mode
(spatial $\times$ spectral).
This resulted in a resolution of 22~\kms\ across the entire wavelength
range of 3900--11,000~\AA.

\subsection{Standard Observing Procedure}

Unless there was a hardware malfunction, we would observe several dome
flats at the beginning of each night (and occasionally at the end). We
would also observe emission-line calibration lamps (``arcs'') at both
the beginning of the night and 
often at the position of each object. Our final calibrations relied on
observing standard stars throughout the night at a variety of
airmasses. The goal was to obtain at least one standard star (in the
case of single-beam spectrographs; both blue and red standard stars
for double-beam spectrographs) at an airmass near 1.0 and at least one
at an airmass comparable to or higher than the highest airmass of any
SN observed during that night.  The standard stars were typically from
the catalogs of \citet{Oke83} and \citet{Oke90}, with the cool,
metal-poor subdwarfs and hot subdwarfs calibrating the red and blue
sides, respectively. 

Most observations were made at the parallactic angle to reduce
differential light loss \citep{Filippenko82}. Exceptions were usually
at an airmass $< 1.2$ or when the slit was positioned at a specific
angle to include a second object (the host-galaxy nucleus, a trace star, a
second SN, etc.).  In August~2007, LRIS was retrofitted with an
atmospheric dispersion corrector \citep[ADC;][]{Phillips06}.  With the
ADC, differential light loss is substantially reduced regardless of
position angle, even at high airmass.


\section{Data Reduction}\label{s:data}

All data were reduced in a similar, consistent manner by only a
handful of people. Two people were responsible for reducing nearly
half of the data while the work of only five people account for over
90~per~cent of the spectral reductions presented here.
There are slight differences for each
instrument, but the general method is the same.  Previous descriptions
of our methods can be found in \citet{Matheson00:93j},
\citet{Li01:00cx}, \citet{Foley03}, \citet{Foley07}, and
\citet{Matheson08}, but the discussion below supersedes them.


\subsection{Calibration}\label{ss:calibration}

Despite differences between instruments, the general procedure for
transforming raw, two-dimensional spectrograms into fully reduced,
wavelength and flux calibrated, one-dimensional spectra is similar for
all of our data.  We will discuss differences in the procedure for the
various instruments below.  The general prescription is as follows. 

\begin{enumerate}

\item  Correct for bias using an overscan region and trim the
  two-dimensional images to contain only the region with sky data.
  Our data do not typically show a bias pattern and do not have large
  dark currents.  Therefore, we do not subtract bias frames, which
  would increase noise.

\item  Combine and normalise flat-field exposures.  We pay particular 
attention to masking emission lines from the flat-field lamps and absorption
features from the air between the flat-field screen/dome and the
detector.  The normalizing function is generally a low-order spline.

\item  Correct pixel-to-pixel variations in our spectra using our
flat-field exposures.

\item  Extract the one-dimensional spectra.  We use local background
subtraction, attempting to remove as much host-galaxy contamination as
possible.  The spectra are typically optimally extracted using the
prescription of \citet{Horne86}.

However, prior to mid-1997, we did not use the optimal extraction for
our Kast data but, rather, employed ``standard'' extractions (i.e.,
not optimally weighted).  While this typically had minimal impact on
the signal-to-noise ratio (S/N) achieved (most SNe observed with Kast
are quite bright, thus standard and optimal extractions yield similar
noise levels in the final spectrum), one possible effect this has on
our spectra from this era is that they may be spectrophotometrically
inaccurate at the \about5~per~cent level.  This is due to
time-variable spatial focus variations that existed across the CCDs in the
Kast spectrograph.  By using the optimal extraction
since then, we have significantly mitigated the effects of these
variations in our data.

\item  Calibrate the wavelength scale.  Using arc-lamp spectra, we fit
the wavelength scale with a low-order polynomial and linearise the
wavelength solution.  We then make small shifts in the wavelength
scale to match the night-sky lines of each individual spectrum to a master
sky template.

\item  Flux calibrate the spectra.  We fit splines to the continua of
our standard-star spectra, producing a sensitivity function that maps
CCD counts to flux at each wavelength.  These sensitivity functions
are then applied to each individual SN spectrum.

\item  Correct for telluric absorption.  Using our standard-star
spectra, we interpolate over atmospheric absorption regions, providing
an estimate of the atmospheric absorption at a particular time and
airmass.  Then, accounting for the differences in airmass, we apply
these corrections to our spectra, allowing for slight wavelength shifts
between the ``A'' and ``B'' telluric absorption bands.

\item  Remove cosmic rays (CRs) and make other minor cosmetic changes.
  In the remaining one-dimensional spectra there may be unphysical
  features due to CRs, chip gaps, or bad, uncorrected pixels.  We
  interpolate over these features.

\item  Combine overlapping spectra.  For instruments with both a red
and blue side (or multiple orders in the case of ESI) we combine the
spectra, scaling one side to match the other in the wavelength region
where the spectra overlap.  For multiple, successive observations of
the same object we combine the spectra to achieve the highest S/N in
the resulting spectrum, weighting each spectrum appropriately (usually
by exposure time).

\end{enumerate}

Through mid-1997, we used our own VISTA and Fortran
routines to complete all of the above steps.  For about a year after
that we used a combination of generic-purpose IRAF\footnote{IRAF: The
  Image Reduction and Analysis Facility is distributed by the National
  Optical Astronomy Observatory, which is operated by the Association
  of Universities for Research in Astronomy (AURA), Inc., under cooperative
  agreement with the National Science Foundation (NSF).} routines and
our own Fortran routines for our spectral reductions. Since about
mid-1998, we have performed our reductions using both generic-purpose
IRAF routines and our own IDL scripts.  Step 1 is achieved with either
IRAF or IDL depending on the instrument. Steps 2--5 are generally
performed with IRAF, while steps 6--9 are performed in IDL.

We consider the resulting spectra ``fully reduced.''  However, for a
subsample of our spectra where we have multi-filtered host-galaxy
photometry at the position of the SN and SN photometry near the time
the spectrum was obtained, we can make additional corrections to
obtain an accurate absolute flux scale as well as account for
host-galaxy contamination (see Sections~\ref{ss:specphot} and
\ref{ss:galaxy}).

%

\subsubsection{Kast on the Lick 3~m}

The Kast spectrograph has large amplitude, variable fringing on the red-side
CCD.  We observe red-side dome flats at the position of each object and we
apply these flats to each object individually.  As the dome moves into
place to take flats, we also obtain a red-side arc exposure.  Using
this arc spectrum we shift the wavelength solution derived from our afternoon
arc exposures (which typically have more lines and are observed with a
0.5\arcsec\ slit, yielding higher-resolution lines) and apply those
wavelength solutions to the appropriate SN observations.  However, we
still apply a small wavelength shift based on the night-sky lines
later in the reduction process.

\subsubsection{Stover on the Nickel 1~m}

The Stover spectrograph does not have the ability to rotate the slit
with respect to the sky; thus, all spectra obtained with this
instrument were observed with a fixed sky position angle of
$0^\circ$.  When observations were at relatively large airmasses (as
they were for some of the spectra presented here) this caused their
continuum shape to be unreliable.  The spectra in our dataset from the
Stover spectrograph have been previously published
\citep{Wells94,Li01:00cx}, and while strange spectrophotometric calibration
issues when using this instrument with our setup and reduction
routines have been noted by \citet{Leonard02:99em}, \citet{Li01:00cx} find no
such problems.

\subsubsection{LRIS on the Keck 10~m}

The blue side of LRIS has two CCDs offset in the spatial direction
(allowing a full spectrum to be on a single CCD).  We typically
position our objects on the slit so they will be centred on one CCD,
ignoring the other CCD completely.  However, in some observations
circumstances dictated that objects be on the other CCD.  Each CCD
must be calibrated separately (different flat-field response functions,
sensitivity functions, etc.).  The LRIS flat-field lamp is not
particularly hot, providing few photons at the bluest wavelengths of
LRIS.  We therefore mask this region in the flat-field response,
leaving the bluest portions uncorrected for pixel-to-pixel variations.

When our dataset was obtained, the red-side CCD of the spectrograph had
large fringes.  We account for these fringes by applying dome flats
obtained during the afternoon or morning.  We occasionally obtained 
internal flats at the position of an object, but we have found these
to typically be worse for removing pixel-to-pixel variations than the
nightly dome flats.  However, there are rare instances where they were
used instead of dome flats.

\subsubsection{DEIMOS on the Keck 10~m}\label{sss:deimos}

The long slit for DEIMOS is slightly tilted, producing slightly
different wavelengths for a pixel in a given column.  This tilts the
night-sky lines, providing additional sampling of the lines.  Since our
typical procedure is not adaptable to tilted sky lines (our background
subtraction would produce dipoles for every sky line), we implement a
modified version of the DEEP2 DEIMOS
pipeline
\citep{DEEP2,DEEP:pipeline}\footnote{{\tt\url{http://astro.berkeley.edu/~cooper/deep/spec2d/}}.} 
to rectify and  
background-subtract our spectra.  The pipeline bias-corrects,
flattens, traces the slit, and fits a two-dimensional wavelength
solution to the slit by modeling the sky lines.  This final step
provides a wavelength for each pixel.  The slit is then sky subtracted
(in both dimensions) and rectified, producing a rectangular
two-dimensional spectrum where each pixel in a given column has the
same wavelength.  From this point we resume our normal
procedure, starting with extracting the spectrum (step 4).  Since the
spectrum has already been sky subtracted, we do not attempt any
additional sky subtraction (which would only increase the noise)
unless the SN is severely contaminated by its host galaxy.

\subsubsection{ESI on the Keck 10~m}

The CCD of ESI has several large defects which we mask before
starting our reductions.  These produce \about50~\AA\ gaps in our
spectra, usually near 4500--4600~\AA.  They can also affect our
measurement of the trace, but this was rarely a problem for the 
low-redshift, relatively bright SNe presented in this paper.  ESI observes
10 orders; we reduce each order individually and stitch the orders
together at the end (step 9), using our standard-star observations
to determine the scaling and overlap regions.  We weight the spectra
in the overlap by their variance in each pixel before combining.  For
ESI we do not linearise the wavelength solution, but instead rebin to
a common velocity interval, thus producing pixels of different sizes in
wavelength space.

\subsubsection{Other Instruments}

In addition to the aforementioned instruments, our dataset contains a
few spectra which were obtained by observers exterior to the BSNIP
group at observatories aside from Lick and Keck.  These data come from
the Low Dispersion Survey Spectrograph 3 \citep[LDSS-3;][]{Mulchaey05}
mounted on the 6.5~m Clay Magellan II telescope, the R.C.~spectrograph
mounted on the Kitt Peak 4~m telescope, and the Double Spectrograph
mounted on the Hale 5~m telescope at Palomar Observatory
\citep{Oke82}.

\subsubsection{Additional Reduction Strategies}

Occasionally our standard procedures produce nonoptimal spectra.  In
these cases we augment our procedures to produce higher-quality
spectra.

For some spectra we perform a CR cleaning of the two-dimensional
spectra before extraction (step 4).  This procedure is done in IRAF
and detects pixels that have significantly more counts than their
surrounding pixels, replacing them with the local median. Since this
procedure has the potential to remove real spectral features, it is
not automatically performed on every spectrum.

We can obtain better sky subtraction on some spectra by performing a
two-dimensional sky subtraction.  This procedure fits each pixel in
the spatial direction with a polynomial or spline function (usually
constrained to the region near the SN position) and subtracts that fit
from each pixel in that column.  We have found, however, that local
sky subtraction generally produces better results.

On rare occasions, we have multiple dithered images of a single
object.  With these images we can (after proper scaling) subtract one
from another to remove residual fringing and sky lines.  We can also
shift the spectra spatially and combine the two-dimensional spectra to
increase the S/N of the object.  This can produce better traces.

For objects without a defined trace across the entire chip, we would
create a trace function for the object either using the trace of a
nearby object such as the host-galaxy nucleus or of a bright star
(often an offset star) taken in the previous exposure at the same
position as the SN.

\subsection{Spectrophotometry}\label{ss:specphot}

Using our standard reduction procedure outlined in
Section~\ref{ss:calibration}, the {\it relative} spectrophotometry of our data is
usually quite accurate.  However, there are many ways in which the
spectrophotometry may be corrupted.  First, there are
achromatic effects such as clouds that affect the {\it absolute}
spectrophotometry.  Absolute spectrophotometry is not necessary for
many spectroscopic studies (although we will discuss absolute
spectrophotometry in more detail in
Section~\ref{sss:abs_specphot}); accurate relative
spectrophotometry, however, may be important.  There are
many reasons why the relative photometry of a spectrum may be
incorrect, but variable atmospheric absorption, nonparallactic slit
angles leading to differential light losses \citep{Filippenko82}, and
incorrect standard-star 
spectrophotometry can all contribute significant errors. As shown
below, after rigorous testing we find that the relative
spectrophotometry of the 
BSNIP data is accurate to \about0.05--0.1~mag across most of the
wavelengths covered by the spectra.

Occasionally, events external to the normal operations of the
telescope and data reduction can result in questionable
spectrophotometry.  Instrument failures (e.g., a broken shutter) or
environmental effects (e.g., nearby wild fires) are the most
troublesome.  There is no clear way to fully correct the
spectrophotometry in these cases.  Using our detailed records as well
as those of Lick Observatory, we have identified several spectra where
the spectrophotometry may be affected by these external factors and
exclude them from any estimates of the fidelity of our
spectrophotometry.  Including spectra obtained with the Stover
spectrograph, which does not have a rotator and so nearly all spectra
were not observed at the parallactic angle, we have flagged 88 spectra
as having possibly troublesome spectrophotometry.

\subsubsection{Relative Spectrophotometry}\label{ss:relspecphot}

Two of the key attributes of the BSNIP sample are the large wavelength
range and the consistent and thorough reduction procedures.  The
spectra in the sample likely have similar systematic (and hopefully
small) uncertainties.  The large wavelength range makes the spectra ideal 
for comparing near-ultraviolet and near-infrared features in a single 
spectrum, but such investigations will be limited by the accuracy of our
spectrophotometry.  Since most of the spectra in our sample have
corresponding \bvri light curves \citep{Ganeshalingam10:phot_paper},
we can test the spectrophotometry of a spectrum by comparing synthetic
colours from the spectrum to those of the light curves at the time that
the spectrum was obtained.  In fact, this has previously been
performed on some of the data presented herein (at a somewhat less
rigorous level) by \citet{Poznanski02}.

For this test, we examine only the spectra of objects that have
corresponding filtered light curves.  To assure that our estimates of
the SN colours from the photometry are accurate, we further limit the
sample to spectra that have a light-curve point within 5~d of when the
spectrum was taken.

We use the light-curve fitter ``Multi-colour Light Curve Shape''
\citep[MLCS2k2,][]{Jha07} to model the filtered light curves, allowing
us to interpolate between data points. We fit each filter individually
to provide the largest degree of flexibility in each, and all of the fits 
are inspected to ensure that a good fit is obtained. In cases where the
MLCS2k2 fit does not adequately reflect the data and the data are well 
sampled, we use a cubic spline with a Savitzky-Golay smoothing filter
\citep{Savitzky64}.

We estimate the uncertainties in the model light curve by running a series 
of Monte Carlo simulations. For each data point, we randomly draw from a 
Gaussian distribution with mean given by the reported magnitude and 
$\sigma$ by the photometric uncertainty to produce a simulated data point.
 Each simulated light curve is refit. This process is repeated 50 times 
and the scatter in the derived light curves is taken as the uncertainty in 
the model. This process is applied to objects with MLCS2k2 and spline fits.

To determine the synthetic photometry from the spectra, we convolve
each spectrum with the Bessell filter functions \citep{Bessell90}.  We
calibrate our photometry by measuring the spectrophotometry of the standard 
star BD+17$^{\circ}$4708 \citep{Oke83} and applying zero-point offsets
to match the 
standard photometry.  We then apply these offsets to the synthetic
photometry derived from the SN spectra.  The Bessell filter functions
have approximate wavelength ranges of 3700--5500, 4800--6900,
5600--8500, and 7100--9100~\AA\ for $B$, $V$, $R$, and $I$,
respectively.  Most of our spectra fully cover the \bvri bands.

There are several effects which may reduce the accuracy of our
spectrophotometry.  By far, the most important is galaxy
contamination.  Although our reduction process removes as much galaxy
light as possible from a SN spectrum (see
Section~\ref{ss:calibration}), some of our SN spectra are still
contaminated by galaxy light.  The measured synthetic colours from 
galaxy-contaminated spectra will likely be vastly different from the
SN colours even if our spectrophotometry is excellent.  For spectra
with multi-colour template images of the host galaxy and multi-colour
light curves concurrent with the spectrum, we can correct for galaxy
contamination to a large degree (see Section~\ref{ss:galaxy}).
However, this correction relies on excellent relative
spectrophotometry.

\begin{figure*}
\begin{center}
\includegraphics[width=\textwidth]{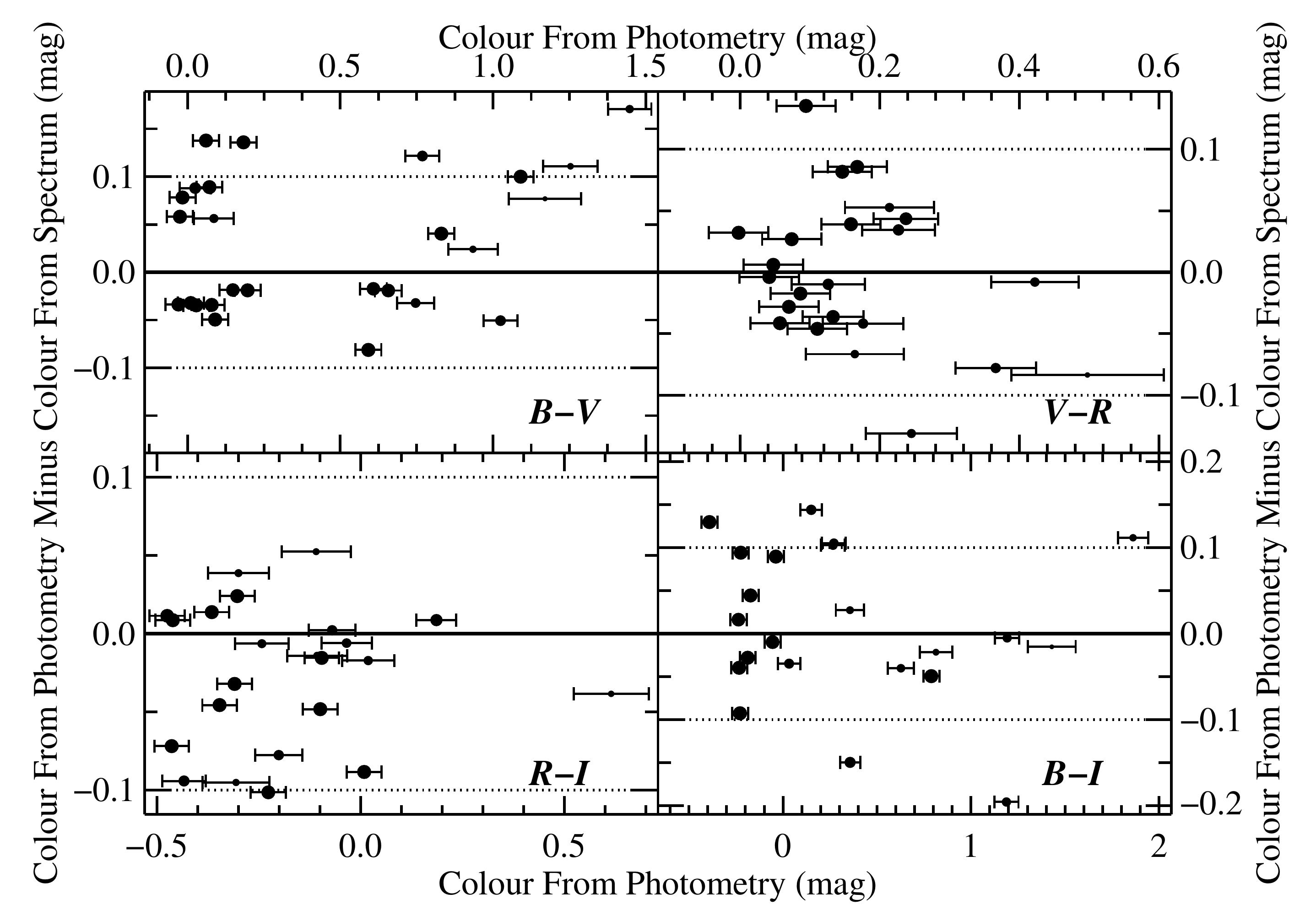}
\caption[Comparison of synthetic to measured colours with no host
contamination]{Comparison of synthetic colours derived from our
  spectra to those measured from light curves at the same epoch.  
  Only spectra of objects where there is no obvious galaxy contamination 
  at the position of the SN (as determined from late-time imaging) are 
  included.  Clockwise
  from the upper-left panel, we present the $B-V$, $V-R$, $B-I$, and
  $R-I$ colours.  The size of each circle represents the size of the
  photometric uncertainty, with larger circles representing smaller
  uncertainty. The dotted lines in each panel are residuals of $\pm
  0.1$~mag.}\label{f:specphot_lc} 
\end{center}
\end{figure*}

We have selected a subsample of SNe that are relatively isolated from
their host galaxy, so their spectra should have minimal
galaxy contamination.  These objects all have template images (taken
after the SN had faded) that indicate minimal galaxy light.  A sample
of spectra of objects from this low-contamination sample of SNe is
constructed to test the fidelity of our relative spectrophotometry.
For this sample, we require that the spectra have $t < 30$~d and that
the spectrum was obtained at the parallactic angle or at an airmass
$\le 1.2$.  We present the synthetic and photometric colours for the
low-contamination sample in Figure~\ref{f:specphot_lc}.  Although the
number of spectra in this sample is limited, they span a large range
of colour.

\begin{figure*}
\begin{center}
\includegraphics[width=\textwidth]{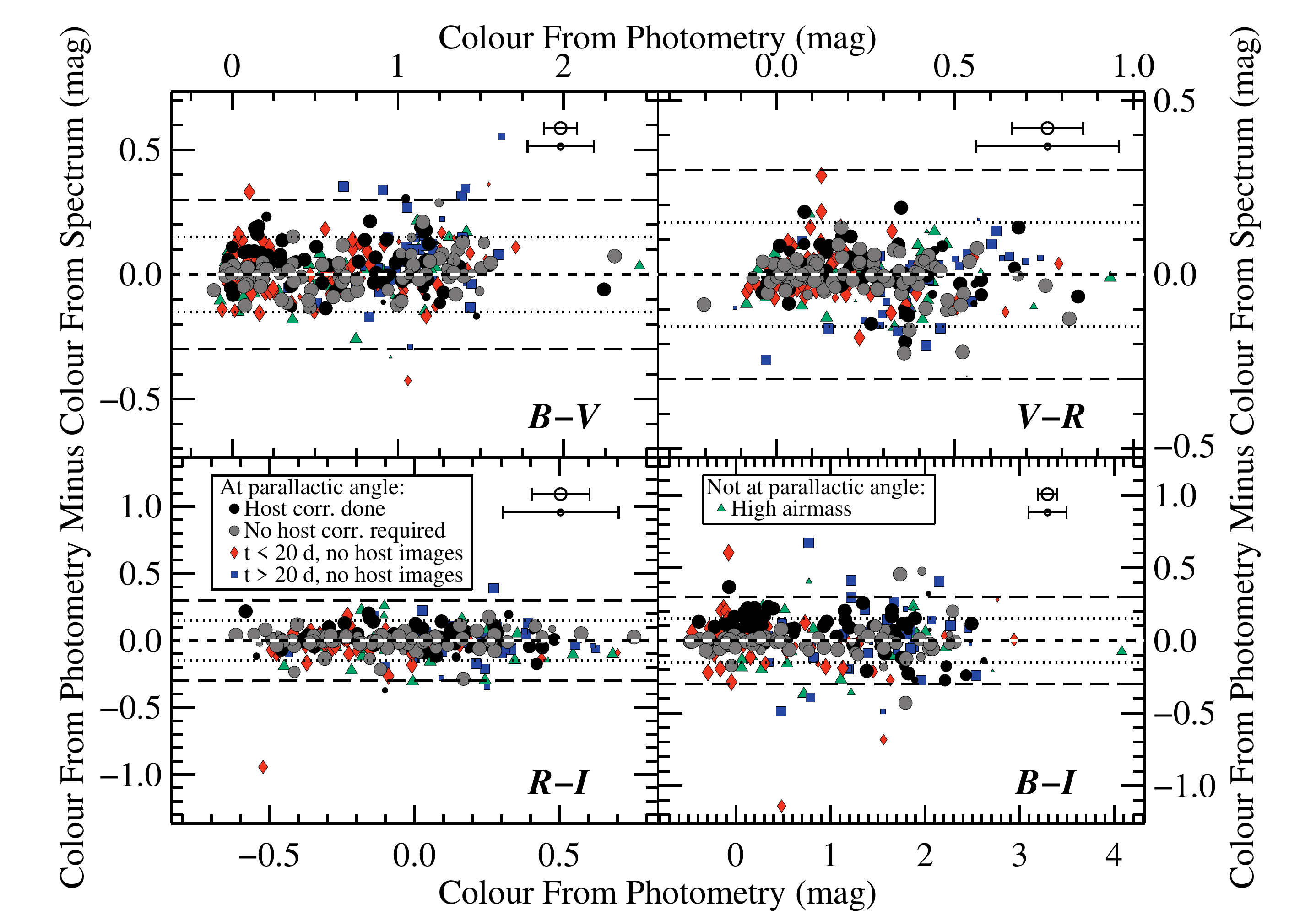}
\caption[Comparison of synthetic to measured colours]{Comparison of
  synthetic colours derived from our spectra to 
those measured from light curves at the same epoch.  Clockwise from
the upper-left panel, we present the $B-V$, $V-R$, $B-I$, and $R-I$
colours.  The green triangles, blue squares, red diamonds, grey
circles, and black circles represent (respectively) spectra not observed 
at the parallactic angle and at high airmass; spectra with $t > 20$~d
observed at the parallactic angle, but lacking host-galaxy images to
perform host-galaxy subtraction; spectra with $t < 20$~d observed
at the parallactic angle, but lacking host-galaxy images to perform
host-galaxy subtraction; spectra observed at the parallactic angle,
with host-galaxy images, but no host-galaxy subtraction is required; 
and spectra observed at the parallactic angle that have been corrected
for host-galaxy contamination.  The size of each symbol
represents the size of the photometric uncertainty, with larger
symbols representing smaller uncertainty.  Representative sizes (0.1
and 0.2~mag) are shown with error bars in the upper-right corner of
each panel. The dotted and long-dashed lines in each panel are
residuals of $\pm 0.15$~mag and $\pm 0.3$~mag,
respectively.}\label{f:specphot}  
\end{center}
\end{figure*}

We present a comparison of synthetic colours derived from our
low-contamination and possibly contaminated spectra to those measured
from light curves at the same epoch in Figure~\ref{f:specphot}.  An
estimate of the uncertainty in the spectrophotometry can be made by
examining the $\chi^{2}$ per degree of freedom (dof) of the residual
of the synthetic to photometric colours.  The uncertainty in the
photometric colours is measured by examining the residuals of the
photometry measurements near the epoch of the spectrum relative to the
model.  The uncertainty in the relative spectrophotometry is the
uncertainty added to each point which causes the residuals of the
synthetic to photometric colours to have $\chi^{2} / \dof = 1$.  If
$\chi^{2} / \dof \le 1$ with only photometric uncertainties, then the
spectrophotometry does not have uncertainties larger than the
photometry itself.  We present estimates of the uncertainties in
Table~\ref{t:specphot}.

\begin{table*}
\begin{center}
\caption{Relative Spectrophotometric Accuracy for the BSNIP Sample}\label{t:specphot}
\begin{tabular}{lccccr}
\hline\hline
\multicolumn{1}{c}{ } &
\multicolumn{4}{c}{Additional Uncertainty to Achieve $\chi^{2} / \dof = 1$} &
\multicolumn{1}{c}{\# of Spectra} \\
\cline{2-5}
Subsample & $B-V$ (mag) & $V-R$ (mag) & $R-I$ (mag) & $B-I$ (mag) & (for $V-R$) \\
\hline
Low contamination           & 0.057 & 0.000 & 0.008 & 0.067 &  23 \\
\hline
All spectra              & 0.095 &  0.055 &  0.096 &  0.170 &  306 \\
Not parallactic          & 0.089 &  0.048 &  0.107 &  0.158 &   48 \\
Parallactic              & 0.097 &  0.056 &  0.094 &  0.171 &  258 \\
Gal.\ sub.\ -- no corr.\ & 0.088 &  0.053 &  0.073 &  0.140 &   67 \\
Gal.\ sub.\ -- corr.\    & 0.057 &  0.042 &  0.055 &  0.100 &   81 \\
No gal.\ sub.; $t > 20$~d & 0.151 &  0.075 &  0.108 &  0.224 &   47 \\
No gal.\ sub.; $t \leq 20$~d & 0.093 &  0.060 &  0.133 &  0.223 &   63 \\
\hline
${\rm Airmass} \leq 1.1$    & 0.081 &  0.050 &  0.061 &  0.154 &   24 \\
$1.1 < {\rm Airmass} \leq 1.3$ &  0.078 &  0.017 &  0.067 &  0.118 &   37 \\
$1.3 < {\rm Airmass} \leq 1.5$ &  0.076 &  0.048 &  0.056 &  0.080 &   37 \\
${\rm Airmass} > 1.5$    & 0.064 &  0.060 &  0.067 &  0.126 &   50 \\
\hline
${\rm S/N} < 20$         & 0.104 &  0.093 &  0.088 &  0.195 &   18 \\
$20 \leq {\rm S/N} < 30$    & 0.077 &  0.026 &  0.074 &  0.152 &   16 \\
$30 \leq {\rm S/N} < 40$    & 0.065 &  0.055 &  0.061 &  0.099 &   27 \\
$40 \leq {\rm S/N} < 50$    & 0.073 &  0.020 &  0.077 &  0.088 &   31 \\
${\rm S/N} \geq 50$         & 0.065 &  0.036 &  0.036 &  0.102 &   56 \\
\hline
Reduced by T.\ Matheson  & 0.108 &  0.065 &  0.115 &  0.211 &    6 \\
Reduced by R.\ Chornock  & 0.065 &  0.050 &  0.062 &  0.101 &   34 \\
Reduced by R.\ Foley     & 0.058 &  0.041 &  0.040 &  0.107 &   43 \\
Reduced by J.\ Silverman & 0.085 &  0.054 &  0.081 &  0.144 &   51 \\
Reduced by T.\ Steele    & 0.071 &  0.022 &  0.034 &  0.043 &   11 \\
\hline\hline
\end{tabular}
\end{center}
\end{table*}
\normalsize

For the low-contamination sample, the spectrophotometry has a typical
additional uncertainty of $\le 0.07$~mag across the entire spectrum
(i.e., $B-I$), with no additional uncertainty required for $V-R$ and
very little additional uncertainty (0.008~mag) required for $R-I$
across a large range of colours.  Our entire sample is only slightly
worse, with the additional uncertainty in $V-R$ being 0.055~mag.

This implies that the accuracy of the relative flux calibration for
the low-contamination sample is difficult to assess since the
uncertainties from the photometry are enough to account for the
majority of the scatter in the synthetic colours (and the entire
scatter for the wavelength region spanning from $V$ to $R$).
Nonetheless, we can place limits on the accuracy based on the
additional uncertainty required and the standard deviation.  From
this, we find that the low-contamination sample is accurate to
5.2--6.9~per~cent, 0.0--5.8~per~cent, 0.7--4.5~per~cent, and
6.0--9.0~per~cent for the wavelength regions spanning $B$ to $V$, $V$
to $R$, $R$ to $I$, and $B$ to $I$, respectively.  For the sample of
objects corrected for galaxy contamination, the additional errors are 
similar to those of the low-contamination sample (5.3--6.5~per~cent,
3.9--4.8~per~cent, 4.9--5.1~per~cent, and 4.5--9.2~per~cent for the
wavelength regions listed above), but lower than those for the entire
sample (8.8~per~cent, 5.1--6.2~per~cent, 7.3--8.9~per~cent, and
12.7--15.6~per~cent), indicating that the galaxy-contamination
correction works well at least for broad-band colours.

Additionally, we have split our sample by various spectral attributes.
The spectrophotometry does not depend significantly on airmass.  It
does depend significantly on S/N, but the spectrophotometry does not
improve as S/N increases beyond ${\rm S/N} \approx 20$~pixel$^{-1}$. The 
additional uncertainties also depend slightly on the individual who reduced
the spectra.  However, this trend may be the result of observation and 
reduction techniques slowly improving over time.

We have also calculated the mean and standard deviations of the
difference between the synthetic colours derived from our spectra and
those measured from light curves for the various subsamples.  All
subsamples have a mean that is $< 0.6$ standard deviations from zero, with
nearly all being $< 0.3$ standard deviations from zero.  The means for
the subsamples are also typically $< 0.02$~mag from zero, with no
clear bias in any particular subsample. Furthermore, there are very few
significant outliers in any colour, with only 2--5~per~cent of the
spectra (depending on the colour) $>2\sigma$ away from zero.

In summary, our relative spectrophotometry is excellent.  In
particular, objects with little galaxy contamination or those where we
are able to correct for galaxy contamination have extremely good
relative spectrophotometry.  This is achieved simply through our
reduction methods and the relatively simple host-galaxy contamination
correction outlined below; there is no spectral warping of any kind to
achieve these results.

\subsubsection{Absolute Spectrophotometry}\label{sss:abs_specphot}

As mentioned above, there are many achromatic effects which can affect
our absolute spectrophotometry.  We can correct for these effects if
we have concurrent photometry.  For these cases, we determined the
synthetic photometry of our spectra and applied a multiplicative
factor to scale the synthetic photometry to match our true
photometry.  This scaling is a byproduct of correcting for
host-galaxy contamination, as described in Section~\ref{ss:galaxy}.

\subsection{Host-Galaxy Contamination}\label{ss:galaxy}

SNe generally do not exist in isolation.  The vast majority occur within
galaxies, sometimes close to or on top of complex regions such as
spiral arms or \ion{H}{II} regions.  With photometry, one can correct
for this by obtaining a template image after the SN has faded (or in
some cases, before the star explodes), and subtracting the template
from the image with the SN, leaving only the SN.  Although this approach is
also feasible with  
spectroscopy (obtaining a spectrum at the
position of the SN after it has faded), it is not practical.
Spectroscopy time is typically more valued, and reproducing the exact
conditions at the time of the original SN observation is difficult.
We do, however, have methods for reducing the galaxy contamination in
a SN spectrum.

The first method is local
background subtraction, as described in
Section~\ref{ss:calibration}.  Briefly, while extracting the SN
spectrum, we model the underlying background by interpolating between
background regions on either side of the SN.  If the background is
relatively smooth and monotonic between the background regions, this
method works very well.  However, if the SN is near the nucleus of a
galaxy or on a spiral arm or other bright feature, this method can
underestimate the background, leaving galaxy contamination in the SN
spectrum.

We have derived a method for removing the residual galaxy
contamination from our SN spectra.  This approach, which we call
``colour matching,'' requires both SN photometry at the time the
spectrum was obtained, and template colours for the host galaxy at the
position of the SN.  We use the host-galaxy colours to determine the
spectral energy distribution (SED) 
of the host galaxy at the position of the SN.  We then subtract
the host-galaxy SED from the SN spectrum, scaled so that the synthetic
photometry from the galaxy-corrected SN spectrum matches the SN
photometry.  This method was first presented by \citet{Foley10:uv}; we
discuss it in detail below.

\subsubsection{Determining the Host-Galaxy SED}

The parameter space of galaxy SEDs is well known and well behaved,
allowing one to reliably reconstruct galaxy SEDs with broad-band
photometry.  Adopting the approach
described by \citet{Blanton03}, but updated by \citet{Blanton07} to 
include UV wavelengths, and implemented in the IDL software package
\texttt{kcorrect.v4\_1\_4}, we have used our \bvri photometry of the
host galaxy at the position of the SN and the redshifts presented in
Table~\ref{t:obj} to reconstruct the galaxy SED at the position of
the SN.  We perform aperture photometry on galaxy templates obtained
as part of the Lick Observatory SN Search (LOSS) follow-up photometry
effort \citep{Ganeshalingam10:phot_paper} using a 3~pixel (2.4\arcsec, 
similar to our Kast slit size and the typical seeing at Lick Observatory)
aperture and taking the median pixel value of the image to represent
the sky background. Using a 3~pixel aperture for all of our galaxy
templates will represent different physical sizes depending on the
distance to the galaxy.  An aperture significantly different from that
of the slit combined with the seeing could incorporate flux from
stellar populations that do not represent the SED of the galaxy at the
position of SN.  As a check on how aperture size affects measured
galaxy colour, we also used a 4~pixel aperture fixed at the SN
position. We find excellent agreement between the colours derived
using a 3~pixel aperture with a mean difference $\leq 0.02$~mag. For
the typical galaxy with $z < 0.5$ (which includes all redshifts
presented here), the SEDs are recovered to $\la 0.02$~mag in all
filters \citep{Blanton03, Blanton07}.

\subsubsection{Colour Matching}

\paragraph{Motivation}

One approach to subtract galaxy contamination from a SN is to extract
the SN without any local background subtraction, creating a spectrum
that consists of all light at the position of the SN (including galaxy
light) at the time of the spectrum.  If one also has photometry at
that epoch, one can, in principle, scale a galaxy SED to match the
galaxy photometry, scale the spectrum to match the addition of the SN
and galaxy photometry, and subtract the latter from the former to
obtain a SN spectrum \citep[e.g.,][]{Ellis08}.  The main drawbacks of
this method are that (1) one must know the proper point-spread
function (PSF) of the SN and galaxy when the spectrum was obtained,
and (2) if there is a significant amount of galaxy contamination and
the galaxy SED is incorrect, significant errors will be introduced.

When extracting our spectra, we attempt to remove as much galaxy
contamination as possible.  This approach has
the benefit of reducing the galaxy contamination in the SN spectrum
without introducing potential errors associated with an imprecise 
photometrically reconstructed galaxy SED.  Also, considering the lack
of precise observing information for many of our spectra (which date
back over two decades), it would be difficult to estimate the correct
PSF to determine the exact galaxy flux (both SED and amount) entering
our slit for a given observation.

Since the galaxy colours from photometry (which are easier to measure
than the absolute flux entering our slit) determine the galaxy SED, if
our spectrophotometry is well calibrated then simply subtracting the
galaxy SED until the colours of the spectrum match those of the SN
photometry will result in a SN-only spectrum.

We can demonstrate this mathematically.  In general, an observed SN
spectrum is defined by
\begin{equation}\label{e:spec}
  f_{\text{spec}} = A \left ( f_{\text{SN}} + Bf_{\text{gal}} \right ),
\end{equation}
where $f_{\text{spec}}$, $f_{\text{SN}}$, and $f_{\text{gal}}$ are the
vectors of fluxes in the observed spectrum, SN-only spectrum, and
galaxy spectrum, respectively, and $A$ and $B$ are normalisation
factors.  One can think of $A$ as normalizing the spectrum in an
absolute sense to account for slit losses, clouds, and other
achromatic effects.  The parameter $B$ controls the amount of galaxy
contamination, where $B = 0$ if there is no galaxy contamination and
we impose $B \geq 0$. In principle, $B$ could be negative in order to
correct for oversubtraction of galaxy light, but our testing indicates
that allowing $B$ to have negative values produces too much
overfitting of the spectra.

From our image templates, we have $p_{\text{gal}}$, the broad-band
photometry (in flux units) for the host galaxy at the position of the
SN.  Using MLCS2k2 \citep{Jha07} template light curves or spline
interpolations (see Section~\ref{ss:relspecphot}), we are able to
interpolate our SN photometry (independently in each band) to
determine $p_{\text{SN}}$, the broad-band photometry (in flux units)
for the SN at the time the spectrum was obtained.

We can define the function which translates spectra to synthetic
broad-band photometry as $P$, where $P(f_{\text{SN}}) = p_{\text{SN}}$
and $P(f_{\text{gal}}) = p_{\text{gal}}$.  This function is equivalent
to convolving a spectrum with a filter function.  Note that we impose
the first relationship, while the second relationship is required by
our method of determining $f_{\text{gal}}$.

From our spectrum, we are able to determine $p_{\text{spec}} =
P(f_{\text{spec}})$, the broad-band synthetic photometry (in flux
units) of the spectrum, which includes both SN and galaxy light.
These vectors then obey the equation
\begin{equation}\label{e:phot}
  P(f_{\text{spec}}) = A \left ( p_{\text{SN}} + Bp_{\text{gal}} \right ).
\end{equation}

For Equation~\ref{e:phot} to be valid, we make two assumptions.  The
first assumption, which is already noted above, is that our spectra
have accurate relative spectrophotometry.  The second assumption is
that $B$, the relative fraction of the galaxy and SN light, does not
vary strongly with wavelength.  From Section~\ref{ss:relspecphot}, we
have shown that the relative spectrophotometry of our spectra is
accurate to \about0.05--0.1~mag across large wavelength regions,
comparable to the uncertainties of our photometry (after interpolating
to a given date).

Solving for $f_{\text{SN}}$ in Equation~\ref{e:spec}, we have
\begin{equation}\label{e:galsub}
  f_{\text{SN}} = A^{-1} f_{\text{spec}} - Bf_{\text{gal}}.
\end{equation}
With a spectrum spanning at least two bands also covered by SN and
galaxy photometry, one can solve for $A$ and $B$ from
Equation~\ref{e:phot}.  With galaxy photometry, the galaxy SED
($f_{\text{gal}}$) can be properly reconstructed.  It is then simple to
determine the uncontaminated SN spectrum ($f_{\text{SN}}$) from the
galaxy-contaminated, observed spectrum ($f_{\text{spec}}$).  We note
that if $B = 0$, then Equation~\ref{e:galsub} simplifies to merely
scaling the spectrum to match the photometry in an absolute sense.

\paragraph{Testing}

To test this method, we have performed Monte Carlo simulations on six
different spectra with increasing galaxy contamination and appropriate 
photometric errors.  Three of the spectra are linear (in
$f_{\lambda}$) and have negative, zero, and positive slopes
(corresponding to blue, flat, and red spectra). The other three
spectra are SN~2005cf at maximum brightness, \about1~month
after maximum, and \about1~yr after maximum.  
To each of these spectra we added 5 galaxy templates,
those used by the Sloan Digital Sky Survey (SDSS) to perform redshift
cross-correlations, spanning early to late galaxy
types.\footnote{{\tt\url{http://www.sdss.org/dr6/algorithms/spectemplates/}}}
We measured 
the synthetic photometry of the spectra and galaxy templates, and for
each iteration we varied the photometric data randomly using a normal
distribution with width corresponding to the median error in each band
for SNe and galaxies, respectively.  We then performed the
colour-matching technique for the galaxy-contaminated spectra with the
Monte-Carlo-based photometry.

\begin{figure}
\begin{center}
\includegraphics[width=.45\textwidth]{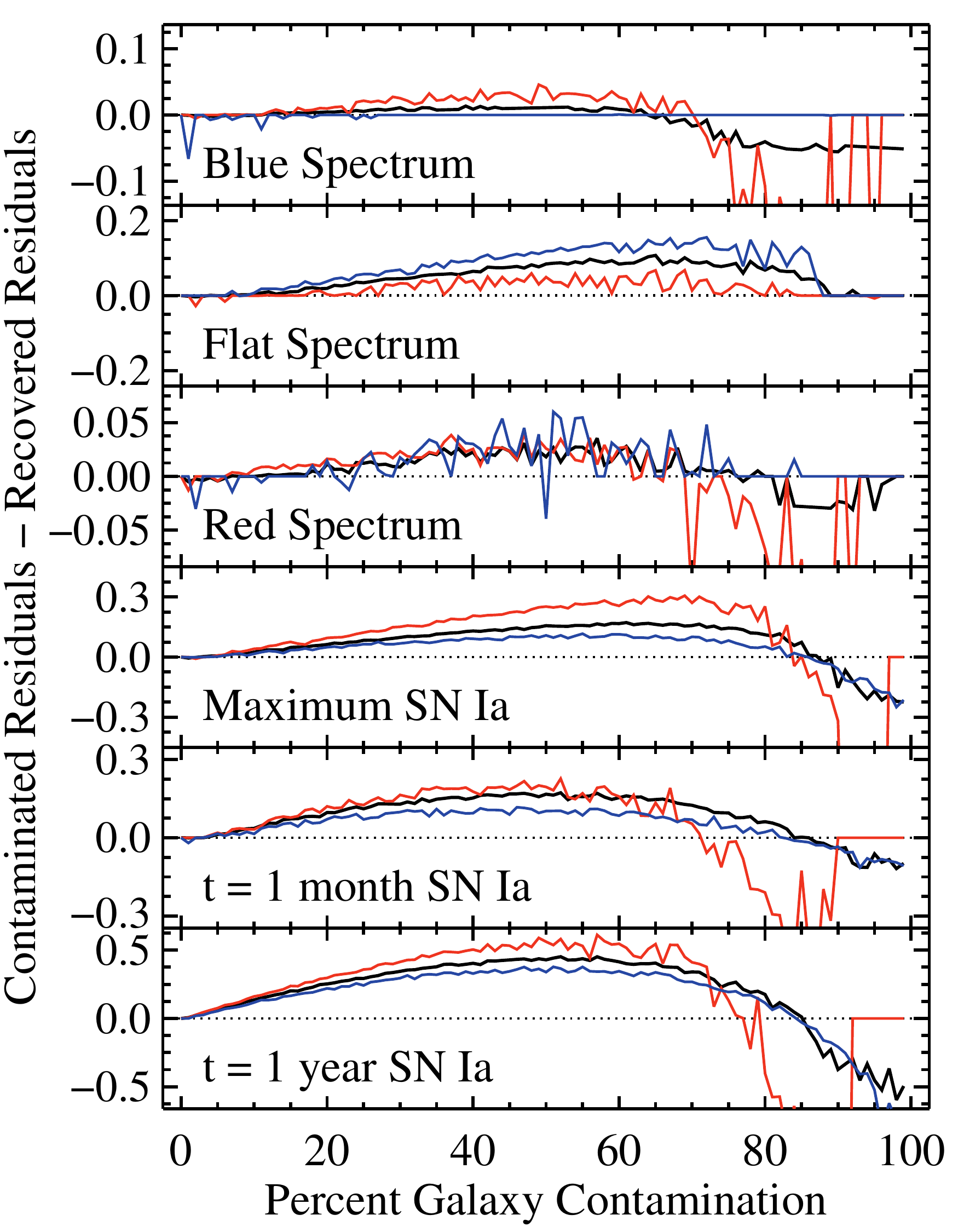}
\caption[Differences between contaminated spectra and recovered spectra after
  colour matching]{Differences between the median of the standard deviation of
  the residuals of contaminated spectra and recovered spectra after
  colour matching 
  for different input spectra and galaxy templates with varying
  amounts of galaxy contamination.  The top through bottom panels
  correspond to input spectra of a blue linear spectrum, a flat linear
  spectrum, a red linear spectrum, SN~2005cf at maximum brightness,
  SN~2005cf \about1~month after maximum, and SN~2005cf
  \about1~yr after maximum, respectively.  Positive values
  imply that our colour-matching technique yields spectra that are
  closer to the input spectra than the contaminated spectra are. The
  horizontal dotted line in each panel represents where the residuals
  of the recovered spectra are equal to those of the contaminated
  spectra. The blue and red lines correspond to the latest and
  earliest galaxy templates, respectively. The black lines correspond
  to the average over 5 galaxy templates.}\label{f:gal_sub_err}
\end{center}
\end{figure}

Our recovered SN spectra were compared to our input spectra, and the
{\it differences} between the standard deviation of the residuals of
the contaminated spectra and the recovered spectra are presented in
Figure~\ref{f:gal_sub_err}.  We see that the residuals for the
recovered spectra are significantly lower (i.e., the recovered spectra
are better at reproducing the input spectra) than the contaminated
spectra for galaxy contaminations $< 70$~per~cent.  The improvement
does depend somewhat on the colour of the SN spectrum and the colour
of the galaxy template, but the differences are relatively small.  At
higher levels of galaxy contamination, the gains are minimal in this
metric, but examining the spectra, it is obvious that this technique
yields impressive results even with very large amounts of galaxy
contamination.

\begin{figure*}
\begin{center}
\includegraphics[width=6.5in]{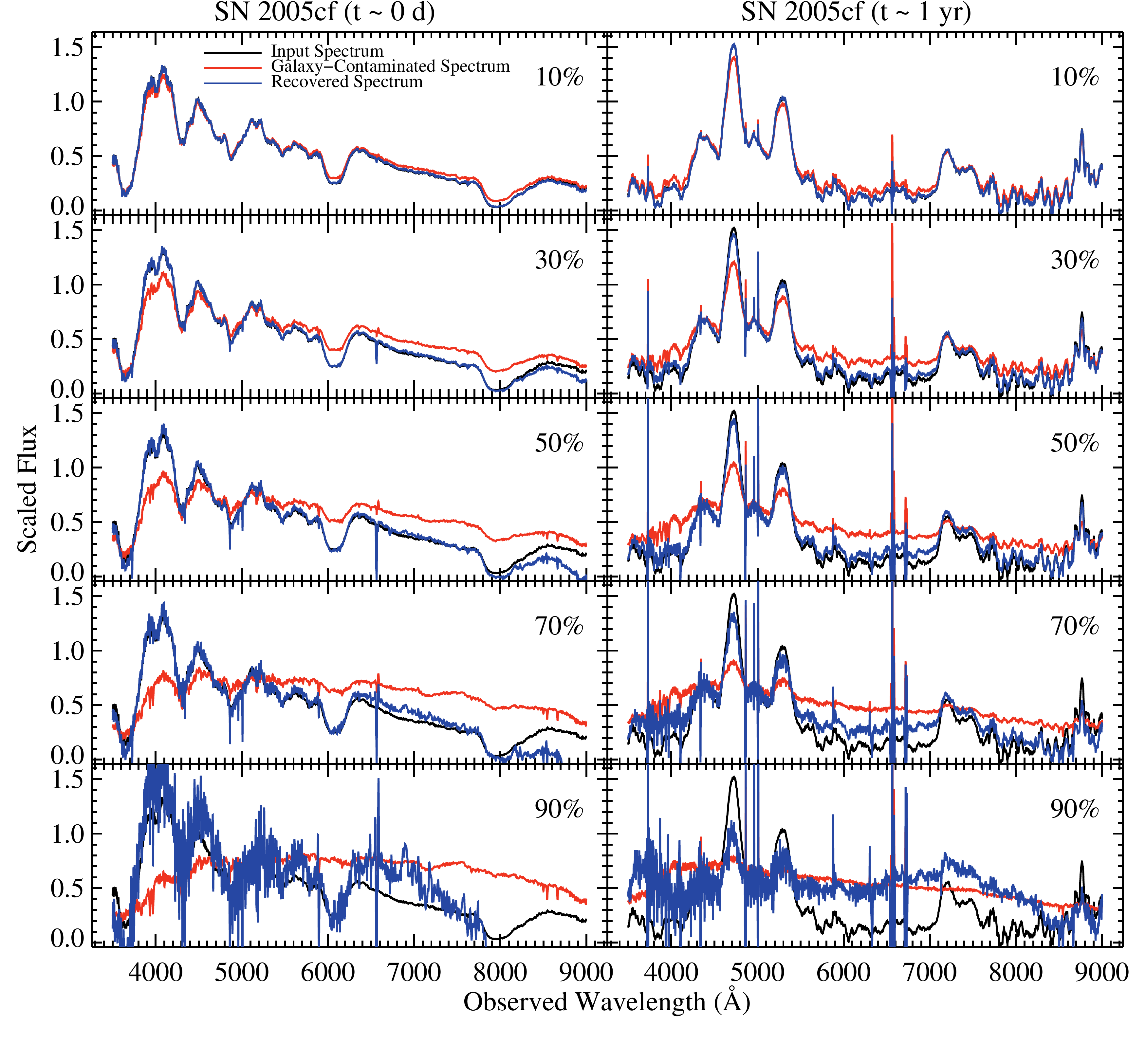}
\caption[SN spectra used for testing the colour-matching method]{SN
  spectra used for testing the colour-matching method.  The 
  left and right columns correspond to SN~2005cf at maximum brightness
  and \about1~yr after maximum, respectively.  Each row
  represents different amounts of galaxy contamination, from
  10~per~cent ({\it top}) to 90~per~cent ({\it bottom}); each panel
  has the 
  percent contamination labelled.  Each panel shows the input spectra
  (black), average galaxy-contaminated spectra (from several
  Monte Carlo realisations, red), and average recovered spectra (from
  several Monte Carlo realisations, blue).  The maximum-light and
  nebular-phase spectra have been contaminated with early-type and
  late-type galaxy templates, respectively.  The spectra are scaled to
  have the same median value. Comparing the black
  (input) spectra to the blue (recovered) spectra gives an indication
  of how well the colour-matching method works. In the top panels, it
  is difficult to see the input spectra because of how closely the
  recovered spectra match the input spectra. However, even at low
  levels of contamination, residuals from narrow emission lines in the
  galaxy spectra are seen in the recovered SN
  spectra.}\label{f:gal_sub_spec}
\end{center}
\end{figure*}

In Figure~\ref{f:gal_sub_spec}, we present our maximum-light and
nebular-phase spectra of SN~2005cf with varying amounts of galaxy
contamination.  At 70~per~cent galaxy contamination, where the
residuals are not very large, we see that the overall shape of the
spectra and spectral lines are well recovered.  Even at 90~per~cent
galaxy contamination, where the contaminated spectra appear to be
simply galaxy spectra, the method is able to recover the overall shape
of the SN spectrum.

The recovered spectra differ most at the ends of the spectral
coverage.  This is due to the galaxy SED reconstruction being
unconstrained beyond these wavelengths.  If we extended our galaxy
photometry beyond \protect\hbox{$BV\!RI$}, this would improve.  The
emission lines of the reconstructed galaxy spectra generally have the
incorrect strength.  This is difficult to model with broad-band
photometry, and these regions of the spectra should be ignored.  The
majority of the differences between the input and recovered spectra
are the result of incorrect galaxy SED reconstruction from errors in
the galaxy photometry.  Improving the galaxy photometry or increasing
the number of bands of galaxy photometry would improve the
reconstruction of the galaxy SED.  As the galaxy contamination
increases, the errors in the reconstructed galaxy SED are amplified.

\paragraph{Implementation}

We have applied this technique to all SN spectra that have (1) \bvri
photometry within 5~d of when the spectrum was taken, and (2) a
wavelength range which spans at least two observed bands.  Spectra
which cover only a single observed band are scaled to match the
photometry at the time of the spectrum. 

The procedure used to subtract galaxy light from an observed spectrum
is as follows.  Using \texttt{kcorrect}, the galaxy SED is
reconstructed from the broad-band galaxy photometry at the position of
the SN. Synthetic photometry is measured from the observed spectrum.
The SN photometry at the time of the spectrum is measured from the
light curves as described in Section~\ref{ss:relspecphot}.  Using
Equation~\ref{e:phot} above, the factors $A$ and $B$ are determined
using a $\chi^{2}$ minimisation technique.  Using
Equation~\ref{e:galsub}, the reconstructed galaxy SED is subtracted
from the observed spectrum to produce the corrected SN spectrum.


\section{Data Management and Storage}\label{s:sndb}

When preparing to present a dataset as large as ours, we required some
sort of internal organised storage and retrieval method. The overall
utility of our dataset will also be greatly increased by having a
user-friendly interface to access the data. In addition to the final
data products, all other information regarding both 
our photometric and spectroscopic samples is stored in our SN Database
(SNDB).  The SNDB holds information about individual SNe (such as
host-galaxy information, type, discovery information, etc.), much of 
which comes from external, online resources.\footnote{For example, IAU
  Central Bureau for Astronomical Telegrams
  ({\tt\url{http://www.cbat.eps.harvard.edu/lists/Supernovae.html}}), 
   NASA/IPAC Extragalactic Database (NED,
   {\tt\url{http://nedwww.ipac.caltech.edu/}}), and Rochester Academy
   of Sciences Bright Supernova List
   ({\tt\url{http://www.rochesterastronomy.org/snimages/}}).} 
The SNDB also contains information regarding individual spectra (such
as observing conditions, instrument, 
resolution, etc.), and individual light curves (number of
points, photometric accuracy, derived light-curve parameters from
various fitting routines, etc.).  A complete list of all fields 
stored in the SNDB can be found in Table~\ref{t:SNDB_fields}.

\begin{table*}
\begin{center}
\caption{SN Database (SNDB) Fields}\label{t:SNDB_fields}
\begin{tabular}{lll}
\hline\hline
SN Name & SNID-Determined Subtype$^\textrm{a}$	 & Host-Galaxy Name \\
Right Ascension & Discovery Date & Host-Galaxy Type	 \\
Declination & Discoverer 	& Host-Galaxy Redshift (and error) \\
Type & Discovery Reference & SN Redshift (and error)	\\
Type Reference & Galactic Reddening  & Other Notes \\
\hline \hline
\multicolumn{3}{c}{Photometry and Light-Curve Information} \\
\hline
\# of Total Photometry Points	 & $\Delta m_{15}(B)$ (and error) & Julian Date of $B$-band Maximum (and error) \\
\# of $B$-band Photometry Points	 & Maximum $B$-band Magnitude (and error) & Plots of MLCS2k2 Fits$^\textrm{b}$ \\
\# of $V$-band Photometry Points	 & $\left(B-V\right)_{B_\textrm{max}}$ (and error) & MLCS2k2 Distance Modulus (and error)$^\textrm{b}$ \\
\# of $R$-band Photometry Points	 & SALT/2 Distance Modulus (and error)$^\textrm{c}$ & MLCS2k2 $A_V$ (and error)$^\textrm{b}$ \\
\# of $I$-band Photometry Points	 & SALT/2 Light-Curve Stretch (and error)$^\textrm{c}$ & MLCS2k2 $R_V$ (and error)$^\textrm{b}$ \\
\# of Unfiltered Photometry Points	 & SALT/2 $c$ (and error)$^\textrm{c}$ & MLCS2k2 $\Delta$ (and error)$^\textrm{b}$ \\
Photometry Data & SALT/2 $m_B$ (and error)$^\textrm{c}$  & MLCS2k2 $m_V$ (and rror)$^\textrm{b}$ \\
Light Curve Reference(s) & SALT/2 $\chi^{2} / \dof$$^\textrm{c}$ & MLCS2k2 $\chi^{2} / \dof$$^\textrm{b}$ \\
\hline \hline
\multicolumn{3}{c}{Spectral Information} \\
\hline
\# of Spectra of a Given SN	 & Exposure Time (s) & SNID Redshift (and error)$^\textrm{a}$ \\
UT Date of Spectrum	 & Position Angle (deg) & SNID Type and Subtype$^\textrm{a}$ \\
Filename	 & Parallactic Angle (deg) & SNID-Determined (Rest-Frame) Age (and error)$^\textrm{a}$ \\ 
Wavelength Range (\AA)	 & (Observer-Frame) Age 
        & SNID $r$lap$^\textrm{a}$ \\
Airmass   & Observer(s) & SNID Best Matching Template$^\textrm{a}$ \\
Seeing (arcsec)	 & Reducer &  Spectral Reference(s)  \\
Spectral Resolution(s) (\AA)	 & Instrument &  Flux Standard Star(s) \\
S/N  &  Flux Correction$^\textrm{d}$ &  \\
\hline \hline
\multicolumn{3}{l}{$^\textrm{a}$See Section~\ref{s:classification} for more information about SNID and its parameters.} \\
\multicolumn{3}{l}{$^\textrm{b}$See \citet{Jha07} for more information about MLCS2k2 and its parameters.} \\
\multicolumn{3}{l}{$^\textrm{c}$See \citet{Guy05} and \citet{Guy07}
  for more information about SALT and SALT2 and their parameters.} \\
\multicolumn{3}{l}{$^\textrm{d}$See Section~\ref{ss:galaxy} for more information about our flux corrections.} \\
\hline\hline
\end{tabular}
\end{center}
\end{table*}
\normalsize

The SNDB contains our entire previously published spectral dataset
(both SNe~Ia and core-collapse SNe) as well as all of the data
presented here.  It also contains photometry and light-curve
information which has been previously published, in addition to
photometric data which have been compiled and refit by Ganeshalingam
\etal (in prep.).

The SNDB uses the popular open-source software stack known as LAMP:
the Linux operating system, the Apache web server, the MySQL relational
database management system, and the PHP server-side scripting
language.  We have also implemented instances of the PHP helper
classes {\tt tar}\footnote{v2.2, Josh Barger ({\tt joshb@npt.com})}
and {\tt JpGraph}\footnote{v1.27.1, Aditus Consulting
  ({\tt\url{http://www.aditus.nu/jpgraph/}})} as well as the
JavaScript libraries {\tt SortTable}\footnote{v2, Stuart Langridge
  ({\tt\url{http://www.kryogenix.org/code/browser/sorttable/}})} and  
{\tt overLIB}\footnote{v4.21, Erik Bosrup
  ({\tt\url{http://www.bosrup.com/web/overlib/}})} to improve the
functionality 
and user friendliness of the SNDB; we are grateful to the authors
of these packages.  The database is stored on machines at UC Berkeley
and multiple backups exist at other locations.

The primary way of accessing the SNDB is via the SN Database Public
Home
Page\footnote{{\tt\url{http://hercules.berkeley.edu/database/index_public.html}}}. 
From here, users can download precompiled datasets and access our
public search page, where they can define various input search
criteria and query the SNDB.  All SNe, spectra, and light curves that
match the search criteria are returned as an HTML table.  The
returned information are also written to a \LaTeX\ table which
is linked from the Search Results Page.  If spectra or
photometry points are returned, users are given the option to download
the actual data or plot the spectra or photometry directly in their
web browser.  If MLCS2k2 light-curve fits are returned, users are
given the option to download a file containing the fits and
probability distributions for each of the light-curve parameters. 

As PI Filippenko's group at UC Berkeley publishes more spectral and
photometric data of SNe of all types in the future, the SNDB will
continually be updated with these newly released data. We hope that
the SNDB and its free, online access will quickly become an invaluable
tool to the SN community for the foreseeable future. 


\section{Classification}\label{s:classification}

Optical spectra are often used to classify SNe \citep[e.g.,][]{Filippenko97,
Turatto03} into four basic types.  SN~II spectra are identified by
strong hydrogen lines which are absent in SN~I spectra.  SN~Ia
spectra are characterised by the presence of a strong \ion{Si}{II}
$\lambda$6355 line typically observed in absorption near 6150~\AA.
SN~Ib spectra lack this \ion{Si}{II} feature but do contain strong
helium lines, and finally SN~Ic spectra lack strong helium lines
and have a \ion{Si}{II} $\lambda$6355 line that is significantly weaker than
those found in SNe~Ia.  We performed automated spectral classification
of our full spectral dataset\footnote{Our full dataset consists of (1)
previously published SN spectra of all types, (2) SN~Ia spectra which
are published here for the first time, and (3) some unpublished SN
spectra of non-Ia types which will be published in the future.} using
the SuperNova IDentification code \citep[SNID;][]{Blondin07}.  Details
of our classification algorithm are presented below.

\subsection{SNID Spectral Templates}

SNID classifies SN spectra by cross-correlating an input spectrum with
a large database of observed SN spectra (known as ``templates'') which
have been deredshifted to the rest frame.  In order to improve the
accuracy of SNID classifications, we decided to create our own set of
SNID spectral templates based on a combination of the default SNID
templates and our own spectral dataset.

\subsubsection{New SNID Subtypes}

We began by downloading SNID~v5.0\footnote{{\tt\url{http://marwww.in2p3.fr/~blondin/software/snid/index.html}}}
which includes a default set of templates consisting of nearby ($z <
0.1$) SNe of all types (Ia, Ib, Ic, II), as well as ``NotSN'' types,
which includes galaxies, active galactic nuclei (AGNs), luminous blue 
variables (LBVs), 
and M stars \citep[see][for the complete default SNID template
set]{Blondin07}. SNID further divides each basic SN type into the
following subtypes: Ia-norm, Ia-91T, Ia-91bg, Ia-csm, Ia-pec, Ib-norm,
Ib-pec, IIb, Ic-norm, Ic-pec, Ic-broad, IIP, II-pec, IIn, and IIL. 

``Norm'' and ``pec'' identify spectroscopically ``normal'' and
``peculiar'' SNe of their respective types.  Detailed descriptions of the
other subtypes can be found in \citet{Blondin07} and \citet{Foley09}.
In addition to these default subtypes, we have added two new SN~Ia
subtypes: Ia-99aa and Ia-02cx.

\begin{figure}
\begin{center}
\includegraphics[width=3.5in]{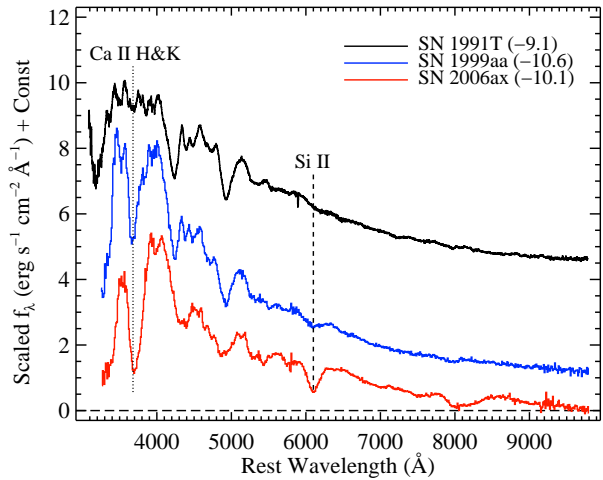}
\caption[Spectra of SN~1991T, SN~1999aa, and the ``normal'' Type~Ia
  SN~2006ax]{Spectra of SN~1991T, SN~1999aa, and the ``normal'' Type~Ia
  SN~2006ax at 9.1, 10.6, and 10.1~d before maximum brightness,
  respectively. All spectra in the figure (as well as all spectra
  plotted in this work) have had their host-galaxy recession
  velocities removed and have been corrected for MW reddening
  according to the values presented in Table~\ref{t:obj} and assuming
  that the extinction follows the \citet{Cardelli89} extinction law
  modified by \citet{ODonnell94}.  Notice how the \ion{Si}{II}
  $\lambda$6355 absorption feature (marked by the dashed line)
  increases in strength from SN~1991T 
  to SN~1999aa to SN~2006ax.  Also note the lack of \ion{Ca}{II}~H\&K
  absorption (marked by the dotted line) in SN~1991T, which is the
  major difference between 
  91T-like and 99aa-like SNe.}\label{f:99aa_comp}
\end{center}
\end{figure}

``Ia-99aa'' SNe have spectra that resemble those of SN~1999aa-like objects
\citep{Li01:pec,Strolger02,Garavini04}. Before maximum brightness, 
99aa-like SNe contain a \ion{Si}{II} $\lambda$6355 absorption 
line that is stronger than those seen in 91T-like
objects, but weaker than those of ``normal'' SNe~Ia.  99aa-like
objects also exhibit prominent \ion{Fe}{II} and \ion{Fe}{III} features at
early epochs, similar to the 91T-like SNe. Moreover, 99aa-like SNe
have strong \ion{Ca}{II}~H\&K absorption, as do normal SNe~Ia, but
quite in contrast with 91T-like SNe which lack this feature; this
is the main spectroscopic difference between 91T-like and 99aa-like 
objects. A comparison of early-time
spectra of a 99aa-like SN, a 91T-like SN, and a ``normal'' SN~Ia is
shown in Figure~\ref{f:99aa_comp}. SN 99aa-like events were previously
included in the Ia-91T subtype in the default set of SNID templates,
but we feel that they {\it perhaps} represent a spectroscopically
distinct subclass and therefore deserve their own subtype in SNID.

Similarly, ``Ia-02cx'' SNe have spectra that resemble those of SN~2002cx-like
objects \citep[e.g.,][]{Li03:02cx,Jha06:02cx,Foley09:08ha}.  These SNe
were previously included in the Ia-pec subtype in the default set of
SNID templates, but again, we believe that they represent their own
subclass of events and should have their own subtype in SNID.  Note
that for our purposes, the ``Ia-pec'' category refers mainly to
SN~2000cx-like objects \citep{Li01:00cx}.

We have made a few further changes to the classification scheme of
SNID.  Namely, Type~IIb SNe \citep[whose spectra evolve from a Type~II
SN to a Type~Ib SN, as in SNe 1987K and 1993J;
see][]{Filippenko88,Filippenko93,Matheson00:93j} are included {\it
  only} in the ``Ib'' SNID type (as opposed to being included in the
``II'' SNID type as well).  We have also added two subtypes to the
``NotSN'' SNID type: quasi-stellar objects (QSOs) 
and carbon stars.  Spectra of these objects
were obtained from the SDSS Data Release 6 Spectral Cross-Correlation
Templates.

\subsubsection{New SNID Templates}

\begin{figure}
\begin{center}
\includegraphics[width=3.5in]{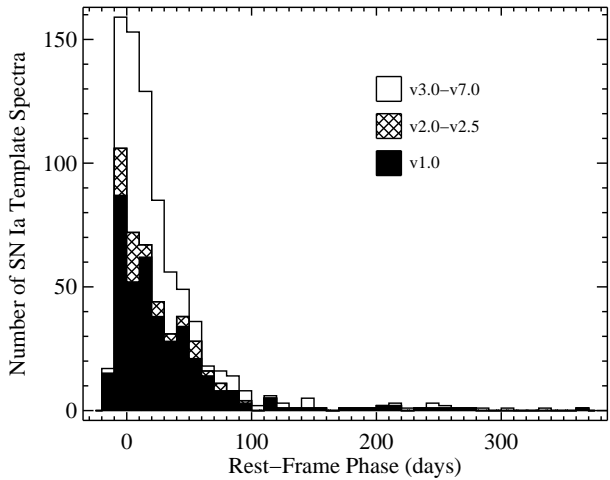}
\caption[Histogram of ages of SN~Ia template spectra]{A histogram of
  the ages of our SN~Ia template spectra 
  separated by SNID template version.}\label{f:template_ages}   
\end{center}
\end{figure}

We began constructing our new set of spectral templates for SNID by
performing a literature search for the best-studied and most canonical
SNe in each subtype (with emphasis on the various SN~Ia subtypes).
Of the objects we deemed to be ``the best'' examples of their
respective subtypes, 30 are already included in the default set of SNID 
v5.0 templates \citep{Blondin07} and 30 are in our full spectral dataset, 
with 23 found in both sets.  These objects make up version 1.0 (v1.0) of 
our new spectral template set.  
Table~\ref{t:snid_templates} contains information about a subset of
the objects included in v1.0, as well as the rest of our final
template set (the full table is available online --- see the Supporting
Information).  
%
%
We present a summary of the
number of each (sub)type of SN included in the final template set in
Table~\ref{t:snid_templates_summary}.  Figure~\ref{f:template_ages}
shows a histogram of the ages of our SN~Ia template spectra.


\onecolumn
\small
\begin{longtable}{llcl}
\caption{SNID v7.0 Spectral Templates} \label{t:snid_templates} \\ [-2ex]
\hline \hline
SN Name & Subtype & Version$^\textrm{a}$ & Age(s)$^\textrm{b}$ \\
\hline
\endfirsthead

\multicolumn{4}{c}{{\tablename} \thetable{} --- Continued} \\
\hline \hline
SN Name & Subtype & Version$^\textrm{a}$ & Age(s)$^\textrm{b}$ \\
\hline
\endhead

\hline \hline
\multicolumn{4}{l}{Continued on Next Page\ldots} \\
\endfoot

\hline \hline
\endlastfoot

SN 1988Z & IIn & 1 & $\cdots$ \\ [.2ex]
SN 1989B$^\textrm{c}$ & Ia-norm & 1 & 
$-6$,$-1$,$4$,$6$,$8$,$10$,[$12$:$14$],[$16$:$25$],$31$,$37$,$49$,50(1)
 \\ [.2ex]
SN 1990H & IIP & 3 & $\cdots$ \\ [.2ex]
SN 1990N$^\textrm{c}$ & Ia-norm & 1 & $-13$,$-6$,$3$,$5$,$15$,$18$,$39$,50(8)
 \\ [.2ex]
SN 1990Q & IIP & 3 & $\cdots$ \\ [.2ex]
SN 1991C & IIn & 2 & $\cdots$ \\ [.2ex]
SN 1991T$^\textrm{d}$ & Ia-91T & 1 & 
$-12$,[$-10$:$-5$],$0$,$7$,$11$,$16$,$19$,$25$,[$46$:$47$],50(5) \\ [.2ex]
SN 1991ao & IIP & 3 & $\cdots$ \\ [.2ex]
SN 1991av & IIn & 2 & $\cdots$ \\ [.2ex]
SN 1991bg$^\textrm{d}$ & Ia-91bg & 1 & 
[$0$:$3$],[$15$:$16$],[$19$:$20$],[$26$:$27$],$30$,[$33$:$34$],[$46$:$48$],50(10)
 \\ [.2ex]
SN 1992A$^\textrm{c}$ & Ia-norm & 1 & 
$-5$,[$-1$:$0$],[$2$:$3$],[$6$:$7$],$9$,$12$,[$16$:$17$],$24$,$28$ \\ [.2ex]
SN 1992H & IIP & 3 & $\cdots$ \\ [.2ex]
SN 1992ad & IIP & 3 & $\cdots$ \\ [.2ex]
SN 1993E & IIP & 3 & $\cdots$ \\ [.2ex]
SN 1993G & IIP & 3 & $\cdots$ \\ [.2ex]
SN 1993J$^\textrm{d}$ & IIb & 1 & 
[$-18$:$-16$],$-11$,[$-5$:$-2$],$1$,[$3$:$7$],[$11$:$13$],$17$,$20$,$24$,[$29$:$30$],$32$,[$35$:$39$],$41$,50(38)
 \\ [.2ex]
SN 1993W & IIP & 3 & $\cdots$ \\ [.2ex]
SN 1993ad & IIP & 3 & $\cdots$ \\ [.2ex]
SN 1994D$^\textrm{d}$ & Ia-norm & 1 & 
[$-12$:$-2$],$0$,[$2$:$3$],[$10$:$17$],$19$,$21$,$24$,$26$,$28$,$30$,$40$,[$43$:$44$],$46$,[$48$:$49$],50(9)
 \\ [.2ex]
SN 1994I$^\textrm{d}$ & Ic-norm & 1 & 
$-6$,[$-4$:$-3$],[$0$:$3$],[$21$:$24$],$26$,$30$,$36$,$38$,$40$,50(1) \\ [.2ex]
SN 1994S & Ia-norm & 3 & $2$ \\ [.2ex]
SN 1994W & IIn & 1 & $\cdots$ \\ [.2ex]
SN 1994Y & IIn & 2 & $\cdots$ \\ [.2ex]
SN 1994ae$^\textrm{d}$ & Ia-norm & 1 & 
[$1$:$4$],$6$,[$9$:$11$],$30$,$36$,$40$,50(6) \\ [.2ex]
SN 1994ak & IIn & 3 & $\cdots$ \\ [.2ex]
SN 1995D$^\textrm{d}$ & Ia-norm & 1 & 
$4$,$6$,$8$,[$10$:$11$],$14$,$16$,$33$,$38$,$43$,50(3) \\ [.2ex]
SN 1995E & Ia-norm & 3 & $-2$ \\ [.2ex]
SN 1995G & IIn & 2 & $\cdots$ \\ [.2ex]
SN 1995J & IIP & 3 & $\cdots$ \\ [.2ex]
SN 1995V & IIP & 3 & $\cdots$ \\ [.2ex]
SN 1995X & IIP & 3 & $\cdots$ \\ [.2ex]
SN 1995ac$^\textrm{d}$ & Ia-91T & 2 & $-6$,$24$ \\ [.2ex]
SN 1996L$^\textrm{c}$ & IIn & 1 & $8$,$34$,$42$,50(4) \\ [.2ex]
SN 1996ae & IIn & 2 & $\cdots$ \\ [.2ex]
SN 1996an & IIP & 3 & $\cdots$ \\ [.2ex]
SN 1996cc & IIP & 3 & $\cdots$ \\ [.2ex]
SN 1997Y & Ia-norm & 3 & $2$ \\ [.2ex]
SN 1997ab & IIn & 2 & $\cdots$ \\ [.2ex]
SN 1997br$^\textrm{d}$ & Ia-91T & 2 & 
[$-9$:$-6$],$-4$,[$8$:$9$],$12$,$17$,$21$,$24$,$38$,$42$,$46$,$49$,50(3)
 \\ [.2ex]
SN 1997da & IIP & 3 & $\cdots$ \\ [.2ex]
SN 1997dd & IIb & 2 & $\cdots$ \\ [.2ex]
SN 1997ef$^\textrm{d}$ & Ic-broad & 1 & 
$-14$,[$-12$:$-9$],[$-6$:$-4$],$7$,[$13$:$14$],$17$,$22$,$24$,$27$,$38$,$40$,$44$,$46$,$48$,50(4)
 \\ [.2ex]
SN 1997eg & IIn & 2 & $\cdots$ \\ [.2ex]
SN 1998A & IIP & 3 & $\cdots$ \\ [.2ex]
SN 1998E & IIn & 2 & $\cdots$ \\ [.2ex]
SN 1998S$^\textrm{d}$ & IIn & 1 & 
$-13$,$-2$,[$1$:$3$],[$10$:$11$],[$13$:$14$],$16$,[$31$:$32$],[$40$:$41$],$44$,[$46$:$47$],50(37)
 \\ [.2ex]
SN 1998bu$^\textrm{d}$ & Ia-norm & 1 & 
[$-3$:$-1$],$1$,[$9$:$14$],[$28$:$44$],50(9) \\ [.2ex]
SN 1998bw$^\textrm{c}$ & Ic-broad & 1 & 
[$8$:$9$],[$12$:$14$],$16$,[$18$:$19$],$21$,$24$,[$26$:$28$],$34$,$37$,$43$,50(5)
 \\ [.2ex]
SN 1998dl & IIP & 3 & $\cdots$ \\ [.2ex]

\hline \hline
\multicolumn{4}{l}{Table abridged; the full table is available online---see Supporting Information.} \\ 
\multicolumn{4}{p{5.8in}}{All spectral templates are solely from our full dataset, unless otherwise noted.} \\
\multicolumn{4}{p{5.8in}}{$^\textrm{a}$Version of new SNID spectral templates when object was added --- 1: v1.0; 2: v2.0--v2.5; 3: v3.0--v7.0.} \\
\multicolumn{4}{p{5.8in}}{$^\textrm{b}$Rest-frame SN age(s), rounded to nearest whole day, in days from $B$-band maximum (for SNe~Ia), from $V$-band maximum (for SNe~Ib/c), or from the estimated date of explosion (for SNe~II). Ages of spectral templates from our dataset are calculated from the light curve references in Table~\ref{t:obj}; ages from the original SNID~v5.0 set of spectral templates are from \citet{Blondin07}. Adjacent ages are listed in square brackets. Spectra whose age exceeds +50~days are grouped together and the number of such spectra are noted in parentheses. Many core-collapse SNe from our full spectral dataset lack age information (though we require SNe~Ia templates to have age information).} \\
\multicolumn{4}{p{5.8in}}{$^\textrm{c}$Spectral templates are from the original SNID~v5.0 set of templates only.} \\
\multicolumn{4}{p{5.8in}}{$^\textrm{d}$Spectral templates are from both our full dataset as well as the original SNID~v5.0 set of templates.} \\
\end{longtable}
\normalsize
\twocolumn



\begin{table*}
\begin{center}
\caption{Summary of SNID v7.0 Spectral Templates}\label{t:snid_templates_summary}
\begin{tabular}{lrlrlrlrlr}
\hline\hline
Ia (Total) &      134 & Ib (Total) &       16  & Ic  (Total) &       14
 & II (Total) &      113 & NotSN (Total) & 29 \\
\hline
Ia-norm &      101 & Ib-norm &        6 & Ic-norm &       10 & IIP &       76
 & AGN & 1 \\
Ia-91T &        3 & Ib-pec &        0 & Ic-pec &        1 & II-pec &        3
 & Gal & 11 \\
Ia-91bg &       16 & IIb &       10 & Ic-broad &        3 & IIn &       34
 & LBV & 3 \\
Ia-csm &        2  & & & & & IIL &        0 & M-star & 7 \\
Ia-pec &        1  & & & & & & & QSO & 4 \\
Ia-99aa &        7 & & & & & & & C-star & 3 \\
Ia-02cx &        4 & & & & & & & & \\
\hline\hline
\end{tabular}
\end{center}
\end{table*}


For each object in v1.0, we examined only spectra that had a S/N of at
least 15~pixel$^{-1}$, a minimum wavelength of less than 4500~\AA, and
a maximum wavelength of greater than 7000~\AA.  We also required that
each SN~Ia have a date of maximum brightness either from published
sources or from Wang \etal (in prep.) so we can accurately
calculate the age of each spectrum.  Finally, each spectrum was
visually inspected by multiple coauthors to be sure they truly
represented their supposed subtype and were relatively free of
host-galaxy contamination.  If a spectrum passed the quantitative
criteria and the by-eye inspection, it was cropped to 3500--10,000~\AA\
(to remove edge artefacts on both ends of the spectra). If a spectrum
did not cover this entire range, 50~\AA\ on both ends of the spectrum
were removed instead.  Finally, the cropped spectrum was made into a
template; the result of this process was v1.0 of our new SNID
templates.

To increase the number of SNe in our template set, we ran SNID (with
our v1.0 templates) on our entire spectral dataset.  
To determine subtypes, we followed similar classification
criteria to those of \citet{Blondin07}, requiring 
that the SNID $r$lap value\footnote{In SNID, the $r$lap value
  is a measure of the strength of the correlation between the
  best-matching 
  spectrum and the input spectrum.} be at least 10 and the 3
best-matching spectra from SNID all be of the same subtype.  We also
ignored any objects that were classified as ``Ia-norm'' or ``IIP''
since we wanted to concentrate only on relatively rare subtypes at
this point and SNID is somewhat biased toward classifying objects as
subtypes that have a large number of templates (such as ``Ia-norm''
and ``IIP'').
The result of this
process was v2.0 of our new SNID templates.  This process was
repeated iteratively, running SNID with the previously created version
of our spectral templates, until no more SNe passed all of the
criteria. It required five additional runs to reach this convergence, 
resulting in v2.5 of our new SNID templates.

We continued creating a new set of spectral templates by running SNID
(with our v2.5 templates) on our full spectral dataset.  
The result of this process was v3.0 of our new SNID
templates.
This
process was repeated iteratively, running SNID with our previously
created version of the spectral templates, until no more SNe were
classified and turned into templates. We finished with the
creation of v7.0, our final set of new SNID
templates, which we use to classify the remainder of our full
spectral dataset (see Table~\ref{t:snid_templates} for information
regarding the entire SNID template set). This v7.0 contains 1543 spectra of
277 SNe, of which 779 spectra and 134 objects are SNe~Ia. Again, we
show a histogram of the ages of all of our SN~Ia spectral templates 
(separated by SNID template version) in Figure~\ref{f:template_ages}.

\subsubsection{Final Verifications}\label{sss:final_ver}

As a sanity check, we perform final classification verifications.
We ran all of our spectra of the objects in v2.5 through SNID
using our v7.0 templates, and then we ran all of our spectra of the
objects in v7.0 through SNID (again using our v7.0 templates).  We
compare the (sub)type of the best-matching template to the 
actual (sub)type of the object, making sure to ignore all templates
of the SN currently being inspected. SNe are not used in this
process if their spectra constitute $> 15$~per~cent of all spectra in the
object's subtype. 

Using only objects from v2.5 (v7.0) we find that SNID, using v7.0 of
our new templates, is able to correctly classify \about97~per~cent
(\about99~per~cent) of SN~Ia spectra as one of the SN~Ia subtypes and
non-Ia SN spectra as one of the non-Ia SN subtypes.  We also find
that SNID is able to correctly classify \about85~per~cent
(\about95~per~cent) of SN~Ia spectra with ages $\le 15$~d past
maximum brightness as the correct subtype.


We compared the classification results of our full dataset using our
v7.0 templates to the results using the default set of SNID
templates.  The average difference between $z_\textrm{gal}$ (the
actual redshift of the host galaxy) and $z_\textrm{SNID}$ (the
redshift of the SN as determined by SNID) was found to decrease using
our v7.0 templates.  Furthermore, the discrepancies between
$t_\textrm{LC}$ (the spectral age derived from photometry) and
$t_\textrm{SNID}$ (the SNID-determined spectral age) improved
drastically when using our v7.0 templates.  Our template set also
markedly suppressed the SNID-determined age bias seen near maximum
brightness and $+30$~d \citep[see][and Section~\ref{ss:verifying} for
more on this bias]{Ostman11}.  Finally, with the significant increase
in the number of templates of peculiar SNe~Ia and our two additional
SN~Ia subtypes, SNID (using our v7.0 template set) can distinguish
between the various spectroscopic subtypes much better than when using
the default templates.  The SNID-determined subtype of
\about15~per~cent (\about26~per~cent) of the spectra (SNe~Ia)
presented here is different when using SNID with our v7.0 templates
versus using SNID with the default templates.

\subsection{Classification of Spectra}\label{ss:classification}

Using our v7.0 SNID templates, we can attempt to classify all of the
spectra in our full dataset using criteria similar to those of 
\citet{Miknaitis07} and \citet{Foley09}.  To do this we execute a
series of SNID runs to separately determine the type, subtype,
redshift, and age of the input spectrum.  For all of the SNID runs we
ignore all templates of the SN currently being inspected so that SNID will
not match an object to itself, and we truncate all spectra at
10,000~\AA\ to avoid any possible second-order light contamination.
Besides these, we use the default parameters of SNID unless
specifically noted below. If a spectrum was obtained within 10$^\circ$
of the parallactic angle (or was obtained at an airmass $< 1.1$) and
corrected for host-galaxy contamination via our colour-matching
technique (as described in Section~\ref{ss:galaxy}), then we use the
galaxy-subtracted spectrum in all of the SNID runs. 
A subset of the results of the classification algorithm presented
below can be found in Table~\ref{t:snid} (the full results are
available online --- see the Supporting Information).
%
%


\setlength{\tabcolsep}{0.04in}
\onecolumn
\footnotesize
\begin{longtable}{lccrrrllllrrr}
\caption{SNID Classification Information} \label{t:snid} \\[-2ex]
\hline \hline
SN Name & Type & Subtype & $z_\textrm{SNID}$$^\textrm{a}$\hspace{.23in} & $t_\textrm{SNID}$$^\textrm{b}$\hspace{.1in} & $r$lap & \multicolumn{7}{l}{Best Match$^\textrm{c}$} \\
\hline
\endfirsthead

\multicolumn{13}{c}{{\tablename} \thetable{} --- Continued} \\
\hline \hline
SN Name & Type & Subtype & $z_\textrm{SNID}$$^\textrm{a}$\hspace{.23in} & $t_\textrm{SNID}$$^\textrm{b}$\hspace{.1in} & $r$lap & \multicolumn{7}{l}{Best Match$^\textrm{c}$} \\
\hline
\endhead

\hline \hline
\multicolumn{13}{l}{Continued on Next Page\ldots} \\
\endfoot

\hline \hline
\endlastfoot

SN 1989A & Ia & Ia-norm & $0.0087$ (0.0014) & $83.3$ (7.4) & 13.1 & sn99dk & (Ia-norm) & $0.0088$ & (0.0044) & $71.83$ & (6.95) & 6 \\
SN 1989B & Ia & Ia-norm & $0.0055$ (0.0039) & $7.5$ (1.4) & 17.4 & sn05ki & (Ia-norm) & $0.0043$ & (0.0031) & $7.96$ & (1.49) & 14 \\
SN 1989B & Ia & Ia-norm & $0.0046$ (0.0041) & $23.6$ (-1.0) & 10.9 & sn99dk & (Ia-norm) & $0.0042$ & (0.0048) & $23.62$ & (0.68) & 2 \\
SN 1989B & Ia & Ia-norm & $0.0032$ (0.0015) & $87.3$ (13.2) & 19.5 & sn08ec & (Ia-norm) & $0.0033$ & (0.0027) & $71.57$ & (9.95) & 16 \\
SN 1989B & Ia & Ia-norm & $0.0029$ (0.0015) & $\cdots$\hspace{.2in} & 15.7 & sn02fk & (Ia-norm) & $0.0035$ & (0.0035) & $148.74$ & (38.33) & 9 \\
SN 1989M & Ia & Ia-norm & $0.0014$ (0.0053) & $\cdots$\hspace{.2in} & 17.2 & sn89B & (Ia-norm) & $0.0043$ & (0.0036) & $-6.30$ & (5.85) & 9 \\
SN 1989M & Ia & Ia-norm & $0.0022$ (0.0054) & $\cdots$\hspace{.2in} & 13.1 & sn92A & (Ia-norm) & $0.0065$ & (0.0041) & $6.30$ & (4.96) & 10 \\
SN 1989M & Ia & Ia-norm & $0.0072$ (0.0018) & $\cdots$\hspace{.2in} & 10.8 & sn90N & (Ia-norm) & $0.0059$ & (0.0053) & $213.40$ & (0.00) & 1 \\
SN 1989M & Ia & Ia-norm & $0.0061$ (0.0000) & $\cdots$\hspace{.2in} & 8.0 & sn90N & (Ia-norm) & $0.0061$ & (0.0073) & $332.50$ & (0.00) & 1 \\
SN 1990G & Ia & Ia-norm & $0.0365$ (0.0047) & $9.3$ (1.1) & 24.2 & sn98bu & (Ia-norm) & $0.0359$ & (0.0022) & $9.30$ & (3.25) & 51 \\
SN 1990M & Ia & Ia-norm & $0.0082$ (0.0000) & $216.9$ (-1.0) & 10.4 & sn98bu & (Ia-norm) & $0.0082$ & (0.0057) & $208.00$ & (6.29) & 2 \\
SN 1990M & Ia & Ia-norm & $0.0077$ (0.0019) & $\cdots$\hspace{.2in} & 12.5 & sn02cs & (Ia-norm) & $0.0091$ & (0.0046) & $31.28$ & (0.00) & 1 \\
SN 1990M & Ia & $\cdots$ & $0.0080$ (0.0042) & $\cdots$\hspace{.2in} & 12.5 & sn91T & (Ia-91T) & $0.0101$ & (0.0042) & $24.80$ & (12.41) & 201 \\
SN 1990M & Ia & $\cdots$ & $0.0081$ (0.0044) & $\cdots$\hspace{.2in} & 13.3 & sn99aa & (Ia-99aa) & $0.0060$ & (0.0042) & $59.43$ & (21.88) & 172 \\
SN 1990M & $\cdots$ & $\cdots$ & $\cdots$\hspace{.33in} & $\cdots$\hspace{.2in} & $\cdots$ &  $\cdots$ & & & & & \\
SN 1990O & Ia & Ia-norm & $0.0291$ (0.0025) & $15.1$ (1.1) & 25.6 & sn07qe & (Ia-norm) & $0.0318$ & (0.0024) & $16.00$ & (1.65) & 22 \\
SN 1990O & Ia & $\cdots$ & $0.0287$ (0.0042) & $\cdots$\hspace{.2in} & 19.9 & sn99dq & (Ia-99aa) & $0.0310$ & (0.0026) & $24.09$ & (17.54) & 275 \\
SN 1990O & Ia & Ia-norm & $0.0299$ (0.0017) & $55.1$ (5.2) & 16.1 & sn02bo & (Ia-norm) & $0.0284$ & (0.0032) & $56.35$ & (7.54) & 13 \\
SN 1990N & Ia & Ia-norm & $0.0052$ (0.0054) & $7.1$ (2.8) & 22.5 & sn95D & (Ia-norm) & $0.0030$ & (0.0022) & $8.10$ & (2.93) & 22 \\
SN 1990N & Ia & Ia-norm & $0.0033$ (0.0019) & $18.7$ (3.7) & 18.0 & sn05cf & (Ia-norm) & $0.0019$ & (0.0032) & $18.69$ & (3.20) & 30 \\
SN 1990N & Ia & Ia-norm & $0.0031$ (0.0008) & $43.1$ (4.3) & 16.6 & sn99dk & (Ia-norm) & $0.0021$ & (0.0034) & $44.20$ & (7.14) & 41 \\
SN 1990N & $\cdots$ & $\cdots$ & $\cdots$\hspace{.33in} & $\cdots$\hspace{.2in} & $\cdots$ &  $\cdots$ & & & & & \\
SN 1990N & Ia & Ia-norm & $0.0043$ (0.0013) & $\cdots$\hspace{.2in} & 19.0 & sn94D & (Ia-norm) & $0.0043$ & (0.0031) & $115.09$ & (41.77) & 10 \\
SN 1990R & Ia & Ia-norm & $0.0175$ (0.0010) & $41.3$ (1.4) & 17.5 & sn94D & (Ia-norm) & $0.0178$ & (0.0029) & $43.20$ & (4.29) & 34 \\
SN 1990R & Ia & $\cdots$ & $0.0183$ (0.0042) & $\cdots$\hspace{.2in} & 15.5 & sn99aa & (Ia-99aa) & $0.0332$ & (0.0036) & $26.20$ & (19.30) & 239 \\
SN 1990R & Ia & Ia-norm & $0.0150$ (0.0013) & $91.1$ (2.8) & 18.2 & sn94D & (Ia-norm) & $0.0172$ & (0.0030) & $74.23$ & (12.81) & 14 \\
SN 1990Y & Ia & Ia-norm & $0.0410$ (0.0013) & $16.1$ (2.3) & 12.7 & sn02bg & (Ia-norm) & $0.0397$ & (0.0045) & $-3.65$ & (7.14) & 8 \\
SN 1991B & Ia & Ia-norm & $0.0096$ (0.0021) & $40.5$ (-1.0) & 11.3 & sn05de & (Ia-norm) & $0.0089$ & (0.0045) & $40.49$ & (3.43) & 2 \\
SN 1991B & Ia & Ia-norm & $0.0098$ (0.0009) & $57.1$ (4.2) & 15.3 & sn04ey & (Ia-norm) & $0.0097$ & (0.0035) & $51.49$ & (6.27) & 17 \\
SN 1991B & Ia & $\cdots$ & $0.0133$ (0.0045) & $\cdots$\hspace{.2in} & 14.5 & sn08ds & (Ia-99aa) & $0.0109$ & (0.0033) & $63.44$ & (28.87) & 182 \\
SN 1991K & $\cdots$ & $\cdots$ & $\cdots$\hspace{.33in} & $\cdots$\hspace{.2in} & $\cdots$ &  $\cdots$ & & & & & \\
SN 1991K & Ia & Ia-norm & $0.0189$ (0.0007) & $91.1$ (16.8) & 12.0 & sn07af & (Ia-norm) & $0.0188$ & (0.0043) & $91.06$ & (15.17) & 4 \\
SN 1991M & Ia & Ia-norm & $0.0055$ (0.0010) & $22.5$ (-1.0) & 10.6 & sn01bg & (Ia-norm) & $0.0066$ & (0.0051) & $18.91$ & (2.52) & 2 \\
SN 1991M & Ia & Ia-norm & $0.0066$ (0.0014) & $\cdots$\hspace{.2in} & 9.1 & sn05de & (Ia-norm) & $0.0062$ & (0.0056) & $40.49$ & (7.97) & 89 \\
SN 1991M & Ia & Ia-pec & $0.0096$ (0.0009) & $146.4$ (-1.0) & 14.1 & sn00cx & (Ia-pec) & $0.0091$ & (0.0045) & $146.31$ & (0.06) & 2 \\
SN 1991M & Ia & Ia-norm & $0.0098$ (0.0007) & $\cdots$\hspace{.2in} & 10.1 & sn90N & (Ia-norm) & $0.0089$ & (0.0054) & $213.40$ & (0.00) & 1 \\
SN 1991O & Ia & Ia-91bg & $0.0365$ (0.0028) & $\cdots$\hspace{.2in} & 12.4 & sn06em & (Ia-91bg) & $0.0391$ & (0.0060) & $20.95$ & (0.00) & 1 \\
SN 1991S & Ia & Ia-norm & $0.0556$ (0.0021) & $38.5$ (7.1) & 8.7 & sn94D & (Ia-norm) & $0.0556$ & (0.0051) & $43.20$ & (11.50) & 124 \\
SN 1991T & Ia & Ia-91T & $0.0029$ (0.0012) & $-5.6$ (-1.0) & 8.6 & sn95ac & (Ia-91T) & $0.0053$ & (0.0057) & $-5.61$ & (2.39) & 2 \\
SN 1991T & Ia & Ia-91T & $0.0037$ (0.0015) & $\cdots$\hspace{.2in} & 14.5 & sn97br & (Ia-91T) & $0.0042$ & (0.0037) & $-7.40$ & (0.00) & 1 \\
SN 1991T & Ia & Ia-norm & $0.0041$ (0.0015) & $\cdots$\hspace{.2in} & 16.2 & sn94ae & (Ia-norm) & $0.0052$ & (0.0037) & $9.40$ & (0.00) & 1 \\
SN 1991T & Ia & $\cdots$ & $0.0081$ (0.0042) & $\cdots$\hspace{.2in} & 14.8 & sn98es & (Ia-99aa) & $0.0057$ & (0.0038) & $78.70$ & (28.43) & 143 \\
SN 1991T & Ia & Ia-norm & $0.0061$ (0.0011) & $87.3$ (6.1) & 14.6 & sn94D & (Ia-norm) & $0.0070$ & (0.0040) & $87.19$ & (13.58) & 12 \\
SN 1991T & $\cdots$ & $\cdots$ & $\cdots$\hspace{.33in} & $\cdots$\hspace{.2in} & $\cdots$ &  $\cdots$ & & & & & \\
SN 1991T & Ia & $\cdots$ & $0.0082$ (0.0078) & $\cdots$\hspace{.2in} & 23.1 & sn94D & (Ia-norm) & $0.0080$ & (0.0028) & $115.09$ & (91.53) & 49 \\
SN 1991T & $\cdots$ & $\cdots$ & $\cdots$\hspace{.33in} & $\cdots$\hspace{.2in} & $\cdots$ &  $\cdots$ & & & & & \\
SN 1991T & $\cdots$ & $\cdots$ & $\cdots$\hspace{.33in} & $\cdots$\hspace{.2in} & $\cdots$ &  $\cdots$ & & & & & \\
SN 1991am & Ia & Ia-norm & $0.0604$ (0.0022) & $16.4$ (2.9) & 16.6 & sn89B & (Ia-norm) & $0.0602$ & (0.0030) & $13.40$ & (3.79) & 21 \\
SN 1991ak & Ia & Ia-norm & $0.0114$ (0.0010) & $41.5$ (2.0) & 15.9 & sn94D & (Ia-norm) & $0.0115$ & (0.0031) & $43.20$ & (4.26) & 31 \\
SN 1991ak & Ia & Ia-norm & $0.0112$ (0.0016) & $57.5$ (1.5) & 19.3 & sn02cr & (Ia-norm) & $0.0100$ & (0.0023) & $57.47$ & (3.98) & 13 \\
SN 1991ak & Ia & $\cdots$ & $0.0124$ (0.0042) & $\cdots$\hspace{.2in} & 20.2 & sn08ds & (Ia-99aa) & $0.0104$ & (0.0023) & $63.44$ & (26.22) & 202 \\
SN 1991at & Ia & Ia-norm & $0.0429$ (0.0021) & $36.4$ (5.1) & 12.2 & sn04dt & (Ia-norm) & $0.0416$ & (0.0036) & $31.97$ & (8.38) & 12 \\
SN 1991as & Ia & $\cdots$ & $0.0135$ (0.0042) & $\cdots$\hspace{.2in} & 17.4 & sn00cx & (Ia-pec) & $0.0140$ & (0.0034) & $88.93$ & (77.90) & 78 \\
SN 1991ay & Ia & Ia-norm & $0.0487$ (0.0016) & $14.5$ (1.4) & 17.8 & sn95D & (Ia-norm) & $0.0471$ & (0.0033) & $16.10$ & (4.14) & 31 \\
SN 1991bd & Ia & Ia-norm & $0.0144$ (0.0044) & $23.6$ (-1.0) & 11.6 & sn99dk & (Ia-norm) & $0.0139$ & (0.0049) & $23.62$ & (1.36) & 2 \\
SN 1991bc & Ia & Ia-norm & $0.0223$ (0.0026) & $17.7$ (2.9) & 24.0 & sn90N & (Ia-norm) & $0.0219$ & (0.0024) & $17.70$ & (4.24) & 33 \\
SN 1991bc & Ia & $\cdots$ & $0.0233$ (0.0039) & $\cdots$\hspace{.2in} & 14.9 & sn91T & (Ia-91T) & $0.0202$ & (0.0035) & $75.20$ & (22.67) & 259 \\

\hline \hline
\multicolumn{13}{l}{Table abridged; the full table is available online---see Supporting Information.} \\ 
\multicolumn{13}{l}{The entries in this table match one-to-one with the entries in Table~\ref{t:spec_info}.} \\
\multicolumn{13}{l}{$^\textrm{a}$The redshift uncertainty is in parentheses.} \\
\multicolumn{13}{p{6.6in}}{$^\textrm{b}$Phases of spectra are in rest-frame days. The phase uncertainty is in parentheses. Phase uncertainties of 0 imply that only one template was a ``good" match.} \\
\multicolumn{13}{l}{$^\textrm{c}$The best matching SNID template in the form: ``template SN'' (subtype)  $z_\textrm{SNID}$ (error) $t_\textrm{SNID}$ (error) N\_good\_matches} \\
\end{longtable}
\normalsize
\twocolumn

\subsubsection{SNID Type}

If the object's redshift is known {\it a priori} (from the host
galaxy, usually via NED), we force SNID to use this redshift;
otherwise we do not use any redshift prior.  For the first attempt to
determine a type, the minimum $r$lap value is set to 10.  A spectrum is
determined to be of a given type if the fraction of ``good''
correlations that correspond to this type exceeds 50~per~cent. In
addition, we require the best-matching SN template to be of this same
type. If the spectrum's type cannot be determined using these criteria
we perform another SNID run, this time using a minimum $r$lap value of
5. This resulted in 1232 of our \totspec\ spectra receiving a SNID
type. If the type of the input spectrum is successfully determined
(using either $r$lap value), an attempt is made to determine its
subtype.  

\subsubsection{SNID Subtype}

Again, we adopt the host-galaxy redshift when available and the minimum
$r$lap used to determine if the subtype is the same as was used to
successfully determine the type (either 5 or 10).  We also force SNID
to only consider templates of the previously determined SN type. A spectrum
is determined to be of a given subtype if the fraction of ``good''
correlations that correspond to this subtype exceeds 50~per~cent. In
addition, we require the best-matching SN template to be of this same
subtype. If the spectrum's subtype cannot be determined using these
criteria, {\it and} a minimum $r$lap of 10 was used, we perform another
SNID run, this time using a minimum $r$lap value of 5. This resulted in
1098 of our \totspec\ spectra receiving a SNID subtype. Regardless of
whether SNID determines a subtype, a third run is executed to
determine the redshift.

\subsubsection{SNID Redshift}

The SNID redshift is calculated by taking the median of all ``good''
template redshifts and the redshift error is the standard deviation of
these redshifts. If a subtype has been successfully determined, we
force SNID to only use templates of that subtype; otherwise we only
use templates of the previously determined type.  For this SNID run no
{\it a priori} redshift information is used.  1232 of our \totspec\
spectra received a SNID redshift. If a redshift is successfully
determined in this run, a fourth run is executed to calculate the age
of the spectrum.

\subsubsection{SNID Age}\label{sss:snid_age}

The SNID age is calculated by taking the median of all ``good''
template ages that have an $r$lap value larger than 75~per~cent of the
$r$lap value of the best-matching template. The age error is the
standard deviation of these ages. Only if a subtype has been
successfully determined do we attempt to calculate an age. We also
require that the age error be less than either 4~d or 20~per~cent of
the SNID-determined age (whichever is larger). 
For this run, we again force SNID to adopt the host-galaxy
redshift when available.  If no host-galaxy redshift is known, we use
the previously determined SNID redshift instead. 849 of our \totspec\
spectra received a SNID age. 

\subsection{Verifying Redshifts and Ages from SNID}\label{ss:verifying}

It has been shown previously that SNID-determined redshifts correlate
extremely well with actual redshifts of SN host galaxies
\citep[e.g.,][]{Matheson05,Foley09,Ostman11}. 
The SNID-determined redshifts (using v7.0 of our templates) agree
well with the host-galaxy redshifts of our data.  The dispersion about
the one-to-one correspondence is only \about0.002 for the 1184 spectra
for which the redshift is known and which SNID determined were SNe~Ia
and calculated a redshift. This is as good as or better than what
has been found previously using much smaller samples of
higher-redshift SNe \citep{Matheson05,Foley09,Ostman11}.  However, 
the majority of the largest outliers appear
to have SNID redshifts that are lower than the host-galaxy redshifts. 
The normalised median absolute deviation \citep[i.e., a measure of the
precision of our redshifts;][]{Ilbert06}, defined as
\begin{equation}
\sigma \equiv 1.48 \times \textrm{ median} \left[
  \frac{\left|z_\textrm{SNID} - z_\textrm{gal}\right|}
  {1 + z_\textrm{gal}}\right],
\end{equation}
is 0.002.

There is only one spectrum that is a significant outlier when
comparing $z_\textrm{SNID}$ to $z_\textrm{gal}$.  It is
360~d past maximum brightness (according to the light curve) and we
only have a small number of SN~Ia templates that are this old (only 3
older than 300~d). The relative lack of good matches with old SN~Ia
spectra, as opposed to much younger spectra at much higher redshifts,
is most likely the cause of the erroneous redshift (and age) from
SNID. However, aside from this extreme case, the SNID-determined
redshifts are quite accurate. Only 8~per~cent (3~per~cent) of the
objects are more than 2$\sigma$ (3$\sigma$) away from the one-to-one
correspondence.

\begin{figure*}
\begin{center}$
\begin{array}{c}
\includegraphics[width=5in]{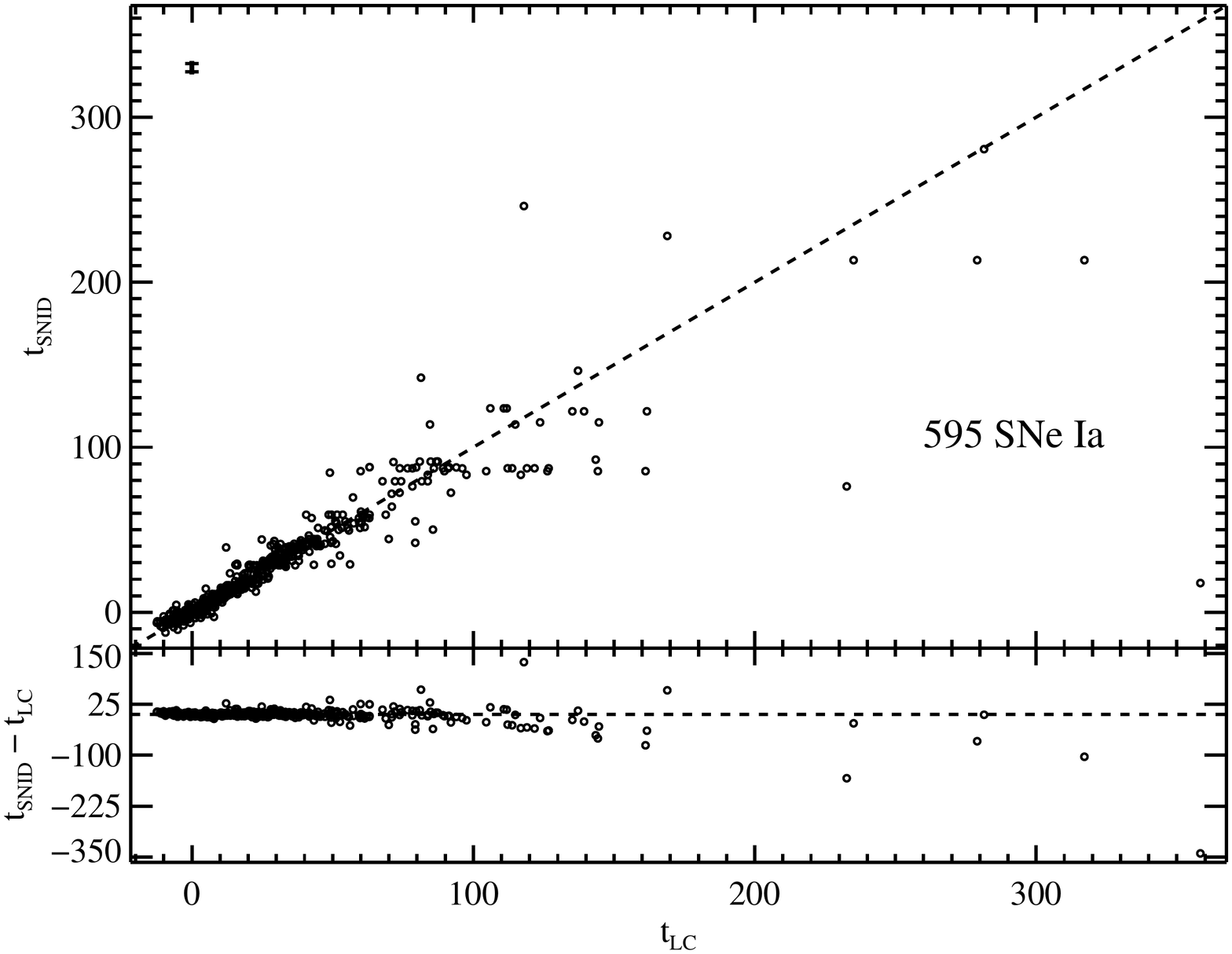} \\
\includegraphics[width=5in]{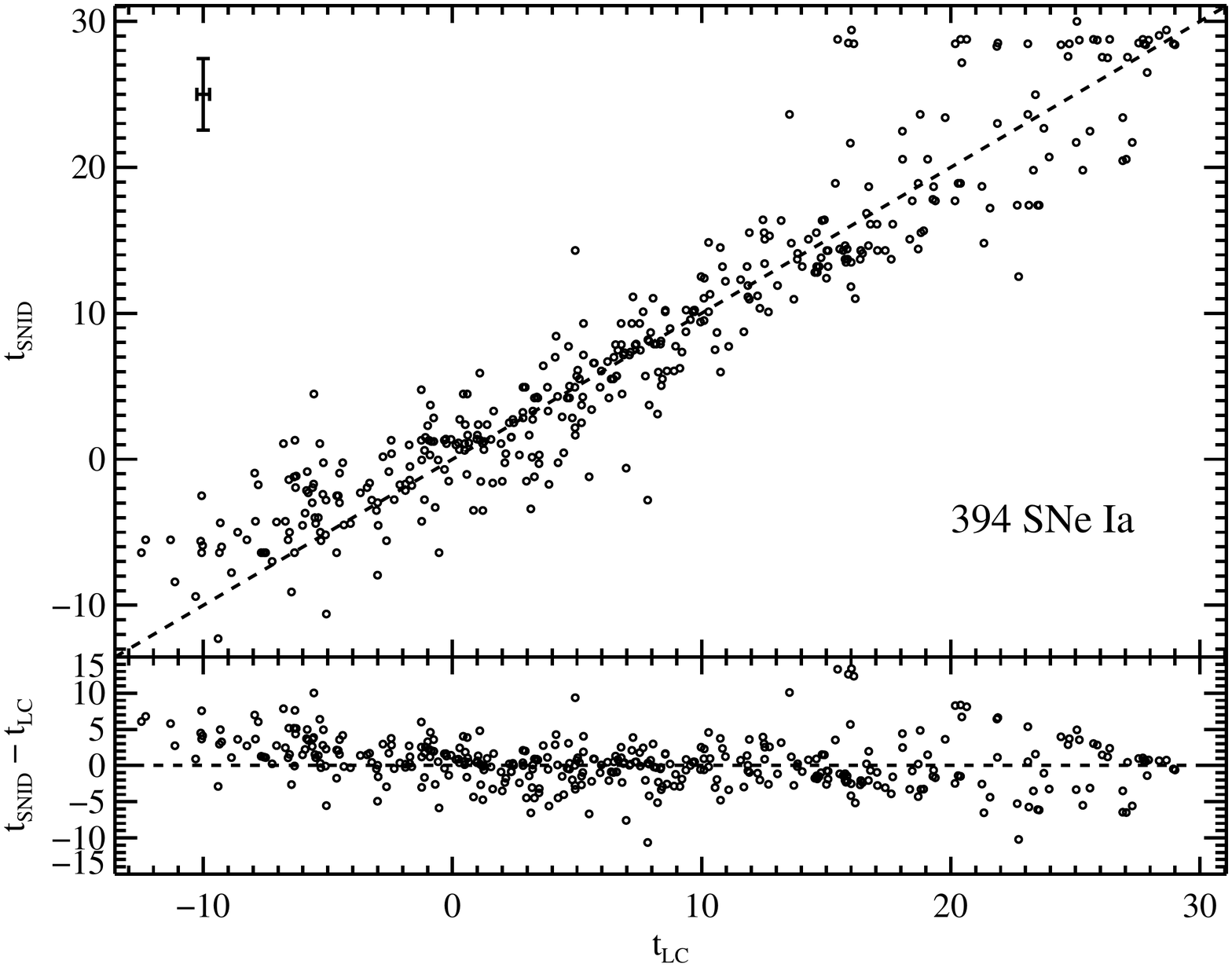}
\end{array}$
\caption[Comparison of light-curve ages and
  SNID-determined ages]{Comparison of rest-frame light-curve ages 
  ($t_\textrm{LC}$) and SNID-determined ages ($t_\textrm{SNID}$) using 
  our SNID classification scheme: all 595 spectra which SNID
  determined were SNe~Ia and that have both SNID-determined ages and
  light-curve ages ({\it top}), and a zoom-in on the 394 SN~Ia spectra with
  $t_\textrm{LC} \le 50$~d and $t_\textrm{SNID} \le 30$~d ({\it bottom}). 
  The median error bar in both directions for the 
  entire sample is shown in the top left of each plot.  The
  bottom of both plots shows 
  the residuals versus $t_\textrm{LC}$.}\label{f:snid_comp}
\end{center}
\end{figure*}

The original SNID template spectra have ages which have been corrected
for the $1 / (1 + z)$ time-dilation factor that we expect to observe in an
expanding universe
\citep[e.g.,][]{Riess97:timedilation,Foley05,Blondin08}.  Thus, SNID
templates 
should have ages in the rest frame of the SNe.\footnote{When creating
  our own SNID templates, we transformed our SN ages to the
  rest frame (using the redshift of the SN or the host galaxy).}
We compare the SNID-determined ages of our SN~Ia spectra to their
actual (rest-frame) ages as derived from their light curves and
redshifts as presented in Table~\ref{t:obj}; the result is shown
in Figure~\ref{f:snid_comp}. There are 595 total spectra that SNID
determined were SNe~Ia (and 409 with light-curve ages $\le 30$~d) which
have both SNID-determined ages and light-curve ages.   The dispersion
about the one-to-one correspondence for the total sample is
\about4.1~d, and the dispersion for the sample with light-curve 
ages $\le 30$~d is \about3.3~d. \citet{Foley09} calculated a
dispersion of 1.8~d for 59 SN~Ia spectra with light-curve ages between
$-11$ and 19.4~d at moderate to high redshift ($0.100 \leq z \leq 0.807$).  We
have 338 spectra in this range and calculate a dispersion of \about
2.8~d for these data. \citet{Ostman11} obtained a dispersion of 4~d
for 127 SN~Ia spectra with light-curve ages between $-11.7$ and 67.9~d
at moderate redshift ($0.03 \leq z \leq 0.32$). We have 529 spectra in
this range and calculate a dispersion of \about3.7~d for  these
data. 

The samples used by \citet{Foley09} and \citet{Ostman11} were at 
higher redshift than our data, and the way in which they determined
spectral ages using SNID was slightly different. \citet{Ostman11}
simply used the median of all ``good'' template ages as the
SNID-determined age. We initially attempted this relatively
straightforward method for our spectra, but we soon found a
significant bias in 
our SNID-determined ages (as compared to ages derived from the SN
light curves).  The bias was causing SNID-determined ages to be
artificially skewed toward about $+30$~d for spectra which are
(according to their photometry) in the range of about 23--33~d.  We
also observed a similar, yet weaker, version of this bias for spectra
whose photometric ages were near maximum brightness as well as spectra
with light-curve ages near \about100~d (still visible in the top panel
of Fig.~\ref{f:snid_comp}).  The bias near maximum brightness is seen
in the higher-redshift data presented by \citet{Ostman11}, and we
suggest that the bias near ages of \about30~d (and perhaps the one
near 100~d as well) {\it would} have been observed in their data if
their dataset had contained spectra near these epochs. 

We thoroughly investigated these so-called ``age attractors'' and
their effect on the dispersions, but
unfortunately found no simple explanation for the apparent bias. In
order to characterise and explain the strongest bias (near photometric
ages of one month past maximum brightness), we used the default set of
SNID templates, instead of our own, and again ran all of our spectra
through our classification routine, only to have the bias appear even 
stronger than before.  We examined the age and light-curve shape
\citep[as characterised by the MLCS2k2 $\Delta$ parameter;][]{Jha07}
distributions of our v7.0 templates, the default set of SNID
templates, and the light-curve ages of our entire dataset, and none of
these showed any deviation at or near $+30$~d that might affect SNID's
age determinations. For example, Figure~\ref{f:template_ages} shows a 
histogram of the ages of all of our SN~Ia spectral templates (separated
by SNID template version), and there does not appear to be any obvious
bias near $+30$~d. Furthermore, we investigated how the strength of
the bias changed with SNID-determined subtype, best-matching subtype,
and $r$lap, but saw no strong correlations.

On the other hand, all reported SNID-determined ages from
\citet{Foley09} required that there were at least 8 good matches (and the
SNID-determined age was calculated from the median age of those top 8
matches), that the age error (defined as the standard deviation of the top
8 matches) was $< 6$~d, and that there was a SNID-determined subtype
(S.~Blondin, 2011, private communication). We also attempted to use this
method of SNID age determination on the BSNIP data but found that an
extremely large fraction of the spectra were not being assigned a SNID
age. 

Various other methods of SNID age determination and verification
were tested (see Table~\ref{t:age_snid_tests} for a summary), and the
method that gave the best compromise between low 
dispersion values and large numbers of spectra being assigned a SNID
age is the one that we ultimately used (described above in
Section~\ref{sss:snid_age}).  
Even though this new method of determining SNID ages reduces the bias
near $+30$~d, the bias near maximum brightness \citep[mentioned above
and seen in][]{Ostman11} is still present. 
It should also be noted that despite the
biases that remain, over 60~per~cent of our SNID-derived ages are
within $1\sigma$ of the one-to-one correspondence with light-curve
age.

\begin{table*}
\begin{center}
\caption{Number of Spectra and Dispersion of Various Methods for Determining SNID Ages}\label{t:age_snid_tests}
\begin{tabular}{lcccccccccccc}
\hline\hline
 & \multicolumn{2}{c}{BSNIP} & \multicolumn{2}{c}{BSNIP} & \multicolumn{2}{c}{BSNIP} & \multicolumn{2}{c}{BSNIP} & \multicolumn{2}{c}{ O11$^\textrm{e}$} & \multicolumn{2}{c}{ F09$^\textrm{f}$} \\
 & \multicolumn{2}{c}{(final)$^\textrm{a}$} & \multicolumn{2}{c}{(alternate)$^\textrm{b}$} & \multicolumn{2}{c}{ (O11)$^\textrm{c}$} & \multicolumn{2}{c}{ (F09)$^\textrm{d}$} & & & & \\
\hline
Age Range (d) & $N$ & $\sigma$ (d) & $N$ & $\sigma$ (d) & $N$ & $\sigma$ (d) & $N$ & $\sigma$ (d) & $N$ & $\sigma$ (d) & $N$ & $\sigma$ (d) \\
\hline
$-20 \leq t \leq 360$ & 595 & 4.1 & 869 & 5.0 & 860 & 5.4 & 315 & 3.2 & $\cdots$ & $\cdots$ & $\cdots$ & $\cdots$ \\
$-20 \leq t \leq 30$   & 409 & 3.3 & 580 & 3.9 & 572 & 4.1 & 264 & 2.9 & $\cdots$ & $\cdots$ & $\cdots$ & $\cdots$ \\
$-12 \leq t \leq 68$   & 529 & 3.7 & 757 & 4.5 & 748 & 4.7 & 314 & 3.2 & 127 & 4 & $\cdots$ & $\cdots$ \\
$-11 \leq t \leq 19$   & 338 & 2.8 & 451 & 3.3 & 444 & 3.4 & 233 & 2.4 & $\cdots$ & $\cdots$ & 59 & 1.8 \\
\hline \hline
\multicolumn{13}{p{4.6in}}{``O11'' = \citet{Ostman11}, ``F09'' = \citet{Foley09}.} \\
\multicolumn{13}{p{4.6in}}{$^\textrm{a}$The method described in Section~\ref{sss:snid_age}.} \\
\multicolumn{13}{p{4.6in}}{$^\textrm{b}$Similar to the method described
  in Section~\ref{sss:snid_age}. The SNID age is calculated by taking
  the median of all ``good'' template ages that have an $r$lap value
  larger than 75~per~cent of the $r$lap value of the best-matching
  template. However, no other requirements are imposed.} \\
\multicolumn{13}{p{4.6in}}{$^\textrm{c}$The method used by 
  \citet{Ostman11} applied to the BSNIP data. This simply uses the
  median of all ``good'' template ages as the SNID-determined age.} \\
\multicolumn{13}{p{4.6in}}{$^\textrm{d}$The method used by
  \citet{Foley09} applied to the BSNIP data. It requires
  that there are at least 8 good matches (and the SNID-determined age
  is calculated from the median age of those top 8 matches), that the age
  error (defined as the standard deviation of the top 8 matches) is
  $< 6$~d, and that there is a SNID-determined subtype (S.~Blondin,
  2011, private communication).} \\
\multicolumn{13}{p{4.6in}}{$^\textrm{e}$Values reported by
  \citet{Ostman11} using data with $0.03 \leq z
  \leq 0.32$.} \\
\multicolumn{13}{p{4.6in}}{$^\textrm{f}$Values reported by
  \citet{Foley09} using data with $0.10 \leq z \leq
  0.81$.} \\
\hline\hline
\end{tabular}
\end{center}
\end{table*}

\subsection{Classification of Objects}

After classifying (or attempting to classify) all of the spectra in
our dataset using the method described in
Section~\ref{ss:classification}, we use the SNID information for all
spectra of a given object to determine the SN's (sub)type. 

\subsubsection{Classification Accuracy}\label{sss:snid_acc}

To investigate the accuracy of our SN~Ia classification scheme, we
attempt to find correlations between the accuracy of our SN~Ia subtype
determination and the $r$lap value of the best-match template, the
SNID-determined age of the input spectrum, and the S/N of the input
spectrum. We find that our accuracy is similarly correlated with both
$r$lap and S/N, which is unsurprising since S/N is one of the factors
which determines the $r$lap value during a SNID run.  Thus, we attempt
to correlate our classification accuracy with only $r$lap and spectral
age, simultaneously.

\begin{figure}
\centering
\includegraphics[width=3.5in]{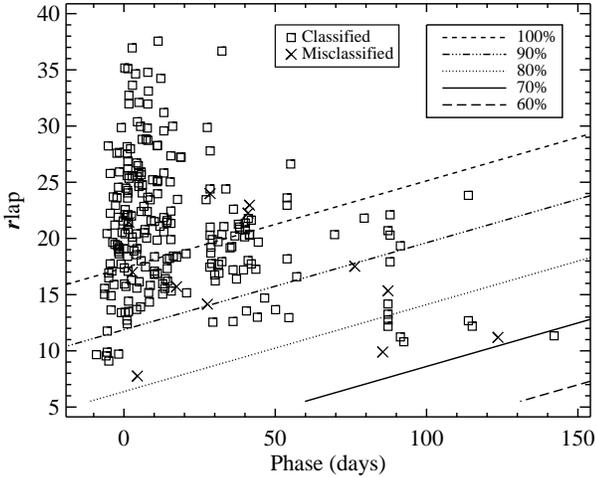}
\caption[Accuracy of SNID-determined SN~Ia subtypes]{Accuracy of our
  SNID-determined SN~Ia subtypes (using our 
  classification scheme described in Section~\ref{ss:classification})
  versus SNID-determined (rest-frame) phase (in days) and $r$lap.
  The $\square$ are correctly classified spectra from our v7.0 SNID
  templates; $\times$ are misclassified spectra from our v7.0 SNID
  templates.  The contours (from bottom to top) represent 60~per~cent,
  70~per~cent, 
  80~per~cent, 90~per~cent, and 100~per~cent accuracy. The few spectra
  which are misclassified but above the 100~per~cent contour are due
  to the fact that we fit all v7.0 template spectra and thus a handful
  of misclassified objects formally have accuracies above
  100~per~cent.}\label{f:snid_accuracy}
\end{figure}

To do this, we compare the subtype we determined using our
classification scheme (Section~\ref{ss:classification}) for each SN~Ia 
template spectrum in v7.0 to the actual subtype of the template
object.  If our multiple SNID runs correctly classified a given
spectrum, we assigned it an ``accuracy percentage'' ($P$) of 1, and if
it was misclassified it received an accuracy classification of 0. We
then used the $r$lap value of the best-matching template ($r$lap) and
the SNID-determined (rest-frame) age ($t$, in days) of each spectrum
to fit a two-dimensional surface of the form 
\begin{equation}\label{e:perc_acc}
P = c_1 + c_2 \times t + c_3 \times r\textrm{lap}
\end{equation}
to the SNID classifications of our v7.0 SN~Ia template spectra.  This
function was fit on a grid of phases from $-20$ to 200~d (in steps of
2~d) and $r$lap values from 5 to 40 (in steps of 1).  Our best-fitting
values for the constants were $c_1 = 0.68 \pm 0.04$, $c_2 = -0.0014
\pm 0.0002$, and $c_3 = 0.018 \pm 0.002$, and our resulting contours
are shown in Figure~\ref{f:snid_accuracy}. The few spectra which are
misclassified but lie above the 100~per~cent contour are simply ones
where even though they are misclassified, their age and $r$lap values
formally imply an ``accuracy percentage'' of $> 100$~per~cent. Using
our values for $c_1$,  $c_2$, and $c_3$, we can now calculate an
``accuracy percentage'' ($P$) for {\it any} SN~Ia spectrum that has a
SNID-determined age and $r$lap (by using Equation~\ref{e:perc_acc}).

%

As expected, our accuracy increases with increased $r$lap value since
$r$lap is a measure of the strength of the correlation of the input
spectrum with the best-matching SNID template.  In addition, our
accuracy decreases with increased age since, as noted
earlier, as some subtypes of SNe~Ia evolve and age their optical
spectra begin to resemble those of ``normal SNe~Ia.'' 

\subsubsection{Final Object Classification}\label{sss:final_class}

For SNe with multiple spectra, we must consider each spectrum's
classification when obtaining a final classification for a given
object. To do this we first determine an object's type by counting the
total number of spectra of each type for the given object.  The object
is then classified as the type with the highest count.  If there is
more than one type with the same highest count, we compare the spectra
of those types only and use the type of the spectrum with the largest
$r$lap (though we add a ``?'' to the type to denote our uncertainty
regarding the classification). 

For each object whose definite type has been determined (i.e., no
trailing ``?'') and that {\it is not} a SN~Ia, we assign a subtype by
counting the total number of spectra of each subtype for the given
object.  The SN is then classified as the subtype with the highest
count.  If there is more than one subtype with the same highest count,
we cannot accurately determine the subtype and thus classify the
object as simply the previously determined type.

For each object that we have determined is a definite SN~Ia (i.e., no
trailing ``?''), we calculate the ``accuracy percentage'' ($P$) for
each spectrum of that object using Equation~\ref{e:perc_acc}.  We then
combine the ``accuracy percentages'' of all of the spectra of the
given SN to calculate the {\it maximum} probability that it is of each
subtype.  For example, if a spectrum has an ``accuracy percentage'' of
$P$, then that is a measure of the probability that the spectrum is of
subtype $m$ (where subtype $m$ is the subtype determined by our
classification scheme in Section~\ref{ss:classification}).  This means
that $1-P$ is the probability that this spectrum is any of the other
SN~Ia subtypes, and it is in fact the {\it maximum} probability that
the spectrum is of subtype $n$ for any $n \ne m$.  Therefore, we can 
combine the ``accuracy percentages'' for multiple spectra of different
subtypes to calculate a maximum probability for each subtype, 
\begin{equation}
\widetilde{P}_m = \frac{\prod_{n=m}P_n \times \prod_{n \ne
    m}\left(1-P_n\right)}{k_m},
\end{equation}
where $\widetilde{P}_m$ is the maximum probability the given object is
of subtype $m$, the first product is over all spectra whose subtype
($n$) is equal to $m$, the second product is over all spectra whose
subtype ($n$) is not equal to $m$, and $k_m$ is a normalisation
constant for subtype $m$ defined as
\begin{equation}
k_m = \prod_{n=m}P_n \times \prod_{n \ne m}\left(1-P_n\right) +
\prod_{n=m}\left(1-P_n\right) \times \prod_{n \ne m}P_n.
\end{equation}
Once we calculate the maximum probability for each subtype, we 
classify the given object as the subtype with the largest such
probability.

SNID is merely a tool, albeit a useful one for determining spectral
subtypes of SNe, but ultimately humans classify SN spectra.  Thus, we
visually inspected any objects in our dataset that were classified as 
``Ia'' or ``Ia?'' (with no subtype) or as any of the non-SN~Ia
subtypes, in addition to any object that had spectra that were
classified as more than one SN~Ia subtype.  From these visual
inspections we changed a handful of the final object classifications
from what our SNID-based classification algorithm would have
yielded. We also forced all objects that were v7.0 SNID templates to
be their actual subtype.  These final (sub)type determinations can be
found in Table~\ref{t:obj}; we also present a summary of the final
(sub)type determinations from SNID, as well as our adjusted
classifications, in Table~\ref{t:snid_summary}. 


\begin{table}
\begin{center}
\caption{Summary of Final SN Classifications}\label{t:snid_summary}
\begin{tabular}{lr||lr}
\hline\hline
\multicolumn{2}{c||}{SNID} & \multicolumn{2}{c}{Adjusted} \\
\hline
(Sub)Type & \# & (Sub)Type & \# \\
\hline
Ia &       94 & Ia &       37 \\
Ia? &        2 & Ia? &        0 \\
Ia-norm &      430 & Ia-norm &      459 \\
Ia-91T &        3 & Ia-91T &        7 \\
Ia-91bg &       25 & Ia-91bg &       46 \\
Ia-csm &        0 & Ia-csm &        1 \\
Ia-pec &        0 & Ia-pec &        1 \\
Ia-99aa &        2 & Ia-99aa &       13 \\
Ia-02cx &        7 & Ia-02cx &       10 \\
Ic &        1 & Ic &        0 \\
Ic-norm &        1 & Ic-norm &        0 \\
II &        1 & II &        0 \\
IIP &        3 & IIP &        0 \\
IIn &        1 & IIn &        0 \\
Gal &        2 & Gal &        0 \\
Unknown &       10 & Unknown &        8 \\
\hline\hline
\end{tabular}
\end{center}
\end{table}


Of all of the v7.0 template objects (which include SNe~Ia as well as
other SN types), 95~per~cent were classified as 
the correct type by SNID and 86~per~cent were classified as the
correct subtype by SNID.  We were unable to determine a type (subtype)
for 2~per~cent (7~per~cent) of the v7.0 template objects using our
aforementioned SNID-based classification scheme.  Of solely the v7.0
template objects that are SNe~Ia, 99~per~cent received the correct
type 
classification from SNID and 92~per~cent the correct subtype
classification, and we could determine a type and subtype from SNID
for all of the v7.0 template SNe~Ia. These are more informative and
more accurate percentages than those from the simple sanity checks
that were discussed in Section~\ref{sss:final_ver}.

Of the \totobj\ SNe~Ia that are presented in this paper, our
SNID-based algorithm classifies 96.8~per~cent of them as SNe~Ia,
1.5~per~cent as other SN types, and we are unable to classify
1.7~per~cent.  Over half of the spectra of objects that are classified
by SNID as non-Ia types or that are unclassified are relatively old
and/or noisy, and thus it is not surprising that our classification
scheme failed on these observations.  A few spectra of these objects
are heavily contaminated by host-galaxy light and lacked sufficient
photometry for us to apply our galaxy-correction algorithm
(Section~\ref{ss:galaxy}), and so again it is reasonable that these are
not correctly classified. When it was not clear from our own spectra
that these objects were SNe~Ia, we relied on previous spectral
classifications from the literature. In addition, some of the objects
which were classified as non-Ia types are well-known, peculiar SNe~Ia
(including a few of our v7.0 SNID templates), but we do not have
enough templates of other objects of their subtype at similar ages for
SNID to get good matches to our observations.  This is most likely
responsible for many of SNID's ``Ia'' (with no definitive subtype) and
``Ia?'' classifications as well, and is partially why we visually
reinspected these objects.

\subsubsection{Relative Fractions of Subtypes}\label{sss:fractions}

Even though our SN~Ia spectral sample is not complete by any rigorous
definition and suffers from multiple observational biases, it is still
illuminating to calculate the relative fractions of SN~Ia subtypes (as
determined by SNID) in our dataset.  Of the objects for which SNID
determines a SN~Ia subtype, 92.1~per~cent are normal, 5.4~per~cent are
91bg-like, 0.6~per~cent are 91T-like, 0.4~per~cent are 99aa-like, and
1.5~per~cent are 02cx-like. These fractions are listed in 
Table~\ref{t:fractions}, along with those from other SN~Ia
studies. For the purposes of comparing these 
percentages to values in the literature, we will follow \citet{Li11a} and
combine 99aa-like objects with 91T-like objects, yielding 1.0~per~cent
for this group.  If, as we did when creating our new SNID templates in
order to get more accurate subtype classifications, we require spectra
to have ages less than 15~d past maximum brightness and S/N $>
15$~pixel$^{-1}$, then we find that of these objects 90.7~per~cent are
normal, 6.5~per~cent are 91bg-like, 1.9~per~cent are 91T/99aa-like,
and 0.9~per~cent are 02cx-like.

\begin{table*}
\begin{center}
\caption{Percentages of SN~Ia Subtypes}\label{t:fractions}
\begin{tabular}{lcccccc}
\hline\hline
 & BSNIP & BSNIP & G10 & L11$^\textrm{b}$ & O11 & F09 \\
 & (Total) & (Cut)$^\textrm{a}$  & & & & \\ 
\hline
$\overline{z}$ &  0.0283 & 0.0219 & 0.0194 & 0.0132 & 0.17 & 0.35 \\
Ia-norm & 92.1\phantom{$^\textrm{c}$} & 90.7\phantom{$^\textrm{d}$} & 80.6 & 70 & 97.9\phantom{$^\textrm{e}$} & 81--96 \\
Ia-91bg & \phantom{0}5.4\phantom{$^\textrm{c}$} & \phantom{0}6.5\phantom{$^\textrm{d}$} & 10.9 & 15 & \phantom{0}0.0\phantom{$^\textrm{e}$} & 0 \\
Ia-91T/99aa & \phantom{0}1.0$^\textrm{c}$ & \phantom{0}1.9$^\textrm{d}$ & \phantom{0}5.5 & \phantom{0}9 & \phantom{0}1.4$^\textrm{e}$ & 4--19 \\
Ia-02cx & \phantom{0}1.5\phantom{$^\textrm{c}$} & \phantom{0}0.9\phantom{$^\textrm{d}$} & \phantom{0}2.4 & \phantom{0}5 & \phantom{0}0.0\phantom{$^\textrm{e}$} & 0 \\
Ia-pec & \phantom{0}0.0\phantom{$^\textrm{c}$} & \phantom{0}0.0\phantom{$^\textrm{d}$} & \phantom{0}0.6 & \phantom{0}0 & \phantom{0}0.7\phantom{$^\textrm{e}$} & 0 \\
\hline \hline
\multicolumn{7}{p{3.4in}}{``G10'' = \citet{Ganeshalingam10:phot_paper},
  ``L11'' = \citet{Li11a}, 
``O11'' = \citet{Ostman11}, ``F09'' = \citet{Foley09}.} \\
\multicolumn{7}{p{3.4in}}{$^\textrm{a}$Only spectra with ages less than 15~d
  past maximum brightness and S/N $> 15$~pixel$^{-1}$.} \\
\multicolumn{7}{l}{$^\textrm{b}$A volume-limited sample.} \\
\multicolumn{7}{l}{$^\textrm{c}$0.6~per~cent are 91T-like and
  0.4~percent are 99aa-like.} \\
\multicolumn{7}{l}{$^\textrm{d}$1.0~per~cent are 91T-like and
  0.9~percent are 99aa-like.} \\
\multicolumn{7}{p{3.4in}}{$^\textrm{e}$Even though only 1.4~per~cent of the
  sample from \citet{Ostman11} were classified as Ia-91T/99aa, they
  calculate that 7--32~per~cent of local SNe~Ia should be
  91T/99aa-like.} \\ 
\hline\hline
\end{tabular}
\end{center}
\end{table*}

In the volume-limited sample of \citet{Li11a}, 70~per~cent of
SN~Ia were normal, while 15~per~cent were 91bg-like, 9~per~cent were
91T/99aa-like (only a lower limit, however), and 5~per~cent were
02cx-like. The slightly higher redshift ($\overline{z} = 0.17$) sample
of \citet{Ostman11} detected only a few probable peculiar SNe~Ia, but
they calculate that 7--32~per~cent of local SNe~Ia should be
91T/99aa-like.  The even higher redshift ($\overline{z} \approx 0.35$)
sample of \citet{Foley09} found that 4--19~per~cent were
91T/99aa-like, 
while they had no 91bg-like objects in their
dataset.\footnote{\citet{Foley09}, after correcting for various biases,
  expected only 1--4 91bg-like objects based on the \citet{Li01:pec}
  percentages.} However, an analysis of the companion photometry to
much of the spectroscopic sample presented here
\citep{Ganeshalingam10:phot_paper} finds (using SNID as described in
this work) that 80.6~per~cent of their
SNe~Ia are normal, 10.9~per~cent are 91bg-like, 5.5~per~cent are
91T/99aa-like, 2.4~per~cent are 02cx-like, and 0.6~per~cent are
Ia-pec (i.e., SN~2000cx, which SNID classified herein as a Ia-norm
since we have no other SNID templates of 00cx-like objects).  These
percentages are closer to our fractions of subtypes than the
complete sample presented by \citet{Li11a}. This likely comes
from the fact that for most of the project's lifetime, BSNIP has had a
strong focus on spectroscopically monitoring SNe~Ia that were being
concurrently observed photometrically as part of the sample presented
by \citet{Ganeshalingam10:phot_paper}.

The main differences between our SNID-determined fractions of SN~Ia
subtypes and those found in the complete sample of
\citet{Li11a} is that we classify too many objects as normal
and not enough as 91T/99aa-like and 91bg-like. This can possibly be
explained by the fact that spectra of these peculiar SNe~Ia
(especially 91T/99aa-like objects) resemble spectra of normal 
SNe~Ia within a week or two after maximum brightness
\citep[introducing the so-called ``age bias'';][]{Li01:MC}. Thus, some
of our Ia-norm objects may in fact be 91T/99aa-like SNe~Ia, but the
spectra in our dataset were obtained at too late an epoch
to distinguish between the two
subtypes.  Since there are many more normal SNID templates than
peculiar templates, these objects will ultimately get
classified as normal. Furthermore, it is possible that some objects
have essentially normal spectra and are classified as Ia-norm by SNID,
but have slowly (quickly) declining light curves and are thus
{\it photometrically} classified as 91T/99aa-like (91bg-like) by
\citet{Li11a}. This interesting possibility will be investigated
further in future BSNIP papers.

Many of the non-normal SNe~Ia that we classify in our dataset have
already been noted as peculiar in the literature (either in 
unrefereed circulars or published papers).  However, there are still
a few peculiar classifications that we publish here for the first
time. Details regarding these interesting individual objects can be
found in Section~\ref{ss:reclass}.


\section{The BSNIP Sample}\label{s:sample}

Our SN~Ia spectral dataset consists of a total of \totspec\ spectra of
\totobj\ SNe~Ia observed from 1989 through the end of
2008, representing nearly 470~hr of telescope
time. \unpubspec\ spectra of \unpubobj\ objects are published here for
the first time. 
Plots of all of the fully reduced spectra as well as (for the objects
with multi-band SN and galaxy photometry) galaxy-subtracted spectra
(as discussed in Section~\ref{ss:galaxy}) presented in this work are
available online --- see the Supporting Information. 
Spectral sequences for all objects in our dataset with more than 7
spectra can be found online (see the Supporting Information).
%
%



\subsection{Sample Characteristics}\label{ss:sample_char}

If we remove objects where we have no light-curve information, leaving
only spectra with phase (and light-curve shape) information, the
sample is reduced to \phasespec\ (\lcspec) spectra of \phaseobj\
(\lcobj) SNe~Ia.  If we further restrict ourselves to objects with
relatively well-sampled, filtered light curves, retaining only those
objects with precise distance measurements, our dataset contains
\filtspec\ spectra of \filtobj\ SNe~Ia.  Finally, if we only count
spectra which have reasonable estimates of multi-filtered SN
magnitudes at the time of the spectrum and measurements of the
host-galaxy colours at the position of the SN, providing accurate
flux-calibrated, galaxy-subtracted spectra (see
Section~\ref{ss:galaxy} for more information), our sample contains
\finalspec\ spectra of \finalobj\ SNe~Ia.

\begin{figure*}
\centering$
\begin{array}{cc}
\includegraphics[width=3.45in]{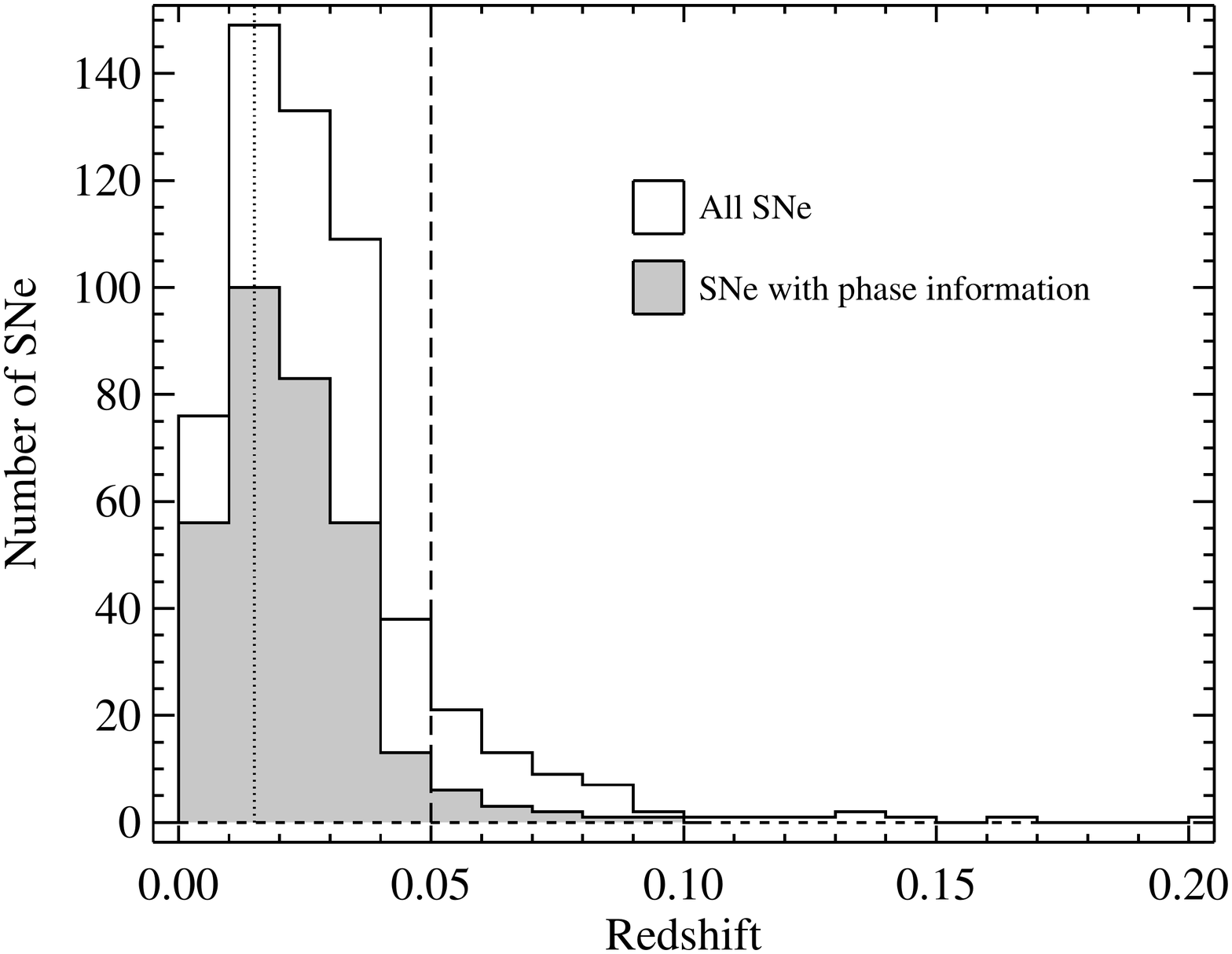} &
\includegraphics[width=3.45in]{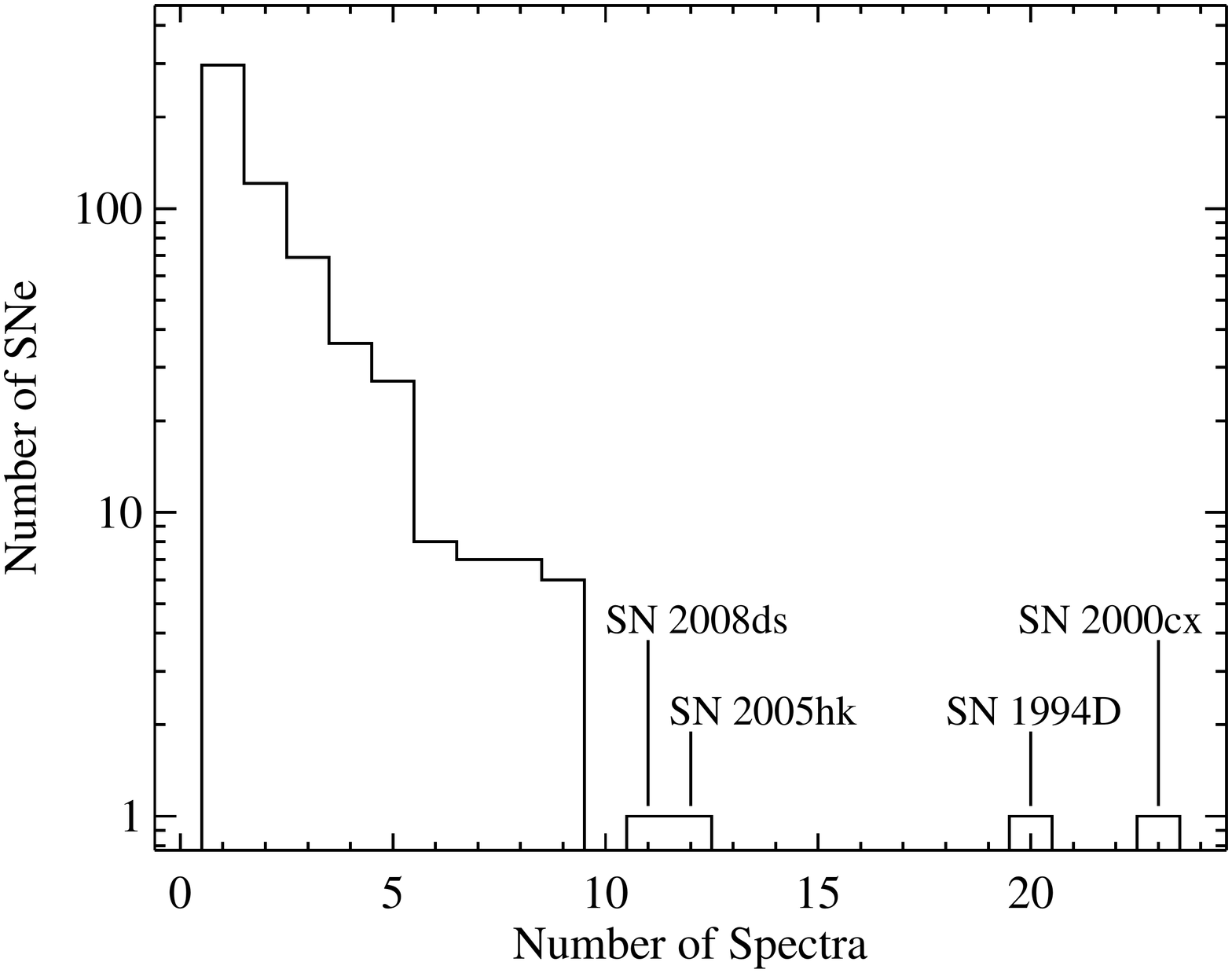} \\
\includegraphics[width=3.45in]{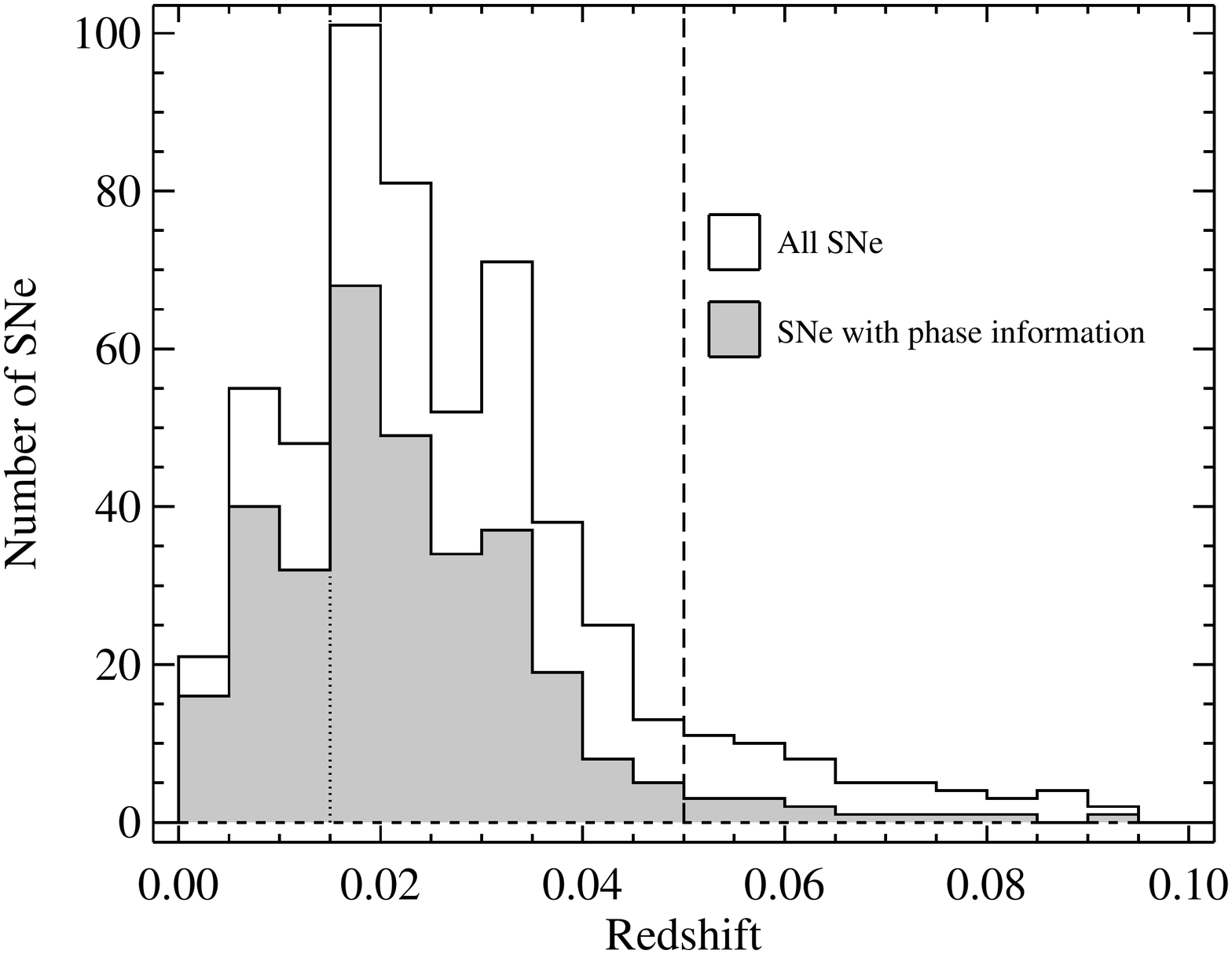} &
 \\
\end{array}$
\caption[Histogram of number of spectra versus number of
  objects]{A histogram of
  the redshifts of all of the SNe~Ia in our 
  sample ({\it top left}) and a zoom-in on those objects with $z \le 0.1$
  ({\it bottom left}). The shaded regions represent objects for which we have phase
  information (i.e., a date of maximum brightness).  The dotted 
  vertical line ($z = 0.015$) is approximately the redshift above which
  peculiar velocities can be ignored in cosmological calculations
  \citep[see, e.g.,][]{Astier06}. The dashed vertical line ($z = 0.05$)
  represents our approximate 90~per~cent cutoff; i.e.,
  \about90~per~cent of our SNe~Ia have redshifts less than 0.05. Our
  average redshift is about 0.0283 and our median uncertainty is 
  0.00004.
  A histogram of the number of spectra versus the number of
  objects in our SN~Ia sample ({\it top right}).  Our average is 
  \specperobj\ spectra per object.}\label{f:specperobj} 
\end{figure*}

This dataset has \mathspec\ times the number of spectra and \mathobj\
times the number of objects as the sample of \citet{Matheson08}.
Due to the scheduling of their telescope time, their dataset consisted 
mainly of well-sampled spectroscopic time series of a handful of
SNe~Ia, averaging 13.5 spectra per object.  By contrast, the BSNIP
sample consists of \specperobj\ spectra per object, showing our
emphasis on the total number of objects rather than the number of
spectra per object.  The histogram of the number of spectra per object
for our sample can be seen in the top-right panel of
Figure~\ref{f:specperobj}.  Thus, our 
sample (emphasizing the total number of objects) and the sample of
\citet{Matheson08} (emphasizing the number of
spectra per object) are complementary. Furthermore, as mentioned in
Section~\ref{s:obs}, our spectra typically cover 3300--10,400~\AA,
compared to the typical 3700--7400~\AA\ wavelength range of the
spectra from \citet{Matheson08}.

All of the SNe~Ia in our dataset have $z \le 0.2$, and the
vast majority (89.6~per~cent) have $z \le
0.05$. The distribution of redshifts for all of our SNe~Ia (as well as
just those with phase information) is shown in the left panels of 
Figure~\ref{f:specperobj}.  The  average redshift of the full
sample of objects presented here is about 0.0283 and the median
uncertainty is 0.00004 (as determined from the redshift uncertainties
reported in NED). About 78~per~cent of our SNe~Ia have $z \ge
0.015$ \citep[which is approximately the redshift 
above which peculiar velocities can be ignored in cosmological
calculations; see, e.g.,][]{Astier06}. 17 of our objects have unknown
host-galaxy redshifts, but all of the spectra of these objects
received a redshift from our SNID classification scheme
(Section~\ref{ss:classification}). 

\begin{figure*}
\begin{center}$
\begin{array}{cc}
\includegraphics[width=3.5in]{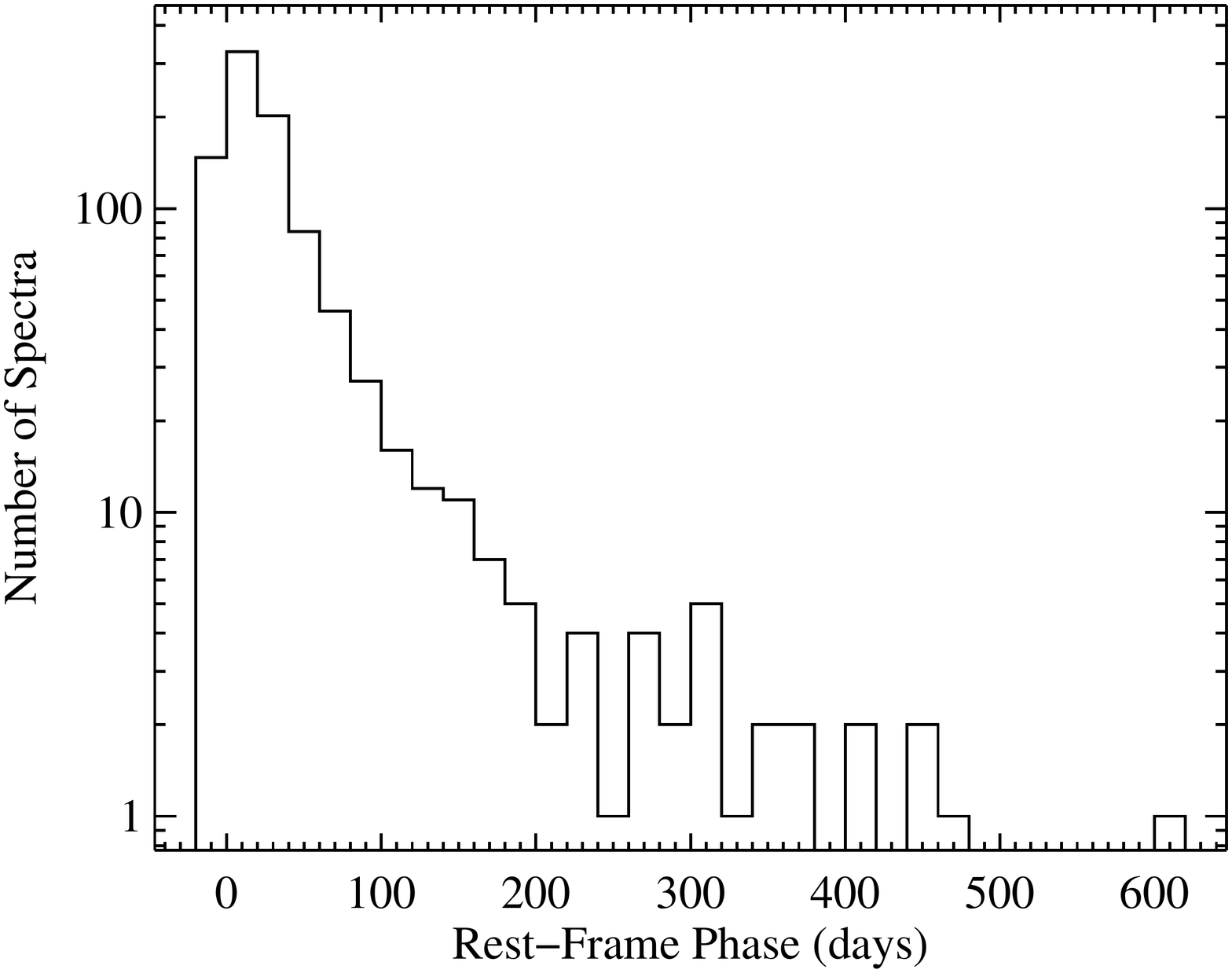} &
\includegraphics[width=3.5in]{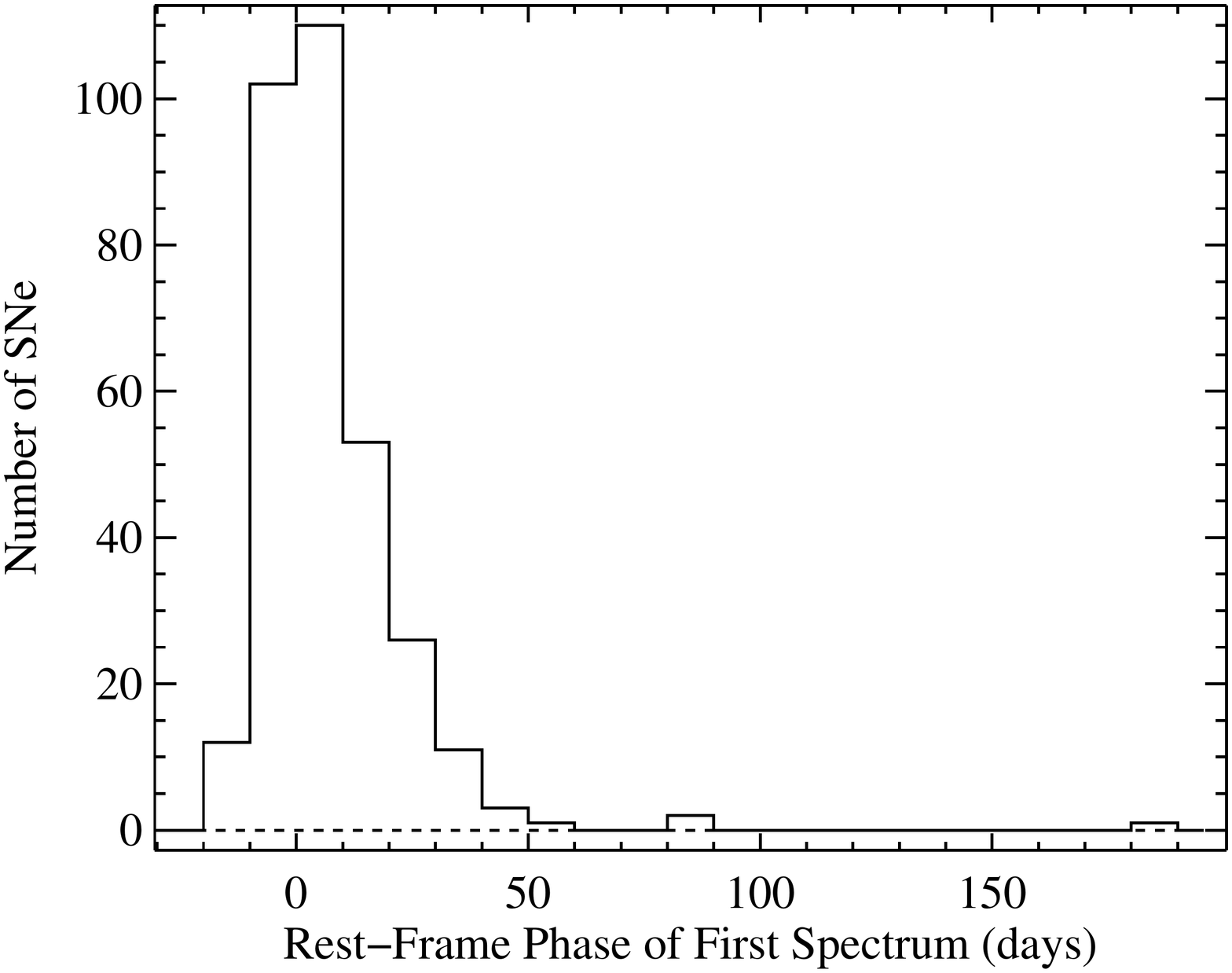} \\
\includegraphics[width=3.5in]{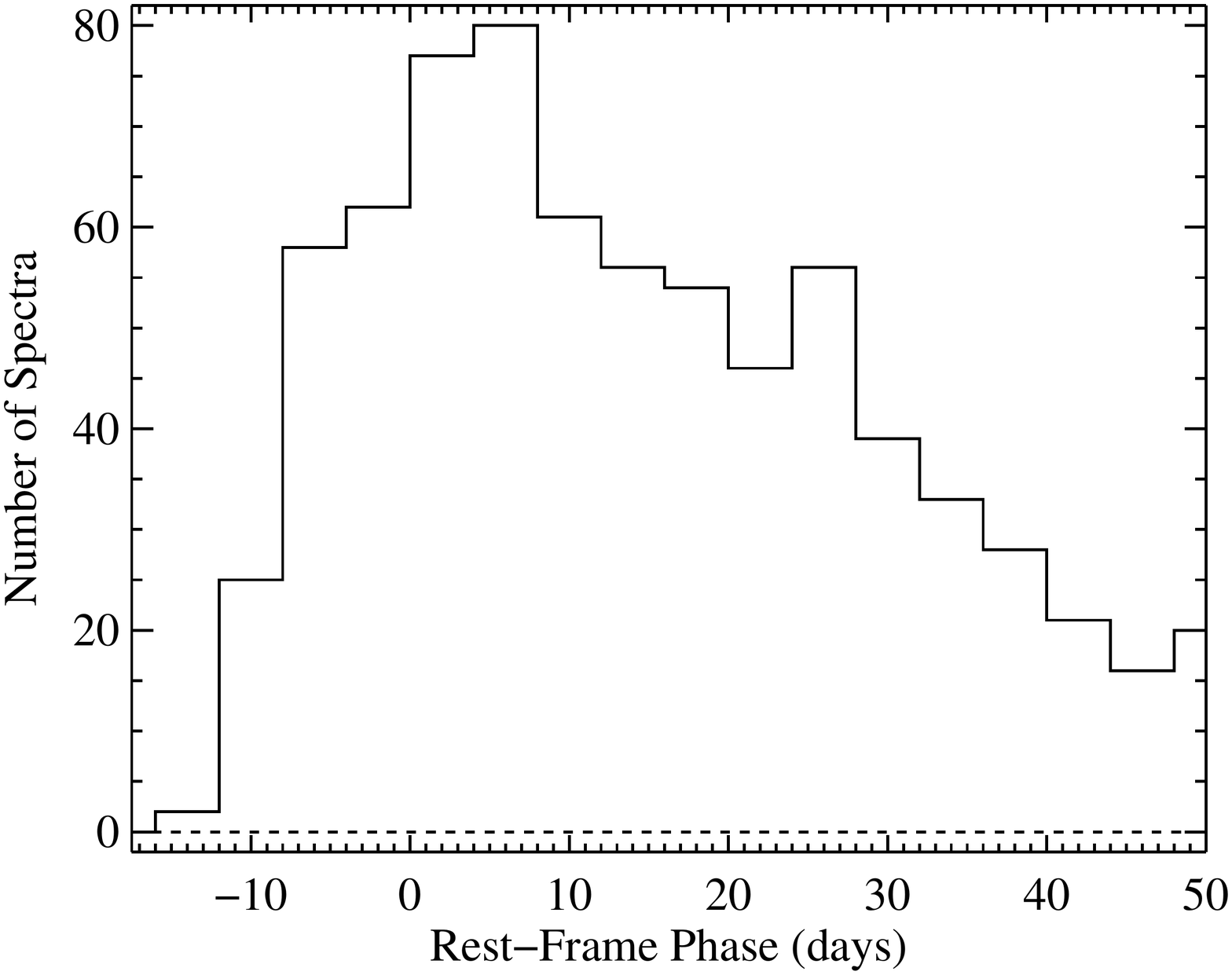} &
\includegraphics[width=3.5in]{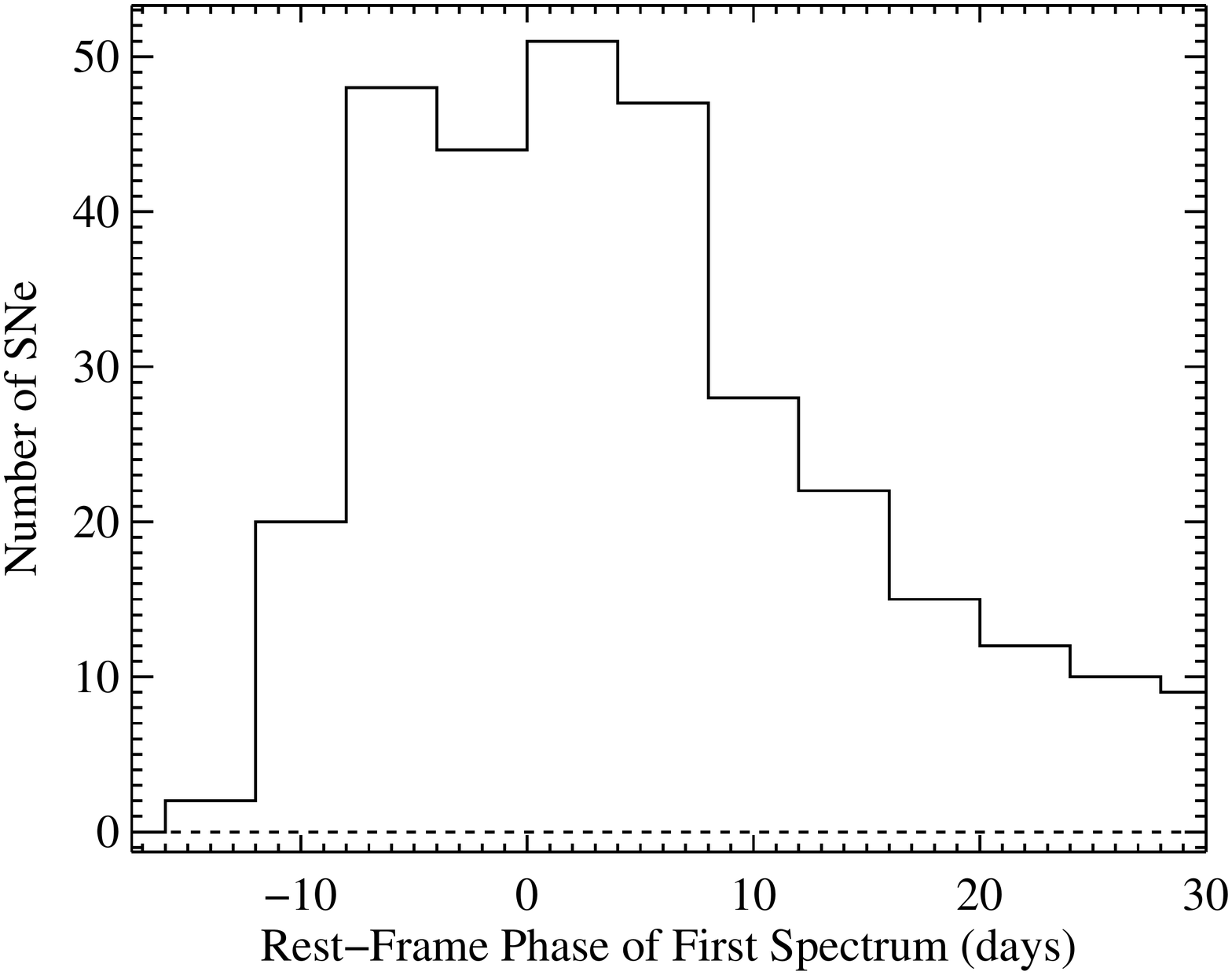} \\
\end{array}$
\caption[Histogram of the phases of all of the spectra]{A histogram of
  the phases of all of the spectra 
  in our sample for which we have phase information ({\it top left}) and a
  zoom-in on those spectra with $t \le 50$~d
  ({\it bottom left}). Our median uncertainty is 0.38~d. A histogram of
  the phase of each SN~Ia (for which we have 
  phase information) at the time of our first spectrum ({\it top right}) and a
  zoom-in on those spectra with $t_\textrm{first} \le 30$~d
  ({\it bottom right}). Our median uncertainty is 0.38~d.}\label{f:phase_hist}
\end{center}
\end{figure*}

As mentioned above, many of our spectra have phase information from
photometric observations; the distribution of phases for all of our
spectra is shown in the left panels of Figure~\ref{f:phase_hist}. We
have 147 spectra of 
114 SNe~Ia before maximum brightness, and 245 (107) spectra of 
181 (96) objects within 7~d (3~d) of maximum brightness.  Our sample
also contains 34 spectra of 20 SNe~Ia older than 180~d past maximum
brightness.  The median uncertainty of our phases is 0.38~d, though
the practical uncertainty is more like 0.5~d
\citep{Ganeshalingam10:phot_paper}. The right panels of
Figure~\ref{f:phase_hist} show the distribution of the phase of each
SN~Ia at the time of its first spectrum.

\begin{figure*}
\centering $
\begin{array}{cc}
\includegraphics[width=3.5in]{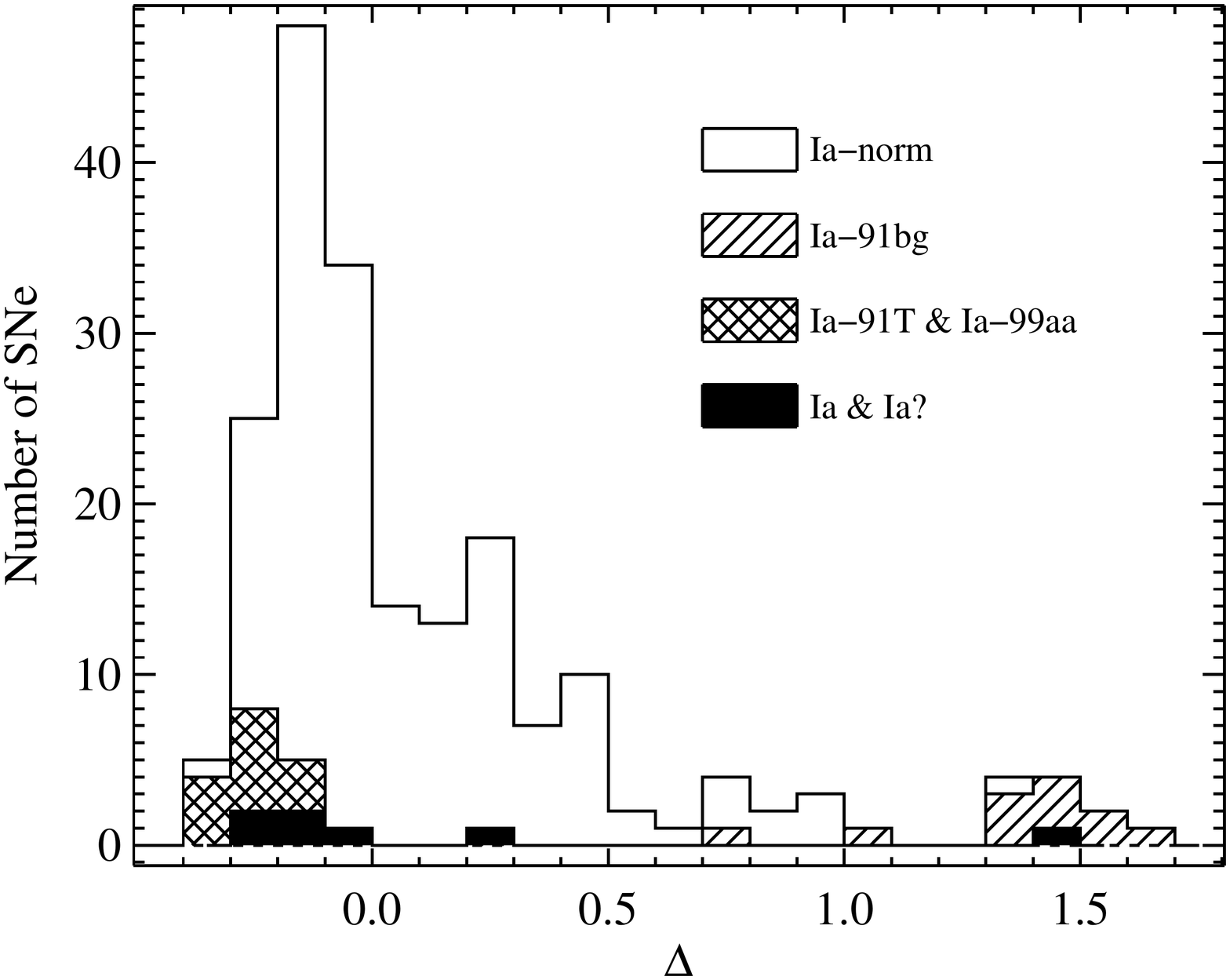} &
\includegraphics[width=3.5in]{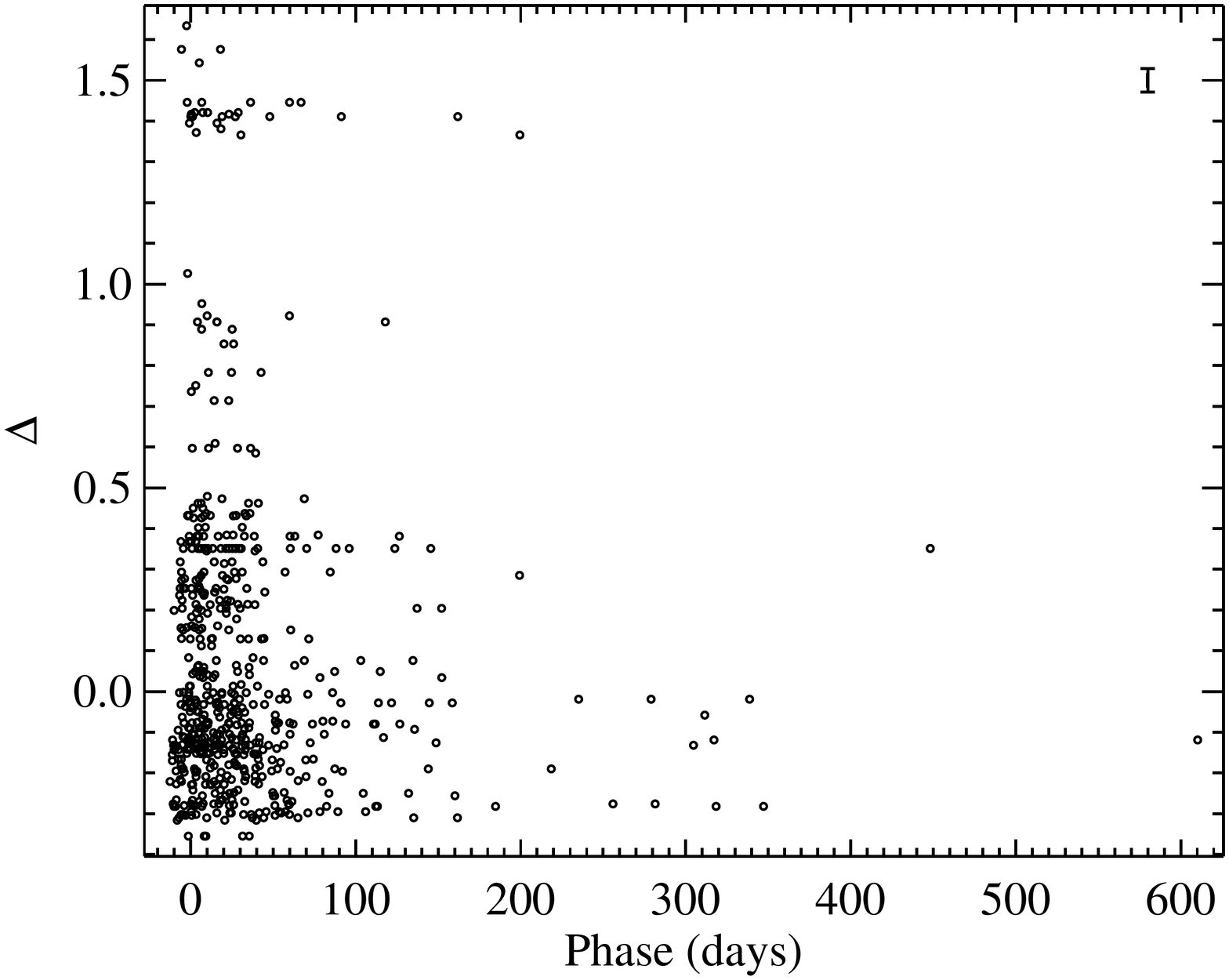} \\
\includegraphics[width=3.5in]{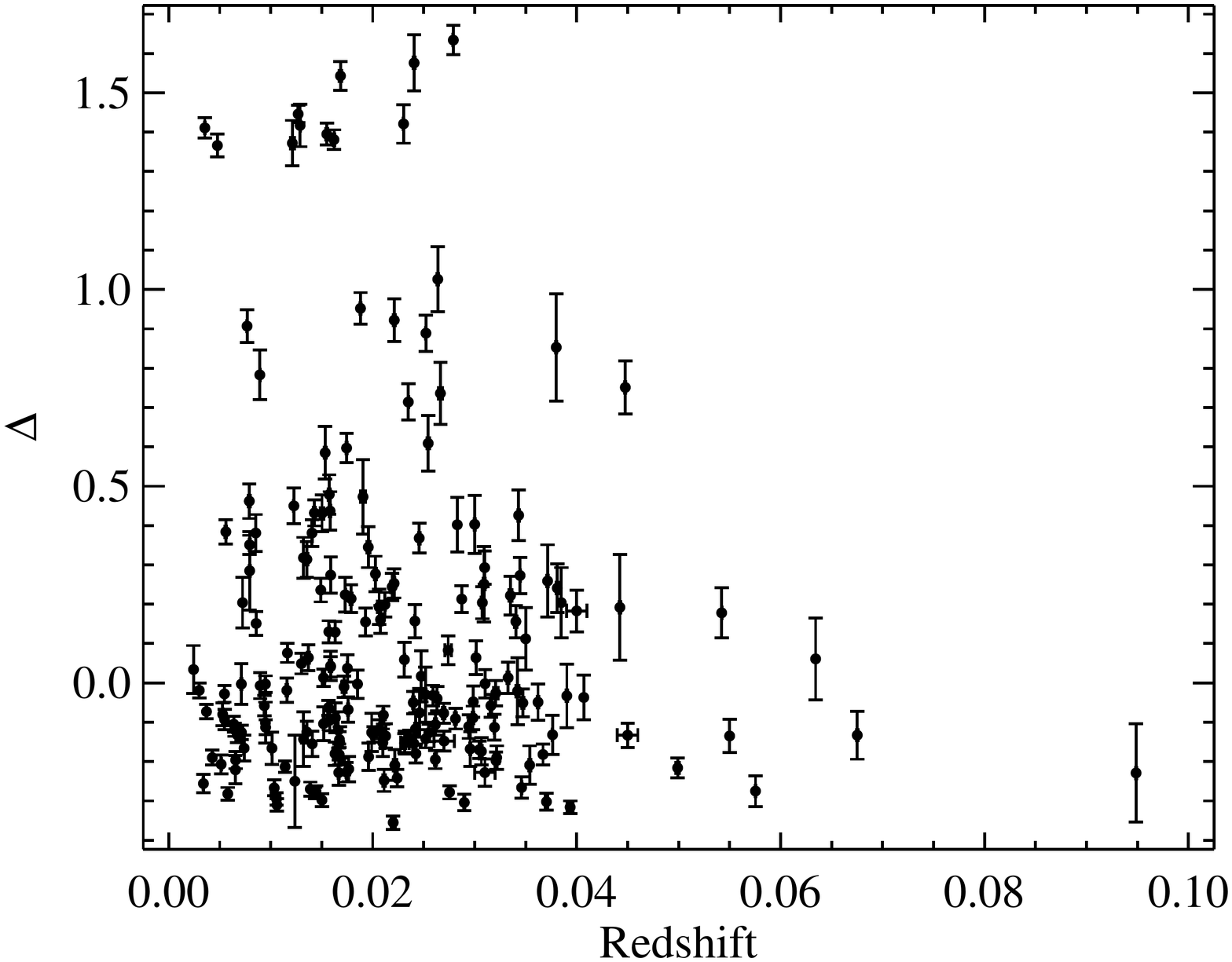} &
\includegraphics[width=3.5in]{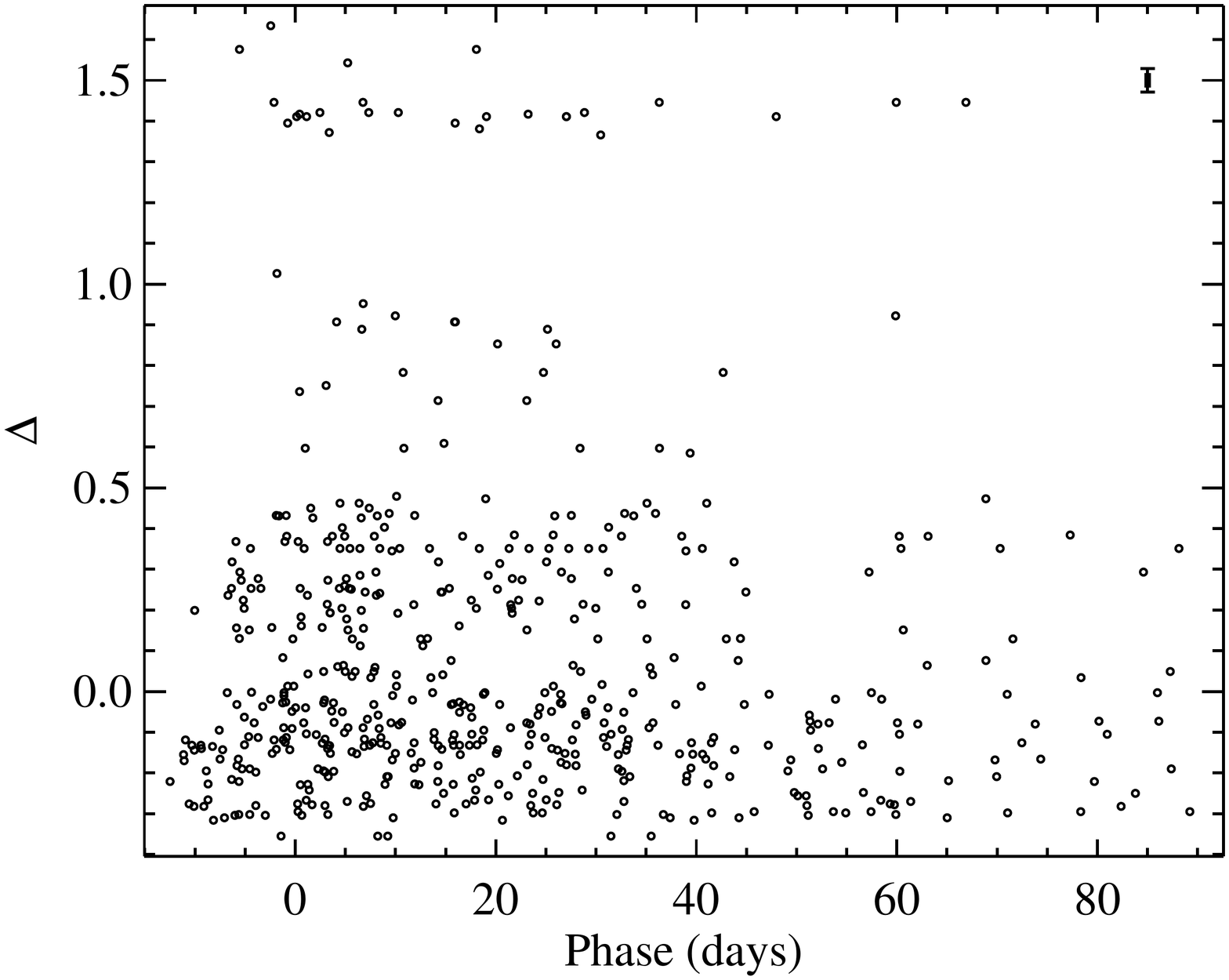} \\
\end{array}$
\caption[Histogram of the MLCS2k2 $\Delta$ parameter]{A histogram of
  the MLCS2k2 $\Delta$ value (which is a 
  measurement of the light-curve shape) of each SN~Ia (for
  which we have light-curve shape information, {\it top left}). Our average $\Delta$
  is about 0.12 and our median $\Delta$ and uncertainty is about
  $-0.03$ and 0.035. The different
  shadings correspond to each object's final SNID classification
  (Section~\ref{sss:final_class}). The
  host-galaxy redshift versus the MLCS2k2 $\Delta$ value of 
  each SN~Ia (for which we have light-curve shape
  information, {\it bottom left}). The phase versus the
  MLCS2k2 $\Delta$ value of each spectrum 
  for which we have light-curve shape information ({\it top right}) and a zoom-in
  on those spectra with $t \le 90$~d ({\it bottom right}).  The median error bar
  in both directions is shown in the top-right corner of each of these
  two panels.}\label{f:delta_hist}
\end{figure*}

As mentioned above, many of our SNe have light-curve shape
information \citep[as characterised here by the MLCS2k2 $\Delta$ 
parameter;][]{Jha07}; the distribution of $\Delta$ values for all of
these objects is shown in the top-left panel of
Figure~\ref{f:delta_hist}.  In the histogram 
we also denote each object's final SNID classification
(Section~\ref{sss:final_class}).  The $\Delta$ values come from
previously published photometric data which have all been compiled and
fit by Ganeshalingam \etal (in prep.).  The average $\Delta$ for
our dataset is \about0.12, and the median $\Delta$ and uncertainty are 
about $-0.03$ and 0.035, respectively. Our SNe~Ia span most of the
standard range of 
$\Delta$ values \citep[about $-0.4$ to 1.6, e.g.,][]{Hicken09}.

The bottom-left panel of Figure~\ref{f:delta_hist} presents the
host-galaxy redshift of 
our SNe~Ia versus their $\Delta$ values.  Our dataset covers most of
the range of $\Delta$ values at the lowest redshifts, but our coverage
decreases at higher redshifts.  SNe~Ia with large $\Delta$ are the
fainter, faster-evolving objects \citep{Jha07}, and thus we have fewer
of those (relative to smaller $\Delta$ objects) in our sample at
higher redshifts. The right panels of Figure~\ref{f:delta_hist}
present the phase of 
each spectrum versus their $\Delta$ values.  Analogous to
the bottom-left panel of the same figure, our sample spans most of the
standard range of $\Delta$ values at early times, but the coverage
begins to drop off at \about40~d past maximum brightness.  Once again, 
we lack large-$\Delta$ objects at the latest times, while we still have
a handful of SNe~Ia with small $\Delta$ values at these epochs.  This
is unsurprising since objects with large $\Delta$ values are fainter
and have faster-evolving light curves, and thus we are not able to follow
them spectroscopically for as long as their small-$\Delta$ brethren.
It is interesting to note that there are relatively few SNe~Ia with
$0.5 \le \Delta \le 0.7$ or $1.1 \le \Delta \le 1.3$; these possible
anomalies will be investigated further in future BSNIP papers.



\subsection{Object (Re-)Classification}\label{ss:reclass}

Some of the SNe~Ia presented here were originally classified as other
types of SNe or remained unclassified prior to this work.  Using our
SNID classification scheme we have reclassified these objects as bona
fide SNe~Ia.  Similarly, there are a few objects in our dataset that,
again after applying our spectral classification procedure, were
found to be examples of some of the peculiar SN~Ia subtypes. All of
the objects for which (re-)classifications were made are described
below and plots of their spectra compared to their best-matching SNID
template can be found online (see the Supporting Information).

\subsubsection{SN~1991O}
This SN was discovered on 1991~Mar.~18, by \citet{SN1991O_disc}, and
classified as a SN~Ia ``about 1--2~months past maximum brightness''
\citep{SN1991O_disc}.  Our SNID classification reveals that SN~1991O
is 91bg-like, most similar to SN~2006em \about21~d past maximum
brightness.


\subsubsection{SN~1993aa}
This SN was discovered on 1993~Sep.~19, by \citet{SN1993aa_disc}, and
classified as a SN~Ia ``probably about 1~month past maximum
brightness'' \citep{SN1993aa_disc}.  Our SNID classification reveals
that SN~1993aa is also 91bg-like, most similar to SN~2007ba \about8~d
past maximum brightness.


\subsubsection{SN~1998cm}
This SN was discovered on 1998~Jun.~10, by \citet{SN1998cm_disc}, and
classified as a SN~Ia ``within a week or two of maximum brightness''
\citep{SN1998cm_disc}.  Our SNID classification reveals that SN~1998cm
is 91T-like, most similar to SN~1997br \about8~d past maximum
brightness. 


\subsubsection{SN~2000J}

This SN was discovered on 2000~Feb.~4, by \citet{SN2000J_disc}. Nearly
6 weeks later it was classified as a SN~II based on the noisy spectrum
presented in this work \citep{SN2000J_type}. However, a SNID fit to
the same spectrum reveals that it is more likely a normal
SN~Ia, most similar to SN~1994D \about54~d past maximum brightness. 


\subsubsection{SN~2001es}

This SN was discovered on 2001~Oct.~7, by \citet{SN2001es_disc}, but
it has remained unclassified until now.  From a SNID fit to one of the
spectra presented in this work, we determine that it is likely a
normal SN~Ia, most similar to SN~2004fz \about22~d past maximum
brightness.  


\subsubsection{SN~2002bp}

This SN was discovered on 2002~Mar.~8, by \citet{SN2002bp_disc}, but
it too has remained unclassified until now.  From a SNID fit to our
spectrum presented here, we determine that it is a SN~2002cx-like
object.  Upon further inspection, it seems to be a better match to the
quite peculiar SN~2008ha \citep{Foley09:08ha} than to the more ``normal''
members of the SN~2002cx-like class \citep[e.g.,][]{Jha06:02cx}.  
We include SNe~2002bp and 2008ha here in our SN~Ia sample even though
there is uncertainty regarding whether SN~2008ha was in fact a SN~Ia
\citep[e.g.,][]{Valenti09}. 


\subsubsection{SN~2004br}
This SN was discovered on 2004~May~15, by \citet{SN2004br_disc}, and
classified as ``an unusual type~Ia supernova, similar to the spectrum
of SN~2000cx'' \citep{SN2004br_type}.  The SNID classification of our
earliest spectrum of SN~2004br reveals that it is likely
99aa-like, most similar to SN~2008ds \about5~d before maximum
brightness. 
We are unable to confidently determine the subtype of the two older
(both $> 2$~weeks past maximum brightness) spectra of SN~2004br in our
dataset. The best-matching template of one of these other spectra is
another 99aa-like SN, and the best-matching template of the other is a
91T-like SN.  Furthermore, Ganeshalingam \etal (in prep.) have 
determined that the MLCS2k2 $\Delta$ parameter of SN~2004br is
$-0.152$.  This spectral and photometric information increases our
confidence in the 99aa-like classification of SN~2004br, though it is
still uncertain.


\subsubsection{SN~2005dh}
This SN was discovered on 2005~Aug.~10, by \citet{SN2005dh_disc}, and
classified as a SN~Ia ``near maximum light''
\citep{SN2005dh_type}. The SNID classifications of both spectra of
SN~2005dh in our sample reveal it to be 91bg-like with the earlier
spectrum being most similar to SN~2006cs \about2~d past maximum
brightness. 
 This object
was specifically noted to have an ``unusually high'' expansion
velocity of 16,000--16,600~\kms\ \citep[based on the minimum of the
\ion{Si}{II} $\lambda$6355 absorption
feature;][]{SN2005dh_type,SN2005dh_type2}.  However, both
\citet{SN2005dh_type} and \citet{SN2005dh_type2} used the host-galaxy
redshift presented by \citet{Falco99}, $z=0.038$, as opposed to the
actual host-galaxy redshift of $z=0.015$ \citep{Adelman08}.  Using the
correct redshift, we calculate a relatively normal expansion velocity
of \about9300~\kms\ for our spectrum of SN~2005dh from a similar
epoch.


\subsubsection{SN~2008Z}
This SN was discovered on 2008~Feb.~7, by \citet{SN2008Z_disc}, and
classified as a Type~Ia \citep{SN2008Z_type}.  The SNID classification
of our earliest spectrum of SN~2008Z reveals that it is
99aa-like, most similar to SN~2008ds at maximum brightness.  
Ganeshalingam \etal (in
prep.) have determined that the MLCS2k2 $\Delta$ parameter of
SN~2008Z is $-0.152$.


\subsubsection{SN~2008ai}
This SN was discovered on 2008~Feb.~13, by \citet{SN2008ai_disc}, and
classified as a Type~Ia \citep{SN2008ai_type} using the earliest
spectrum of this object presented here.  Our SNID
classification of this same spectrum reveals that it is actually 
91bg-like, most similar to SN~2007ba \about5~d past maximum
brightness. 


\subsection{New Redshifts for Individual Objects}\label{ss:redshifts}

Some of the objects in our dataset do not have published spectroscopic
host-galaxy redshifts. Therefore, we have obtained host-galaxy spectra
of several SNe presented in this work in order to determine their
redshift. Furthermore, we have calculated the host-galaxy redshift of
one of these objects with no published redshift based on narrow
features present in our SN spectra. The SNe for which this was done,
their host galaxies, the redshifts themselves, and basic information
about the spectrum from which the redshift was determined can be found
in Table~\ref{t:redshifts}. These redshifts are also listed in
Table~\ref{t:obj}. 

\begin{table*}
\begin{center}
\caption{Previously Unpublished Spectroscopic Host-Galaxy Redshifts}\label{t:redshifts}
\begin{tabular}{lcrccc}
\hline\hline
SN Name & Host & \multicolumn{1}{c}{$cz_\textrm{helio}$} & UT Date & SN/Gal$^\textrm{c}$  & Abs/Emis$^\textrm{d}$ \\
     & Galaxy & \multicolumn{1}{c}{(\kms)$^\textrm{a}$} & of Spectrum$^\textrm{b}$ &  & \\
\hline




SN~2003ah & LOTOSS J044309.01+004553.4 & 10153 (3) & 2008-12-28 & Gal & Emis \\ 
SN~2006mp & MCG~+08-31-29 & 8090 (300) & 2006-11-03 & Gal & Emis \\ 
SN~2008s3$^\textrm{e}$ & 2MASX~J23004648+0734590 & 12300 (300) & 2008-09-08 & Gal & Abs \\ 
SN~2008s5$^\textrm{f}$ & $\cdots$ & 9290 (300) & 2008-09-22 & SN & Emis \\ 
\hline \hline
\multicolumn{6}{l}{$^\textrm{a}$The redshift uncertainty is in parentheses.} \\
\multicolumn{6}{l}{$^\textrm{b}$UT date of the spectrum from which we determined the redshift.} \\
\multicolumn{6}{p{4.8in}}{$^\textrm{c}$ ``Gal'' = Spectrum from which
  we determined the redshift was of the host galaxy itself; ``SN'' = Spectrum from which we determined the redshift was of the SN but contained narrow host-galaxy spectral features.} \\
\multicolumn{6}{p{4.8in}}{$^\textrm{d}$ ``Emis'' = Emission features
  were used to determine the redshift; ``Abs'' = Absorption features were used to determine the redshift.} \\
\multicolumn{6}{l}{$^\textrm{e}$Also known as SNF20080825-006.} \\
\multicolumn{6}{l}{$^\textrm{f}$Also known as SNF20080909-030.} \\
\hline\hline
\end{tabular}
\end{center}
\end{table*}
\normalsize

All of the spectra referred to in Table~\ref{t:redshifts} were
obtained from either Lick or Keck Observatory with the exception of
the spectrum of the host of SN~2003ah. On 2008~Dec.~28, we obtained a
900~s medium-resolution spectrum of the host galaxy of SN~2003ah with
the MagE spectrograph \citep{Marshall08} on the Magellan Clay 6.5~m
telescope. Data reduction was similar to the process described in
Section~\ref{s:data} with the exception of sky subtraction. For this
spectrum, the sky was subtracted from the images using the method
described by \citet{Kelson03}. Further details of MagE data reduction
are described in \citet{Foley09:08ha}. 

Two other objects in the BSNIP sample which lack published
spectroscopic host-galaxy redshifts also deserve special
mention. SN~2001ei was discovered in a faint host
(LOTOSS~J235102.95+271050.6) with no known redshift
\citep{SN2001ei_disc}; however it is likely within the Abell~2666
galaxy cluster ($cz \approx 8040$~\kms), possibly associated with
NGC~7768 ($cz \approx 8190$~\kms). These redshifts are slightly lower
than the SNID-determined redshifts for our two spectra of SN~2001ei (as
listed in Table~\ref{t:snid}). SN~2004fy was discovered in
MCG~+15-1-10, which is probably interacting with NGC~3172 ($cz \approx
6100$~\kms). This matches well with the SNID-determined redshift for
one of our spectra of SN~2004fy, though it is somewhat larger than
that of the other spectrum. Due to the uncertainty of both of these
objects' host redshifts, they are not listed in Table~\ref{t:obj}.


\section{Conclusion}\label{s:conclusions}

In this first BSNIP paper we presented a large, homogeneous set
of low-redshift ($z \le 0.2$) optical spectra of SNe~Ia. \filtspec\
spectra of \filtobj\ 
SNe have well-calibrated light curves with measured distance moduli,
and many of the spectra have had host-galaxy corrections
applied. We also discussed our observing 
and reduction procedures used during the two decades over which we
collected these data, as well as our ``colour matching'' method for
removing residual galaxy contamination.  Our relative
spectrophotometry was shown to be extremely accurate for the vast
majority of our dataset. How the data are currently stored and will
eventually be made accessible to the astronomical community was also
discussed. 

In addition, we described the construction of our own set of SNID
spectral templates as well as our classification scheme which utilises
these new templates.  Using our classification procedure we were able 
to classify for the first time (as well as reclassify) a handful of
objects as bona fide SNe~Ia.  Furthermore, we presented classifications
of objects as members of some of the peculiar SN~Ia subtypes that were
heretofore assumed to be ``normal.''  In total our dataset includes
spectra of nearly 90 spectroscopically peculiar SNe~Ia. We also
determined spectroscopic host-galaxy redshifts of some objects where
these values were previously unknown.

The sheer size of the BSNIP sample and the consistency of our
observation and reduction methods makes this sample unique among all
other published SN~Ia datasets.  In future BSNIP papers we will use
these data to examine the relationships between spectroscopic
characteristics and other observables (such as photometric and
host-galaxy properties).

Our sample is also a preview of coming attractions; new large
transient searches such as the Palomar Transient Factory
\citep[PTF;][]{Rau09,Law09} and Pan-STARRS \citep{Kaiser02} will
compile datasets similar in size to ours in just a few years.

\section*{Acknowledgments}

We would like to thank 
K.~Alatalo, L.~Armus, M.~Baker, M.~Bentz, E.~Berger, M.~Bershady,
A.~Blum, A.~Burgasser, N.~Butler, G.~Canalizo, H.~Chen, M.~Cooper,
C.~DeBreuck, M.~Dickinson, R.~Eastman, M.~Eracleous, S.~Faber, X.~Fan,
C.~Fassnacht, P.~Garnavich, M.~George, D.~Gilbank,  A.~Gilbert,
K.~Glazebrook, J.~Graham, G.~Graves, R.~Green, J.~Greene, M.~Gregg,
M.~Hidas, K.~Hiner, W.~Ho, J.~Hoffman, I.~Hook, D.~Hutchings,
V.~Junkkarinen, L.~Kewley, R.~Kirshner, D.~Kocevski, S.~Kulkarni,
M.~Lehnert, B.~Leibundgut, M.~Malkan, A.~Martel, M.~McCourt,
A.~Miller, E.~Moran, P.~Nandra, J.~Newman, K.~Noeske, C.~Papovich,
C.~Peng, S.~Perlmutter, M.~Phillips, D.~Pooley, H.~Pugh, E.~Quataert,
M.~Rich, M.~Richmond, A.~Riess, S.~Rodney, K.~Sandstrom, W.~Sargent,
K.~Shimasaki, R.~Simcoe, T.~Small, G.~Smith, H.~Smith, H.~Spinrad,
G.~Squires, C.~Steidel, D.~Stern, D.~Stevens, R.~Street, C.~Thornton,
T.~Treu, B.~Tucker, D.~Tytler, W.~van~Breugel, V.~Virgilio,
V.~Viscomi, N.~Vogt, J.~Walsh, D.~Weisz, C.~Willmer, A.~Wolfe, and
J.-H.~Woo  
for their assistance with some of the observations over the last two
decades. We would also like to thank
J.~Choi, M.~Ellison, L.~Jewett, A.~Morton, X.~Parisky, and P.~Thrasher 
for helping to verify some of the information in the SNDB.
Moreover, we thank the referee for comments and suggestions that
improved the manuscript.
We are grateful to the staffs at the Lick and Keck
Observatories for their support. Some of the data presented herein
were obtained at the W. M. Keck Observatory, which is operated as a
scientific partnership among the California Institute of Technology,
the University of California, and the National Aeronautics and Space
Administration (NASA); the observatory was made possible by the
generous financial support of the W. M. Keck Foundation. The authors
wish to recognise and acknowledge the very significant cultural role
and reverence that the summit of Mauna Kea has always had within the
indigenous Hawaiian community; we are most fortunate to have the
opportunity to conduct observations from this mountain. This research
has made use of the NASA/IPAC Extragalactic Data base (NED) which is
operated by the Jet Propulsion Laboratory, California Institute of
Technology, under contract with NASA.
A.V.F.'s group is supported by the NSF grant AST-0908886, DOE grants
DE-FC02-06ER41453 (SciDAC) and DE-FG02-08ER41563, and the TABASGO
Foundation. 
M.M. acknowledges support from Hubble Fellowship grant
HST-HF-51277.01-A, awarded by STScI, which is operated by AURA under
NASA contract NAS5-26555, for the time during which some of this work
was conducted.
KAIT and its ongoing operation were made possible by donations from
Sun Microsystems, Inc., the Hewlett-Packard Company, AutoScope
Corporation, Lick Observatory, the NSF, the University of California,
the Sylvia \& Jim Katzman Foundation and the TABASGO Foundation.
We would like to dedicate this paper to the memory of Marc~J.~Staley,
who never stopped asking the Great Questions.





\bibliographystyle{mn2e}

\bibliography{astro_refs}

\label{lastpage}

\end{document}